\documentclass[]{jfm}
\usepackage{graphicx}
\usepackage{newtxtext}
\usepackage{newtxmath}
\usepackage{natbib}
\usepackage{hyperref}
\usepackage{upgreek}

\hypersetup{
    colorlinks = true,
    urlcolor   = blue,
    citecolor  = black,
}

\newcommand{\RomanNumeralCaps}[1]

\title{Centrifugal instability of Taylor-Couette flow in stratified and diffusive fluids}

\author{Junho Park
  \corresp{\email{junho.park@coventry.ac.uk}}  
 }

\affiliation{Centre for Fluid and Complex Systems, Coventry University, Coventry CV1 5FB, UK}

\begin{document}

\maketitle

\newcommand{\ditto}{\textquotesingle\textquotesingle}

\begin{abstract}
The linear and non-linear dynamics of centrifugal instability in Taylor-Couette flow are investigated when fluids are stably stratified and highly diffusive. 
One-dimensional local linear stability analysis (LSA) on cylindrical Couette flow confirms that the stabilising role of stratification on centrifugal instability is suppressed by strong thermal diffusion (i.e. low Prandtl number $Pr$).
For $Pr\ll1$, it is verified that the instability dependence on thermal diffusion and stratification with the non-dimensional Brunt-V\"ais\"al\"a frequency $N$ can be prescribed by a single rescaled parameter $P_{N}=N^{2}Pr$.
From direct numerical simulation (DNS), various non-linear features such as axisymmetric Taylor vortices at saturation, secondary instability leading to non-axisymmetric patterns or transition to chaotic states are investigated for various values of $Pr\leq1$ and the Reynolds number $Re_{i}$.
Two-dimensional bi-global LSA on axisymmetric Taylor vortices, which appear as primary centrifugal instability saturates nonlinearly, is also performed to find the secondary critical Reynolds number $Re_{i,2}$ at which the Taylor vortices become unstable by non-axisymmetric perturbation. 
The bi-global LSA reveals that $Re_{i,2}$ increases (i.e. the onset of secondary instability is delayed) in the range $10^{-3}<Pr<1$ at $N=1$ or as $N$ increases at $Pr=0.01$. 
Secondary instability leading to highly non-axisymmetric or irregular chaotic patterns is further investigated by the 3D DNS. 
The Nusselt number $Nu$ is also computed from the torque at the inner cylinder for various $Pr$ and $Re_{i}$ at $N=1$ to describe how the angular momentum transfer increases with $Re_{i}$ and how $Nu$ varies differently for saturated and chaotic states.
\end{abstract}



\section{Introduction}
\label{sec:intro}
Thermal diffusion in fluid flows is characterised by the (thermal) Prandtl number $Pr=\nu_{0}/\kappa_{0}$, which denotes the ratio between kinematic viscosity $\nu_{0}$ and thermal diffusivity $\kappa_{0}$. 
The Prandtl number $Pr$ varies from fluids; for instance, we typically consider $\Pran\sim O(10^{3})$ for glycerol, $\Pran\simeq7$ for water, $\Pran\simeq0.7$ for the air, $\Pran\sim O(10^{-1})$ in the Earth's liquid outer core, $\Pran\sim O(10^{-2})$ for liquid metals, $\Pran\sim O(10^{-6})$ or less in the interior of the Sun and stars \citep[][]{Calkins2012,Horn2013,Legaspi2020,Garaud2020}. 
The Prandtl-number dependence has been examined for various flows coupled with heat transfer. 
In convection, \citet{Kerr2000} studied a thermal convection problem with no-slip vertical boundaries and free-slip lateral boundaries and they proposed different scaling laws for the Nusselt number $Nu$ versus the Rayleigh number $Ra$ transitions as $Nu\sim Ra^{2/7}$ for $Pr\sim O(1)$ and $Nu\sim Ra^{1/4}$ for low $Pr\ll 1$.  
\citet{Miquel2020} investigated the convection in a different configuration with internal sources and sinks and they proposed a different scaling law as $Nu\sim Ra^{1/2}Pr^{\chi}$ where the exponent $\chi$ transitions from $\chi\simeq1/2$ for $Pr\leq0.04$ to $\chi\simeq1/6$ for $Pr>0.04$. 
The Prandtl-number dependence has also been explored in studies of stratified flows.
For instance, it is found that turbulence in stably stratified fluids permits high-wavenumber temperature fluctuation that leads to the appearance of small-scale turbulence structures as $Pr$ increases when $Pr>1$ \citep[][]{Legaspi2020}.
The effect of thermal diffusion is examined in the study of exact coherent structures in stratified plane Couette flow in the limits either $Pr\rightarrow0$ or $Pr\rightarrow\infty$, the former in which the thermal diffusion is dominant over stratification and the density variation away from a linear profile of the stratification vanishes and the latter in which the density can be mixed and homogenised by advection and its stratification profile deviates from a linear profile \citep[][]{Langham2020}. 

\begin{figure}
  \centerline{
  \includegraphics[height=3cm]{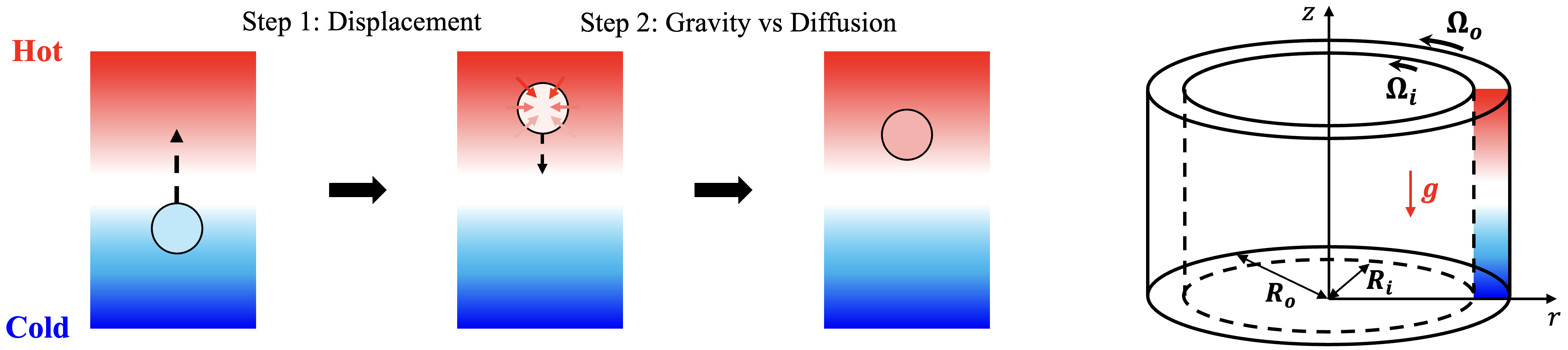}
  }
  \caption{
  (Left) Illustration on how the internal oscillation of a fluid parcel in stably stratified fluid is suppressed by a fast thermal diffusion process. \\
  (Right) Schematic of Taylor-Couette flow with stable temperature stratification.
  }
\label{fig:illust}
\end{figure}
Fluid flows with the stratification and strong thermal diffusion with $Pr\ll1$ have been the subject of interest in geophysics and astrophysics due to their relevance to the interior of the Earth and stars including the Sun \citep[][]{Lignieres1999SPA,Calkins2012,Garaud2020}.
Figure \ref{fig:illust} illustrates a configuration in which a fluid parcel is hypothetically displaced upwards over a length scale $L_{0}$ in a stably stratified fluid with the temperature gradient along the vertical direction of gravity. 
The temperature of the parcel is lower than the surrounding temperature so the parcel absorbs heat at a rate with a characteristic diffusion time scale $\tau_{\mathrm{diff}}=L_{0}^{2}/\kappa_{0}$.
Simultaneously, the stratification acts as a restoring force on the parcel leading to the internal oscillation with the period $\tau_{\mathrm{int}}=1/N_{0}$ where $N_{0}$ is the dimensional Brunt-V\"ais\"al\"a frequency derived from the stratification.
On the one hand, if the thermal diffusivity is low as $\kappa_{0}\ll N_{0}L_{0}^{2}$ (i.e. the diffusion time scale is much larger than the internal oscillation period as $\tau_{\mathrm{diff}}\gg\tau_{\mathrm{int}}$), the restoring force is dominant and the fluid parcel descends to the original position due to the gravitational force, a well-known mechanism for the generation of intertal gravity waves.   
On the other hand, if the thermal diffusivity is high as $\kappa_{0}\gg N_{0}L_{0}^{2}$ (i.e. $\tau_{\mathrm{diff}}\ll\tau_{\mathrm{int}}$), the gravitational force will be weakened due to the rapidly increased temperature of the fluid parcel. 
In this case, the parcel descends and stops at a position higher than the original position (Figure \ref{fig:illust} left).
This implies that strong thermal diffusion can suppress the internal oscillation and the effect of stratification suppressing the vertical fluid motion becomes weak as a consequence.

Such a stratification effect affected by strong thermal diffusion has been considered in prescribing shear instability-driven turbulence and associated angular momentum transport in stellar radiation zones where fluids are stably stratified and the Prandtl number $Pr$ is low.
For instance, for low P\'eclet number $Pe=RePr$ where $Re$ is the Reynolds number, \citet{Lignieres1999SPA} proposed a small-P\'eclet-number approximation that provides an asymptotic form of the Boussinesq equations in the small-$Pe$ limit where thermal diffusion and stratification can be combined and described as a single physical process.   
For vertical shear instability in stratified and highly-diffusive fluids, \citet{Lignieres1999VSI} revealed that instability characteristics for various $N_{0}$ and $Pr\ll1$ can be expressed in terms of a single rescaled number $R_{P}=RiPe$ where $Ri=N_{0}^{2}/S_{0}^{2}$ is the Richardson number with a characteristic shear $S_{0}$. 
Instead of the classical Richardson number criterion for shear instability in stratified flow: $Ri<1/4$, the modified Richardson number criterion $R_{P}<O(1)$ is adopted in the prescription of turbulent effective viscosity in the low $Pe$-limit for simulations of the evolution of stars \citep[e.g.][]{Zahn1992,Mathis2018,Prat2021}. 
For horizontal shear flows in stratified-rotating fluids, \citet{Park2020AA,Park2021AA} reported similar dynamical behaviours such as the stratification effect suppressed by high thermal diffusivity or instability characteristics described by the rescaled parameter $R_{P}$ for shear instabilities in the low-$Pe$ limit. 
The small-P\'eclet-number approximation with the rescaled parameter $R_{P}$ is also used in turbulence simulations of horizontal and vertical shear flows in stratified and highly diffusive fluids \citep[][]{Prat2013,Prat2014} and more attention has been paid recently to the low-$Pr$ and low-$Pe$ regimes in stratified turbulence \citep[][]{Cope2020,Garaud2020,Chang2021,Garaud2024}.
It is noteworthy that P\'eclet numbers of actual astrophysical systems are too high \citep[e.g. $Pe\sim O\left(10^{7}\right)$ with $Re\sim O\left(10^{13}\right)$ and $Pr\sim O\left(10^{-6}\right)$ for the solar tachocline; see also,][]{Garaud2020b} and fully-resolved astrophysical turbulence for such high $Re$ and $Pe$ is not achievable yet from any state-of-the-art simulations.
The current study also cannot address high-$Re$/high-$Pe$ turbulence but still aims to explain the role of strong thermal diffusion at low $Pr$. 

Despite the increasing interest in the effect of strong thermal diffusion on stratified shear flows, the Prandtl-number dependence has not been studied considerably in the context of Taylor-Couette flow, a canonical shear flow between two concentric cylinders rotating independently, in a stably stratified fluid (Figure \ref{fig:illust} right).
Stratification in Taylor-Couette flow tends to suppress the vertical fluid motion and, as a result, the onset of centrifugal instability delays and the vertical length scale of axisymmetric Taylor vortices is reduced as the stratification becomes strong \citep[][]{Withjack1975,Boubnov1995,Caton2000}. 
An interesting phenomenon in stratified Taylor-Couette flow is non-axisymmetric strato-rotational instability (SRI), which occurs due to the resonance between inertia-gravity waves confined between the two cylinders \citep[][]{Molemaker2001,Yavneh2001}.
The SRI has been explored extensively by theoretical investigations \citep[e.g.][]{Rudiger2009,LeDizes2010,Park2013JFM,Leclercq2016,Wang2018,Robins2020}, experiments \citep[e.g.][]{LeBars2007,Ibanez2016,Rudiger2017,Seelig2018,Park2018}, and numerical simulations \citep[e.g.][]{Lopez2020JFM,Meletti2021,Lopez2022}. 
These instability studies considered the Prandtl number of order $O(1)$ for fluids like water or the air or simply $Pr=\infty$ to neglect the effect of thermal diffusion in theoretical analyses, or they considered the analogous Schmidt number $Sc=\nu_{0}/D_{0}$ (where $D_{0}$ is a diffusivity of scalars like density, salinity, etc.) of order $O(100)$ if stratification of water by salt is considered in experiments.
Investigating the dynamics of stratified Taylor-Couette flow under the influence of thermal diffusion, in particular in the low-$Pr$ limit, is important as the results can further be used to provide insights into large-scale flows and propose turbulent viscosity models in multi-physics simulations of astrophysical and geophysical systems such as simulations of the evolution of stars \citep[][]{Richard1999,Dubrulle2005}.
However, Taylor-Couette flow in stratified and highly diffusive fluids is still poorly understood and this motivates our study on stratified Taylor-Couette flow and firstly its centrifugal instability under the influence of strong thermal diffusion at low Prandtl number $Pr\leq1$.

The paper consists of the following sections to investigate the centrifugal instability of Taylor-Couette flow in stratified and diffusive fluids. 
In \S\ref{sec:formulation}, problem formulation and details on numerical methods are provided.  
In \S\ref{sec:LSA}, 1D local linear stability analysis (LSA) is performed to present LSA results on centrifugal instability of cylindrical Couette flow for various parameters. 
In \S\ref{sec:NCI}, 2D and 3D direct numerical simulations are conducted to investigate the non-linear dynamics of centrifugal instability such as saturation, secondary instability or transition to chaotic states. 
In \S\ref{sec:conclusion}, conclusion and discussion are provided. 

\section{Problem formulation and methodology}
\label{sec:formulation}
\subsection{Navier-Stokes equations under the Boussinesq approximation} 
In this study, we consider the Boussinesq approximation in which the reference density $\rho_{0}$ is assumed to be much larger than the density variation $\varrho$ (i.e. $\rho_{0}\gg\varrho$).
The density variation $\varrho$ is assumed to satisfy a linear relation with the total temperature $\vartheta$ as $\varrho/\rho_{0}=-\alpha_{0} (\vartheta-\vartheta_{0})$ where $\alpha_{0}>0$ is the thermal expansion coefficient and $\vartheta_{0}$ is the reference temperature. 
For velocity ${\boldsymbol{U}}=(U_{r},U_{\theta},U_{z})$, temperature variation $\Uptheta=\vartheta-\vartheta_{0}$ and associated pressure variation ${P}$ in the cylindrical coordinates $(r,\theta,z)$, we consider the following continuity, momentum, and energy equations as
\begin{equation}
\label{eq:continuity}
\nabla\cdot{\boldsymbol{U}}=0,
\end{equation}
\begin{equation}
\label{eq:momentum}
\frac{\partial{\boldsymbol{U}}}{\partial t}+{\boldsymbol{U}}\cdot\nabla{\boldsymbol{U}}=-\frac{1}{\rho_{0}}\nabla  P +\alpha_{0}g\Uptheta\vec{e}_{z}+\nu_{0}\nabla^{2}{\boldsymbol{U}},
\end{equation}
\begin{equation}
\label{eq:energy}
\frac{\partial \Uptheta}{\partial t}+{\boldsymbol{U}}\cdot\nabla \Uptheta=\kappa_{0}\nabla^{2}\Uptheta,
\end{equation}
where $g$ is the gravitional acceleration in the vertical direction $z$, $\nu_{0}$ is the kinematic viscosity, $\kappa_{0}$ is the thermal diffusivity, and $\nabla^{2}$ is the Laplacian operator. 
We consider cylindrical Couette flow in a stably stratified fluid as a base state with the base velocity $\boldsymbol{U}_{B}=\left(0,V_{B}(r),0\right)$ and base temperature $T_{B}(z)$ as
\begin{equation}
\label{eq:base}
V_{B}(r)=Ar+\frac{B}{r},~~
A=\Omega_{i}\frac{\mu-\eta^{2}}{1-\eta^{2}},~~
B=\Omega_{i}R_{i}^{2}\frac{1-\mu}{1-\eta^{2}},~~
T_{B}(z)=\frac{\Delta T}{\Delta z}z,
\end{equation}
where $V_{B}(r)$ is the base azimuthal velocity, $A$ and $B$ are the constants as a function of the angular velocities $\Omega_{i}$ and $\Omega_{o}$ and radii $R_{i}$ and $R_{o}$ (where the subscripts $i$ and $o$ denote the inner and outer cylinders, respectively), $\mu=\Omega_{o}/\Omega_{i}$ is the angular velocity ratio, $\eta=R_{i}/R_{o}$ is the radius ratio, $\Delta T$ is the temperature difference along the vertical distance $\Delta z$. 
Throughout this paper, we consider only the case in which the outer cylinder is fixed and only the inner cylinder rotates  (i.e. $\Omega_{o}=0$ and $\mu=0$). 
The temperature gradient $\Delta T/\Delta z$ is assumed to be positive and constant so the fluid is stably stratified and the base temperature $T_{B}$ increases linearly with $z$. 
The base pressure $P_{B}(r,z)$ balances the velocity $V_{B}$ and temperature $T_{B}$ by satisfying the following relations:
\begin{equation}
\label{eq:base_pressure}
-\frac{V_{B}^{2}}{r}=-\frac{1}{\rho_{0}}\frac{\partial P_{B}}{\partial r},~~
0=-\frac{1}{\rho_{0}}\frac{\partial P_{B}}{\partial z}+\alpha_{0}g T_{B}.
\end{equation}
The equations (\ref{eq:continuity})-(\ref{eq:energy}) can be expressed in a non-dimensional form by considering $R_{i}$ as the reference length, $\Omega_{i}^{-1}$ as the reference time, $R_{i}\Omega_{i}$ as the reference velocity, $\rho_{0}R_{i}^{2}\Omega_{i}^{2}$ as the reference pressure, and $\Delta T$ as the reference temperature. 
The distance $\Delta z$ can be chosen arbitrarily without loss of generality, thus we choose $\Delta z=R_{i}$ for simplicity.
The coordinates $z$ and $r$ are considered non-dimensional hereafter, hence the base temperature can simply be expressed as $T_{B}(z)=z$.  
 
We consider perturbation with its velocity $\boldsymbol{u}={\boldsymbol{U}}-\boldsymbol{U}_{B}=\left(u_{r},u_{\theta},u_{z}\right)$, pressure $p=P-P_{B}$ and temperature $T=\Uptheta-T_{B}$. 
By applying the base state and perturbation to (\ref{eq:continuity})-(\ref{eq:energy}) and subtracting the base-state equations (\ref{eq:base})-(\ref{eq:base_pressure}), we obtain the Navier-Stokes equations for perturbation as follows:
\begin{equation}
\label{eq:continuity_ptb}
\frac{\partial u_{r}}{\partial r}+\frac{u_{r}}{r}+\frac{1}{r}\frac{\partial u_{\theta}}{\partial \theta}+\frac{\partial u_{z}}{\partial z}=0,
\end{equation}
\begin{equation}
\label{eq:mom_r_ptb}
\frac{\partial u_{r}}{\partial t}+\Omega\frac{\partial u_{r}}{\partial \theta}-2\Omega u_{\theta}+{N}_{r}=-\frac{\partial p}{\partial r}+\frac{1}{Re}\left(\nabla^{2}u_{r}-\frac{u_{r}}{r^{2}}-\frac{2}{r^{2}}\frac{\partial u_{\theta}}{\partial\theta}\right),
\end{equation}
\begin{equation}
\label{eq:mom_th_ptb}
\frac{\partial u_{\theta}}{\partial t}+\Omega\frac{\partial u_{\theta}}{\partial\theta}+Zu_{r}+{N}_{\theta}=-\frac{1}{r}\frac{\partial p}{\partial\theta}+\frac{1}{Re}\left(\nabla^{2}u_{\theta}-\frac{u_{\theta}}{r^{2}}+\frac{2}{r^{2}}\frac{\partial u_{r}}{\partial\theta}\right),
\end{equation}
\begin{equation}
\label{eq:mom_z_ptb}
\frac{\partial u_{z}}{\partial t}+\Omega\frac{\partial u_{z}}{\partial\theta}+{N}_{z}=-\frac{\partial p}{\partial z}+N^{2}T+\frac{1}{Re}\nabla^{2}u_{z},
\end{equation}
\begin{equation}
\label{eq:energy_ptb}
\frac{\partial T}{\partial t}+\Omega\frac{\partial T}{\partial\theta}+u_{z}+{N}_{T}=\frac{1}{RePr}\nabla^{2}T,
\end{equation}
where $\Omega(r)=V_{B}/r$ is the base angular velocity, $Z(r)=(1/r)\mathrm{d}(r^{2}\Omega)/\mathrm{d}r$ is the base axial vorticity, $N_{r}$, $N_{\theta}$, $N_{z}$ and $N_{T}$ are the non-linear terms as
\begin{equation}
\label{eq:nonlinear_Nr}
{N}_{r}=u_{r}\frac{\partial u_{r}}{\partial r}+\frac{u_{\theta}}{r}\frac{\partial u_{r}}{\partial \theta}+u_{z}\frac{\partial u_{r}}{\partial z}-\frac{u_{\theta}^{2}}{r},
\end{equation}
\begin{equation}
\label{eq:nonlinear_Nth}
{N}_{\theta}=u_{r}\frac{\partial u_{\theta}}{\partial r}+\frac{u_{\theta}}{r}\frac{\partial u_{\theta}}{\partial \theta}+u_{z}\frac{\partial u_{\theta}}{\partial z}+\frac{u_{r}u_{\theta}}{r},
\end{equation}
\begin{equation}
\label{eq:nonlinear_Nz}
{N}_{z}=u_{r}\frac{\partial u_{z}}{\partial r}+\frac{u_{\theta}}{r}\frac{\partial u_{z}}{\partial \theta}+u_{z}\frac{\partial u_{z}}{\partial z},
\end{equation}
\begin{equation}
\label{eq:nonlinear_Nt}
{N}_{T}=u_{r}\frac{\partial T}{\partial r}+\frac{u_{\theta}}{r}\frac{\partial T}{\partial \theta}+u_{z}\frac{\partial T}{\partial z}.
\end{equation}
The parameters $Re$, $N$ and $Pr$ are the Reynolds number, non-dimensional Brunt-V\"ais\"al\"a frequency and Prandtl number, respectively, all of which are defined as
\begin{equation}
\label{eq:nond_numbers}
Re=\frac{R_{i}^{2}\Omega_{i}}{\nu_{0}},~~
N=\sqrt{\frac{\alpha_{0}g}{\Omega_{i}^{2}}\frac{\Delta T}{\Delta z}},~~
Pr=\frac{\nu_{0}}{\kappa_{0}}.
\end{equation}
For convenience in comparison with other literature, we also use a conventional inner cylinder-based Reynolds number $Re_{i}$ defined as
\begin{equation}
\label{eq:nond_numbers_Rei}
Re_{i}=\frac{R_{i}\Omega_{i}d}{\nu_{0}}=Re\left(\frac{1-\eta}{\eta}\right),
\end{equation} 
where $d=R_{o}-R_{i}$ is the gap size between the two cylinders. 
\subsection{Pseudo-spectral formulation for direct numerical simulation}
In this study, spectal-based direct numerical simulations are conducted to solve the equations (\ref{eq:continuity_ptb})-(\ref{eq:energy_ptb}) numerically.
To do so, we decompose the perturbation into the sum of modes using the following Fourier representation
\begin{equation}
\label{eq:Fourier_mode}
\left(
\begin{array}{c}
u_{r}\\
u_{\theta}\\
u_{z}\\
T\\
p
\end{array}
\right)
=
\sum_{j=-M}^{M}
\sum_{l=-K}^{K}
\left(
\begin{array}{c}
\tilde{u}_{jl}(r,t)\\
\tilde{v}_{jl}(r,t)\\
\tilde{w}_{jl}(r,t)\\
\tilde{T}_{jl}(r,t)\\
\tilde{p}_{jl}(r,t)
\end{array}
\right)\exp\left(\mathrm{i}m_{j}\theta+\mathrm{i}k_{l}z\right),
\end{equation} 
where $M$ and $K$ are the cut-off numbers of modes considered in the azimuthal direction $\theta$ and axial direction $z$, respectively, $\tilde{u}_{jl}$, $\tilde{v}_{jl}$, $\tilde{w}_{jl}$, $\tilde{p}_{jl}$ and $\tilde{T}_{jl}$ are the time-dependent mode shapes, $m_{j}=jm$ is the $j$-th multiple of the principal azimuthal wavenumber $m$ and $k_{l}=lk$ is the $l$-th multiple of the principal axial wavenumber $k$.
The above ansatz (\ref{eq:Fourier_mode}) is used for various problems such as the Taylor-Couette formulation in \textit{nsCouette} code \citep[][]{nsCouette} or convection problems \citep[][]{Saltzman1962,Park2021}. 
The formulation is also analogous to semi-linear models, which are applied to centrifugal instability of anti-cyclonic vortices \citep[][]{Yim2020,Yim2023}.
The semi-linear theory allows us to investigate directly non-linear interaction between base flow and an instability mode, and the method is generalised by considering non-linear interaction among multiple instability modes in low-order harmonics while neglecting the triad interaction leading to harmonics of orders higher than the cut-off numbers. 
For each mode with indices $j$ and $l$, we apply the ansatz (\ref{eq:Fourier_mode}) to the equations (\ref{eq:continuity_ptb})-(\ref{eq:energy_ptb}) and obtain  
\begin{equation}
\label{eq:continuity_slm}
\frac{\partial \tilde{u}_{jl}}{\partial r}+\frac{\tilde{u}_{jl}}{r}+\frac{\mathrm{i}m_{j}\tilde{v}_{jl}}{r}+\mathrm{i}k_{l}\tilde{w}_{jl}=0,
\end{equation}
\begin{equation}
\label{eq:mom_r_slm}
\frac{\partial \tilde{u}_{jl}}{\partial t}+\mathrm{i}m_{j}\Omega\tilde{u}_{jl}-2\Omega \tilde{v}_{jl}+\tilde{N}_{r,jl}=-\frac{\partial \tilde{p}_{jl}}{\partial r}+\frac{1}{Re}\left(\tilde{\nabla}_{jl}^{2}\tilde{u}_{jl}-\frac{\tilde{u}_{jl}}{r^{2}}-\frac{2\mathrm{i}m_{j}\tilde{v}_{jl}}{r^{2}}\right),
\end{equation}
\begin{equation}
\label{eq:mom_th_slm}
\frac{\partial \tilde{v}_{jl}}{\partial t}+\mathrm{i}m_{j}\Omega\tilde{v}_{jl}+Z\tilde{u}_{jl}+\tilde{N}_{\theta,jl}=-\frac{\mathrm{i}m_{j}\tilde{p}_{jl}}{r}+\frac{1}{Re}\left(\tilde{\nabla}_{jl}^{2}\tilde{v}_{jl}-\frac{\tilde{v}_{jl}}{r^{2}}+\frac{2\mathrm{i}m_{j}\tilde{u}_{jl}}{r^{2}}\right),
\end{equation}
\begin{equation}
\label{eq:mom_z_slm}
\frac{\partial \tilde{w}_{jl}}{\partial t}+\mathrm{i}m_{j}\Omega\tilde{w}_{jl}+\tilde{N}_{z,jl}=-\mathrm{i}k_{l}\tilde{p}_{jl}+N^{2}\tilde{T}_{jl}+\frac{1}{Re}\tilde{\nabla}_{jl}^{2}\tilde{w}_{jl},
\end{equation}
\begin{equation}
\label{eq:energy_slm}
\frac{\partial \tilde{T}_{jl}}{\partial t}+\mathrm{i}m_{j}\Omega\tilde{T}_{jl}+\tilde{w}_{jl}+\tilde{N}_{T,jl}=\frac{1}{RePr}\tilde{\nabla}_{jl}^{2}\tilde{T}_{jl},
\end{equation}
where $\tilde{N}_{r,jl}$, $\tilde{N}_{\theta,jl}$, $\tilde{N}_{z,jl}$ and $\tilde{N}_{T,jl}$ are the terms convoluted from the non-linear terms (\ref{eq:nonlinear_Nr})-(\ref{eq:nonlinear_Nt}) using the Fourier transform and $\tilde{\nabla}_{jl}^{2}=\partial^{2}/\partial r^{2}+(1/r)\partial/\partial r-k_{l}^{2}-m_{j}^{2}/r^{2}$ is the modal Laplacian operator. 
For the boundary conditions, we consider
\begin{equation}
\label{eq:bc_slm}
\tilde{u}_{jl}=\tilde{v}_{jl}=\tilde{w}_{jl}=\frac{\partial\tilde{T}_{jl}}{\partial r}=0,
\end{equation}
at both cylinders $r=1$ and $r=1/\eta$.

An advantage of using the equations in a modal form is that the equations (\ref{eq:continuity_slm})-(\ref{eq:energy_slm}) can further be simplified by eliminating the pressure $\tilde{p}_{jl}$ as follows:
\begin{equation}
\label{eq:evolution_slm}
\mathcal{A}_{jl}\frac{\partial\tilde{\boldsymbol{q}}_{jl}}{\partial t}=\mathcal{B}_{jl}\tilde{\boldsymbol{q}}_{jl}+\tilde{\boldsymbol{N}}_{jl},
\end{equation}
where $\tilde{\boldsymbol{q}}_{jl}$ is the vector of three variables, $\mathcal{A}_{jl}$ and $\mathcal{B}_{jl}$ are the differential operator matrices and $\tilde{\boldsymbol{N}}_{jl}$ is the vector comprised of the non-linear terms, all of which depend on indices $j$ and $l$ such as whether $l=0$ or not and whether $j=0$ or not. 
We refer to Appendix \ref{app:matrices} for more details on these vectors and matrices. 
For spatial discretisation in the radial direction $r$, a spectral method using the Chebyshev collocation points is considered \citep[][]{Antkowiak2004,Park2012}. 
The vector $\tilde{\boldsymbol{N}}_{jl}$ is computed at each step using a pseudo-spectral method, similar to \citet{Deloncle2008}.
The equation (\ref{eq:evolution_slm}) is written in a spectral form in the azimuthal and axial directions and we use the backward Fourier transform to obtain the vector $\boldsymbol{q}_{jl}$ in the physical space $(r,\theta,z)$. 
$\boldsymbol{q}_{jl}$ is then used to compute the non-linear term $\boldsymbol{N}_{jl}$ in the physical space and apply the forward Fourier transform to $\boldsymbol{N}_{jl}$ to compute $\tilde{\boldsymbol{N}}_{jl}$ in the spectral space $(r,m_{j},k_{l})$.
In this pseudo-spectral approach, we consider the numbers of collocation points $N_{\theta}=2M+1$ and $N_{z}=2K+1$ in the azimuthal and axial directions, respectively. 
The pseudo-spectral approach is similar to that in \citet{Guseva2015} who also considers the axial periodicity with the Fourier method but utilises a fractional time-stepping method by carefully computing an intermediate pressure using the influence-matrix method, which is especially effective in solving equations for the magnetic field. 
On the contrary, we avoid the use of the pressure $\tilde{p}_{jl}$ by considering different operator matrices $\mathcal{A}_{jl}$ and $\mathcal{B}_{jl}$ that depend on the indices $j$ and $l$.
Unlike \citet{Deloncle2008}, the de-alising technique, which is required in turbulence simulations, is not implemented as the Reynolds number $Re_{i}$ considered in this work is not large enough to observe small-scale turbulence.  
For the time advancement, we use a conventional semi-implicit scheme \citep[see e.g.][]{Kim1987} with the Crank-Nicolson method for the linear term and the Adam-Bashforth method for the non-linear term.
For instance, at each step $n$, we find the next step solution $\tilde{\boldsymbol{q}}_{jl}^{(n+1)}$ by solving numerically the discretised version of (\ref{eq:evolution_slm}) as
\begin{equation}
\label{eq:time_marching}
\left(\mathcal{A}_{jl}-\frac{\Delta t}{2}\mathcal{B}_{jl}\right)\tilde{\boldsymbol{q}}_{jl}^{(n+1)}=\left(\mathcal{A}_{jl}+\frac{\Delta t}{2}\mathcal{B}_{jl}\right)\tilde{\boldsymbol{q}}_{jl}^{(n)}+\frac{\Delta t}{2}\left(3\tilde{\boldsymbol{N}}_{jl}^{(n)}-\tilde{\boldsymbol{N}}_{jl}^{(n-1)}\right),
\end{equation}
where $\Delta t$ is the time step.
In this study, the time step $\Delta t$ is fixed to $\Delta t=0.01$, which is found to be sufficiently small for parameters considered in this study.
It is verified that the Courant–Friedrichs–Lewy (CFL) condition for perturbation velocity is satisfied and every DNS demonstrate no numerical divergence with this $\Delta t$.  

\subsection{One-dimensional local linear stability analysis}
\label{sec:LSA_formulation}
By assuming that the perturbation is infinitesimally small, we can neglect non-linear terms and perform 1D local linear stability analysis (LSA) using the normal mode as
\begin{equation}
\label{eq:normal_mode}
\left(
\begin{array}{c}
u_{r}\\
u_{\theta}\\
u_{z}\\
T\\
p
\end{array}
\right)
=
\left(
\begin{array}{c}
\hat{u}(r)\\
\hat{v}(r)\\
\hat{w}(r)\\
\hat{T}(r)\\
\hat{p}(r)
\end{array}
\right)\exp\left(\mathrm{i}m\theta+\mathrm{i}kz-\mathrm{i}\omega t\right)+c.c.,
\end{equation} 
where $c.c.$ denotes the complex conjugate, $\hat{u}$, $\hat{v}$, $\hat{w}$, $\hat{p}$ and $\hat{T}$ are the mode shapes, $m$ is the azimuthal wavenumber, $k$ is the axial wavenumber, and $\omega$ is the complex frequency $\omega=\omega_{r}+\mathrm{i}\omega_{i}$ where $\omega_{r}=\Real(\omega)$ is the temporal frequency and $\omega_{i}=\Imag(\omega)$ is the temporal growth rate. 
Throughout the paper, the non-dimensional wavenumber $k_{d}=kd/R_{i}=k(1-\eta)/\eta$ rescaled by the gap size $d$ is also used for convenience in comparison with other literature. 
The normal mode (\ref{eq:normal_mode}) is applied to the equations (\ref{eq:continuity_ptb})-(\ref{eq:energy_ptb}) with the non-linear terms neglected and after manipulations, we can eliminate the pressure mode shape and obtain the following simplified eigenvalue problem
\begin{equation}
\label{eq:evp}
-\mathrm{i}\omega\mathcal{A}\hat{\boldsymbol{q}}=\mathcal{B}\hat{\boldsymbol{q}},
\end{equation}
where $\hat{\boldsymbol{q}}=\left(\hat{u},\hat{v},\hat{T}\right)^{\mathrm{T}}$ and $\mathcal{A}$ and $\mathcal{B}$ are the operator matrices, which are essentially the same as $\mathcal{A}_{11}$ and $\mathcal{B}_{11}$, respectively, the operator matrices in (\ref{eq:evolution_slm}) with $j=l=1$.
To solve the eigenvalue problem (\ref{eq:evp}), a MATLAB routine eig is used with the following boundary conditions imposed at both cylinders $r=1$ and $r=1/\eta$:
\begin{equation}
\label{eq:bc_LSA}
\hat{u}=\hat{v}=\hat{w}=\frac{\mathrm{d}\hat{T}}{\mathrm{d}r}=0.
\end{equation} 
The Chebyshev spectral method is used for the discretisation in the radial direction $r$ and the number of collocation points $N_{r}$ between 60 and 120 is considered in this study.
The choice of $N_{r}$ depends on how large the parameters such as the Reynolds number $Re$ or the P\'eclet number $Pe=RePr$ are, how the eigenmodes are confined near the boundaries, etc. 
For more details on the LSA and numerical methods, we refer to our previous work \citep[e.g.][]{Park2012,Park2013JFM} that used the same code. 
\subsection{Two-dimensional bi-global linear stability analysis}
For parameters considered in this study, 1D LSA reveals that the axisymmetric mode with $m=0$ is the most unstable one for cylindrical Couette flow.
As demonstrated in the next sections, centrifugal instability develops nonlinearly and saturates leading to axisymmetric Taylor vortices as a new base state. 
The new two-dimensional base state $\bar{\boldsymbol{Q}}(r,z)=(\bar{\boldsymbol{U}},\bar{T})$ is comprised of the new base velocity $\bar{\boldsymbol{U}}(r,z)=\left(\bar{U}(r,z),\bar{V}(r,z),\bar{W}(r,z)\right)$ and the new base temperature $\bar{T}(r,z)$. 
For this Taylor vortex flow, we can analyse its secondary instability through 2D bi-global LSA by considering a non-axisymmetric perturbation $\bar{\textbf{q}}$ with $m\neq0$ expressed in the following ansatz: 
\begin{equation}
\label{eq:normal_mode_global}
\bar{\textbf{q}}=
\left(
\begin{array}{c}
\bar{u}_{r}\\
\bar{u}_{\theta}\\
\bar{u}_{z}\\
\bar{T}\\
\bar{p}
\end{array}
\right)
=
\left(
\begin{array}{c}
\hat{u}_{m}(r,z)\\
\hat{v}_{m}(r,z)\\
\hat{w}_{m}(r,z)\\
\hat{T}_{m}(r,z)\\
\hat{p}_{m}(r,z)
\end{array}
\right)\exp\left(\mathrm{i}m\theta-\mathrm{i}\omega_{m} t\right)+c.c.,
\end{equation} 
where $\hat{u}_{m}$, $\hat{v}_{m}$, $\hat{w}_{m}$, $\hat{T}_{m}$ and $\hat{p}_{m}$ are the 2D global mode shapes and $\omega_{m}$ is the complex frequency of the global mode. 
Similar to the 1D local LSA, an eigenvalue problem can be formulated as
\begin{equation}
\label{eq:evp_biglobal}
-\mathrm{i}\omega_{m}\mathcal{A}_{m}\hat{\textbf{q}}_{m}=\mathcal{B}_{m}\hat{\textbf{q}}_{m},
\end{equation}
where $\mathcal{A}_{m}$ and $\mathcal{B}_{m}$ are the operator matrices detailed in Appendix \ref{app:matrices_2dlsa} and $\hat{\textbf{q}}_{m}=\left(\hat{u}_{m},\hat{v}_{m},\hat{T}_{m}\right)^{\mathrm{T}}$ is the mode shape. 
The matrices $\mathcal{A}_{m}$ and $\mathcal{B}_{m}$ in the 2D bi-global LSA have a much larger size than the size of the matrices $\mathcal{A}$ and $\mathcal{B}$ in (\ref{eq:evp}), thus we use a MATLAB routine eigs that computes only a few eigenvalues near a specified value using the Krylov-Schur algorithm \citep[][]{Stewart2002}. 
To solve the eigenvalue problem (\ref{eq:evp_biglobal}), we consider the following boundary conditions:
\begin{equation}
\label{eq:bc_biglobal}
\hat{u}_{m}=\hat{v}_{m}=\hat{w}_{m}=\frac{\partial\hat{T}_{m}}{\partial r}=0.
\end{equation}
The Chebyshev and Fourier spectral methods are used for discretisation in the radial and axial directions, respectively. 
The Fourier method imposes the periodic boundary condition in the axial direction $z$ given that the axisymmetric Taylor vortices as a new base state are also periodic in $z$ with the wavelength $\lambda_{z}=2\pi/k$. 

\section{Linear centrifugal instability in stratified and diffusive fluids}
\label{sec:LSA}
Linear instability of stratified flows in thermally diffusive fluids, especially with low $Pr$, has been investigated for various planar shear flows in a linear profile \citep[][]{Barker2019,Barker2020,Dymott2023}, a periodic sinusoidal profile \citep[][]{Chang2021,Garaud2024} or a hyperbolic tangent profile \citep[][]{Park2020AA,Park2021AA}.
They consider flows with vertical shear, horizontal shear, or mixed shear (where the `vertical' direction in these studies implies the direction of gravity and stratification) and investigate linear and non-linear properties of shear instabilities. 
In this section, we similarly investigate linear centrifugal instability of Taylor-Couette flow, a horizontally sheared rotating flow, in stratified and diffusive fluids.
The linear analysis results will be followed by non-linear simulation results presented in the next section. 
\subsection{Neutral stability curves}
\begin{figure}
  \centerline{
  \includegraphics[height=4.5cm]{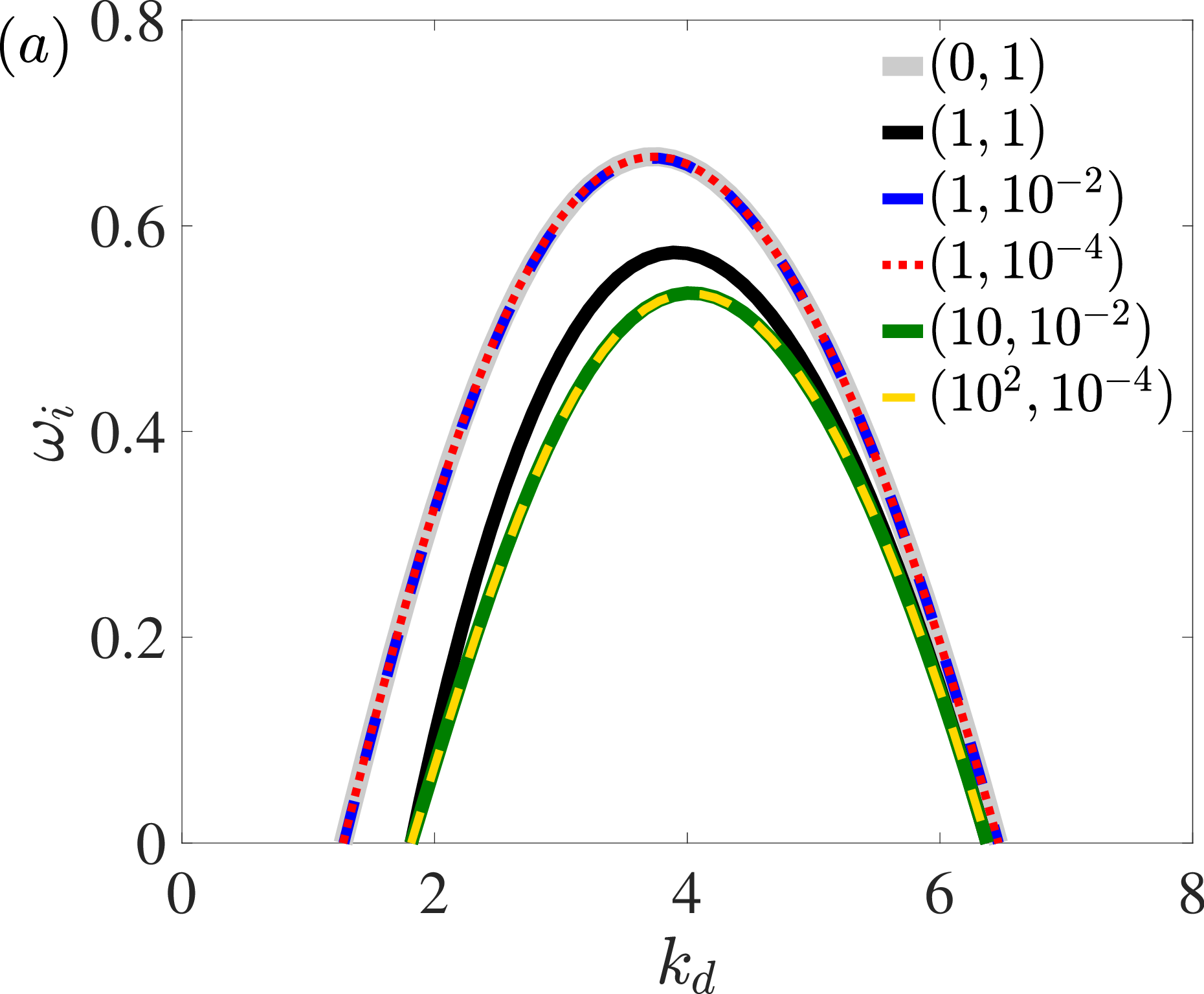}
    \includegraphics[height=4.46cm]{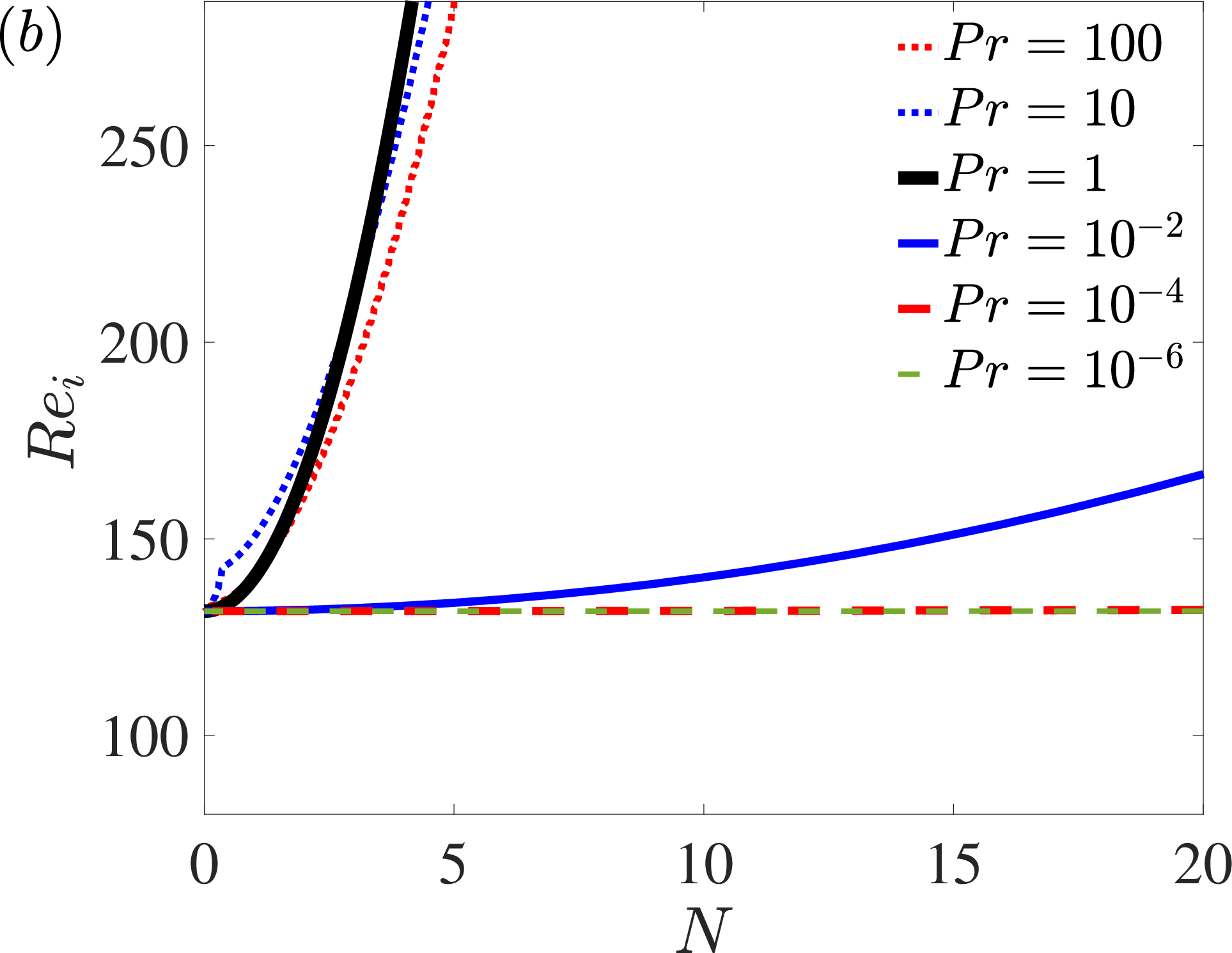}
  }
  \centerline{
  \includegraphics[height=4.5cm]{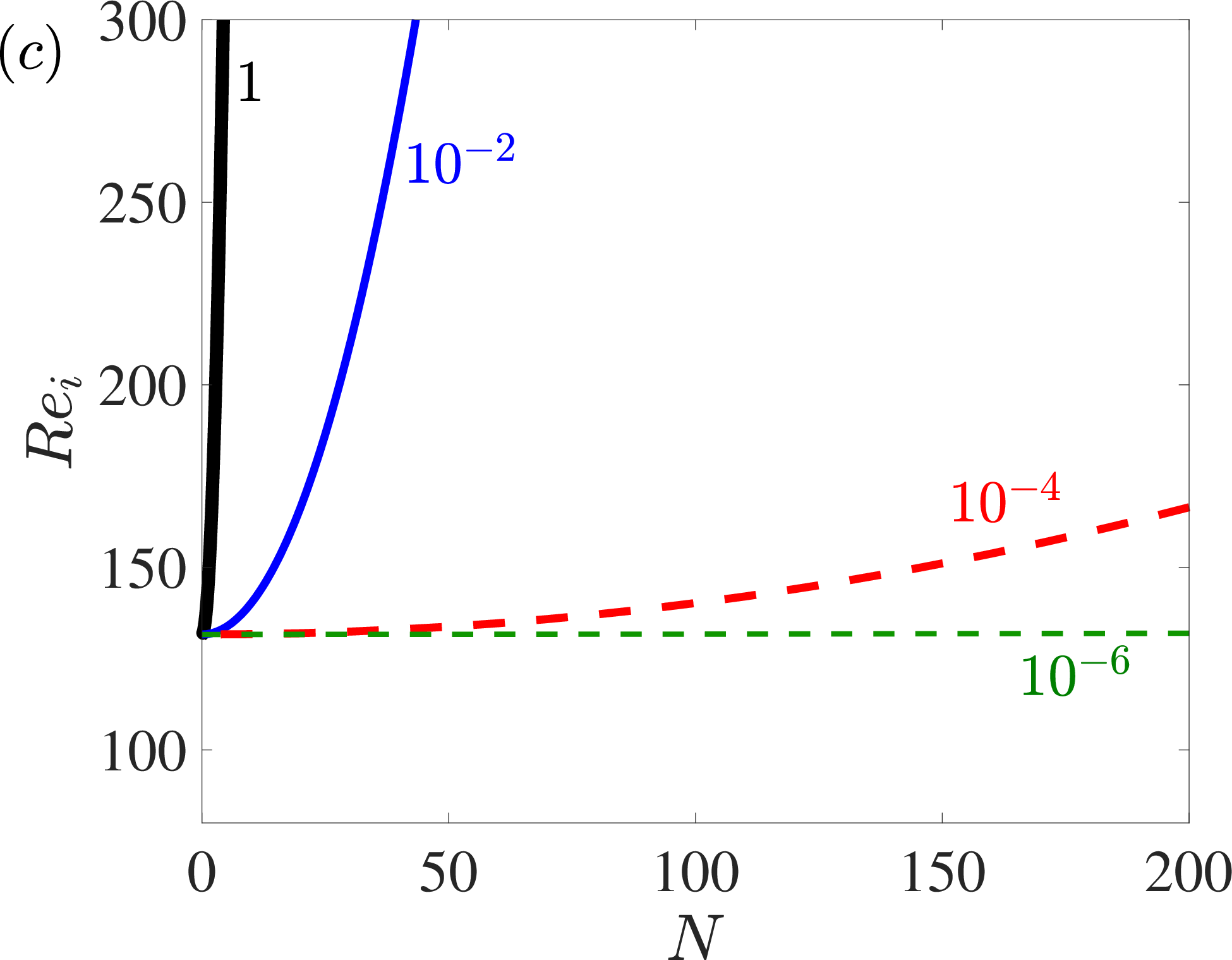}
  \includegraphics[height=4.5cm]{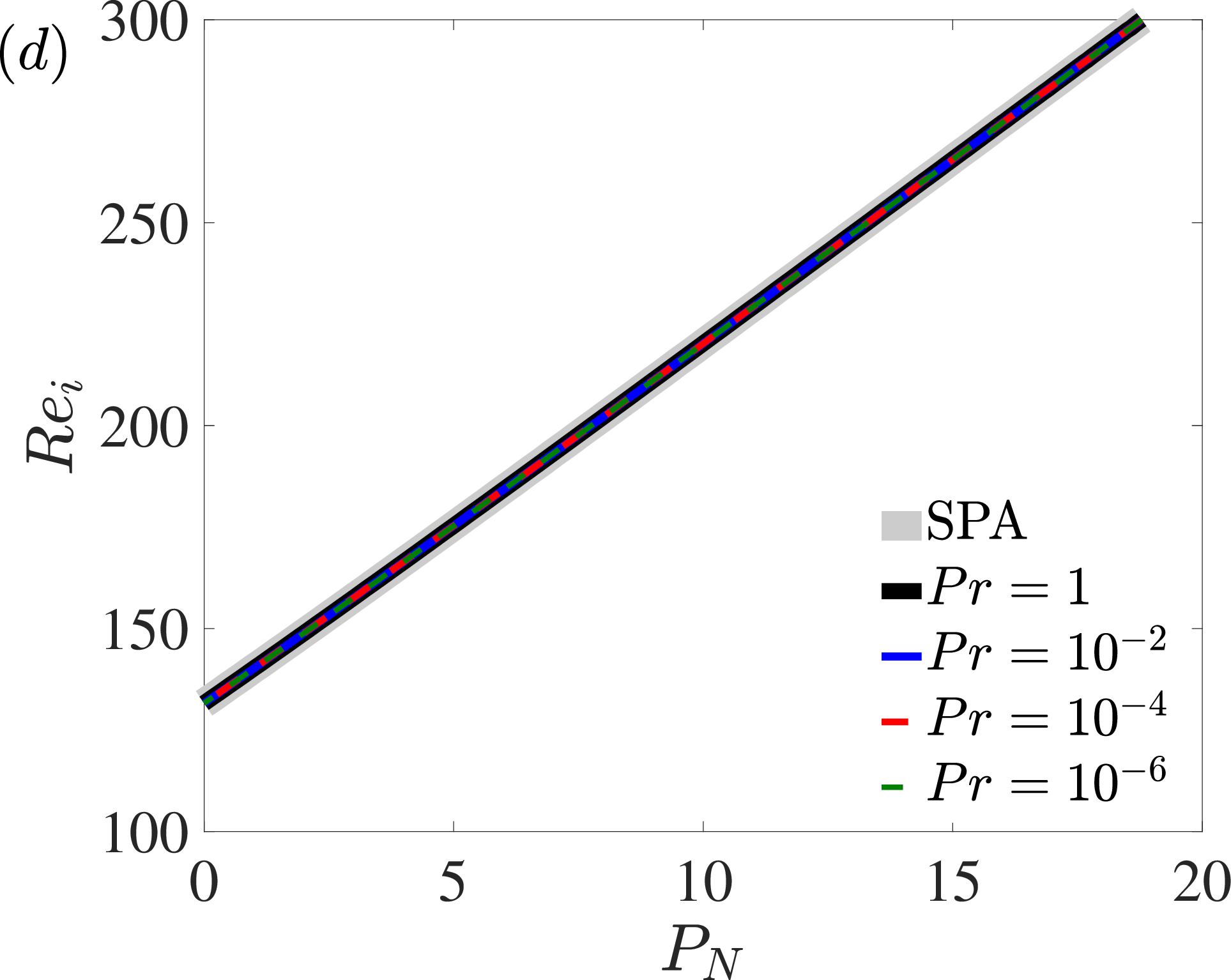}
  }
  \caption{
  (\textit{a}) Growth-rate curves for various sets of $(N,Pr)$ at $\mu=0$, $\eta=0.9$, $Re_{i}=200$ and $m=0$. 
  (\textit{b}) Neutral stability curves in the parameter space $(N,Re_{i})$ for different $Pr$ at $\mu=0$, $\eta=0.9$ and $m=0$. 
  (\textit{c}) The curves for different $Pr$ same as (\textit{b}) but over a wider range of $N$. 
  (\textit{d}) The curves same as (\textit{c}) overlapped due to the rescaled parameter $P_{N}=N^{2}Pr$ on the abscissa. 
  An additional thick grey line is a neutral stability curve obtained from the small-$Pr$ approximation.    
  }
\label{fig:nsc}
\end{figure}
Figure \ref{fig:nsc}(\textit{a}) shows the growth rate $\omega_{i}$ of the axisymmetric mode ($m=0$), which is found to be most unstable, versus the wavenumber $k_{d}$ for different parameter sets of $(N,Pr)$ at $\mu=0$, $\eta=0.9$ and $Re_{i}=200$.
For every case presented in Figure \ref{fig:nsc}(\textit{a}), the corresponding frequency $\omega_{r}$ is zero.
It is known that centrifugal instability reaches its maximum $\omega_{i,\max}$ as $k\rightarrow\infty$ in the inviscid limit $Re\rightarrow\infty$ as the growth rate scales as $\omega_{i,\mathrm{inviscid}}=\omega_{i,\max}-A_{0}/k^{2}$ \citep[][]{Billant2005,Park2017} while the viscous growth rate scales as $\omega_{i,\mathrm{viscous}}=\omega_{i,\mathrm{inviscid}}-A_{1}k^{2}/Re$ \citep[][]{Yim2016} where $A_{0}$ and $A_{1}$ are positive constants that depend on stratification and other parameters. 
Due to this characteristic varying with $k$, the viscous growth rates are positive in a finite range of $k_{d}$ and each curve reaches its peak at a certain wavenumber $k_{d,\max}$. 
As the stratification increases from $N=0$ (grey) to $N=1$ (black) for $Pr=1$, the growth-rate curves descend while the wavenumber $k_{d,\max}$ at the peak of $\omega_{i}$ increases. 
For the fixed $N=1$, as the Prandtl number $Pr$ decreases from $Pr=1$, the growth rate increases and, remarkably, the growth-rate curves for $Pr=10^{-2}$ and $10^{-4}$ overlap with the grey curve, the unstratified case with $N=0$. 
This implies that the centrifugal instability in stratified and highly diffusive fluids with $Pr\ll1$ behaves as the instability in unstratified fluids as the effect of stratification is suppressed by strong thermal diffusion. 
Another remarkable result in Figure \ref{fig:nsc}(\textit{a}) is that the growth-rate curves overlap for $(N,Pr)=(10,10^{-2})$ and $(10^{2},10^{-4})$, the cases with the same $P_{N}=N^{2}Pr=1$.  
The growth rate for other values of $N$ and $Pr\ll1$ is invariant if the rescaled parameter $P_{N}$ is the same. 
Such an invariant feature at the same $P_{N}$ but different $N$ and $Pr$ is similarly reported for both vertical and horizontal shear instabilities in vertically stratified fluids \citep[][]{Lignieres1999VSI,Park2020AA,Park2021AA}.

Figure \ref{fig:nsc}$(b)$ display neutral stability curves, which denote the critical Reynolds number $Re_{i,c}$ at which the growth rate $\omega_{i}$ of the most unstable mode is zero, in the parameter space $(N,Re_{i})$ for different values of $Pr$ at $\mu=0$, $\eta=0.9$ and $m=0$.
For $Pr\geq1$, the neutral stability curves increase rapidly with $N$, a feature similarly found in \citet{Park2017}, and the $Pr=1$ case is more stable than other cases with $Pr>1$ when $N>3$.
This is expected as thermal dissipation is proportional to the diffusivity $\kappa_{0}$, thus lower diffusivity $\kappa_{0}$ (i.e. higher $Pr$) implies less thermal dissipation and more instability.  
In the range $0<N<3$, the situation is more complicated since the $Pr=10$ case is more stable than other $Pr$ cases for the axisymmetric perturbation. 
Although such high-$Pr$ dynamics at a moderate $N$ should further be investigated due to its great importance in geophysical and other contexts (e.g. oceanic flows where the analogous Schmidt number $Sc$ is around 700), the current study will only focus on the highly diffusive regime with low $Pr\leq1$. 
For $Pr<1$, the curves increase very slowly as $N$ increases and it is difficult to see in Figure \ref{fig:nsc}(\textit{b}) the increase of the curves at $Pr=10^{-4}$ and $10^{-6}$ over the range $0\leq N\leq 20$. 
Figure \ref{fig:nsc}(\textit{c}) displays the same neutral stability curves for $Pr\leq1$ over a wider range of $N$ to see how slowly the neutral stability curves increase with $N$ as $Pr$ decreases.  
This increasing trend confirms that, while stratification enhances the stability of stratified Taylor-Couette flow, strong thermal diffusion destabilise by preventing the stabilising role of stratification and making the flow behave like an unstratified flow.
In Figure \ref{fig:nsc}(\textit{d}), we plot again the same neutral stability curves but over the rescaled parameter $P_{N}$ on the abscissa. 
All the curves now overlap each other, even for the case with $Pr=1$, and can be described by a linear relation as $Re_{i,c}= 131.6+8.965P_{N}$ where $Re_{i,c}=131.6$ is the critical Reynolds number for the unstratified case $N=0$ at $m=0$ and $\eta=0.9$, the number agreeing with \citet{DiPrima1984}. 

\begin{figure}
  \centerline{
  \includegraphics[height=3.3cm]{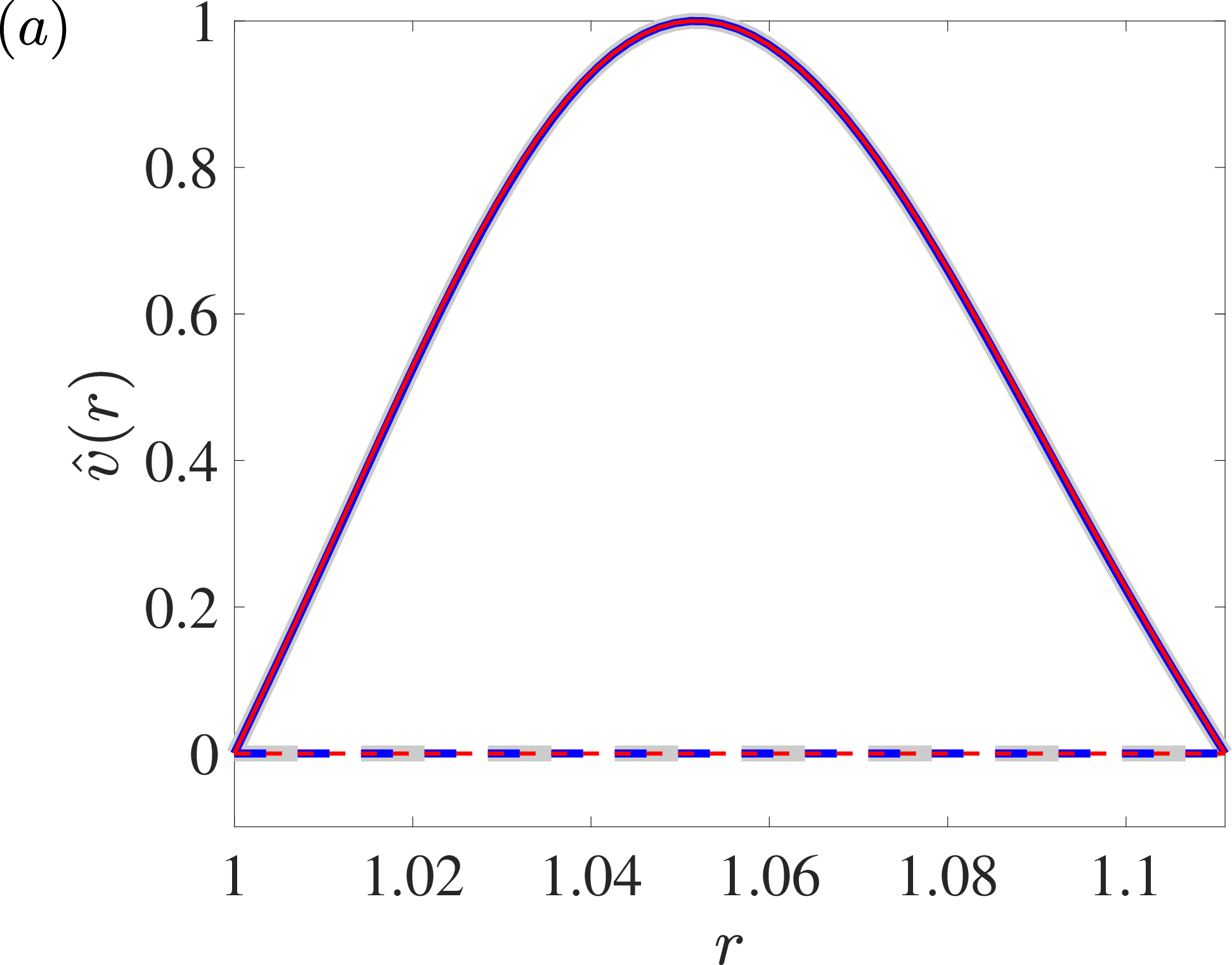}
  \includegraphics[height=3.3cm]{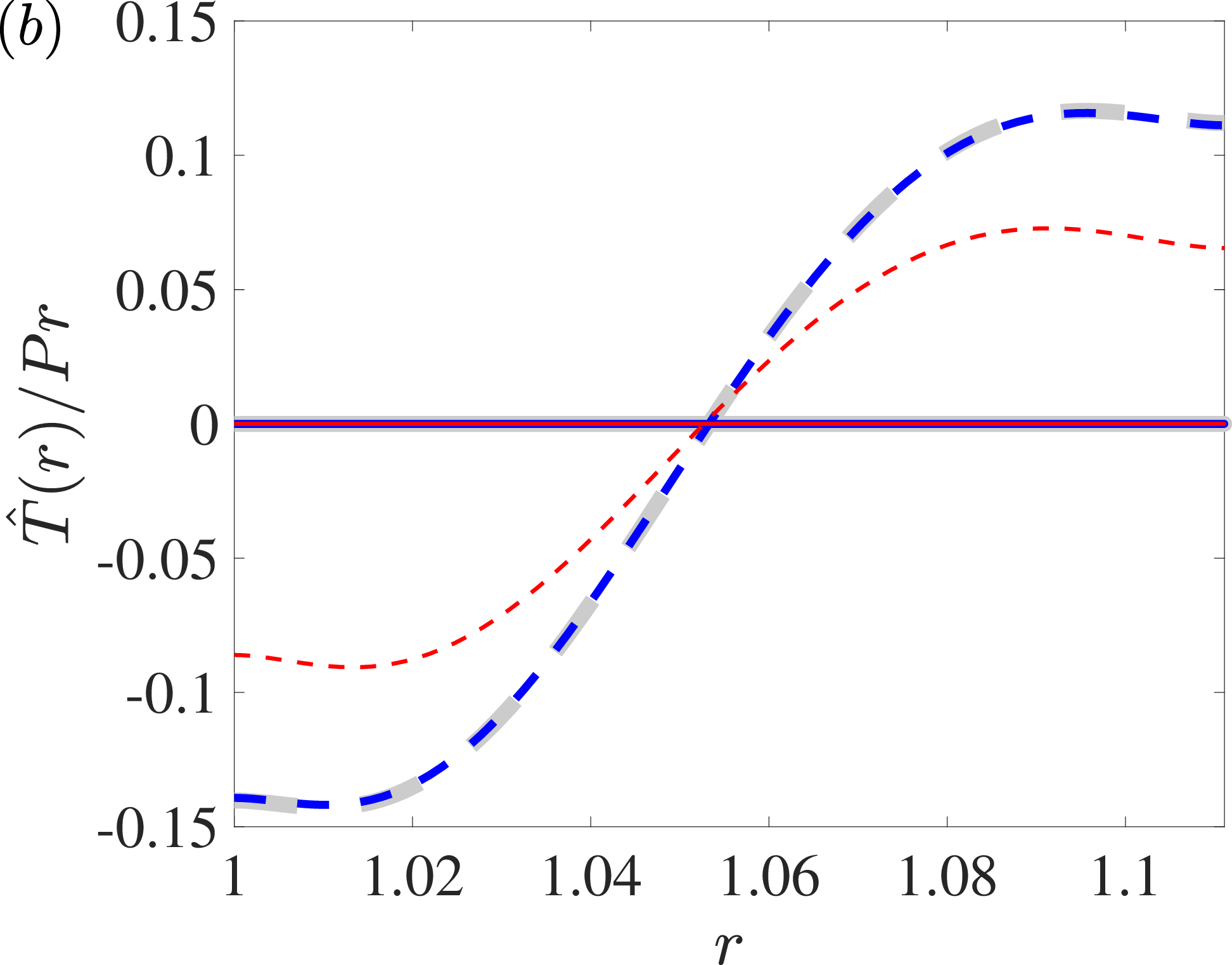}
  \includegraphics[height=3.3cm]{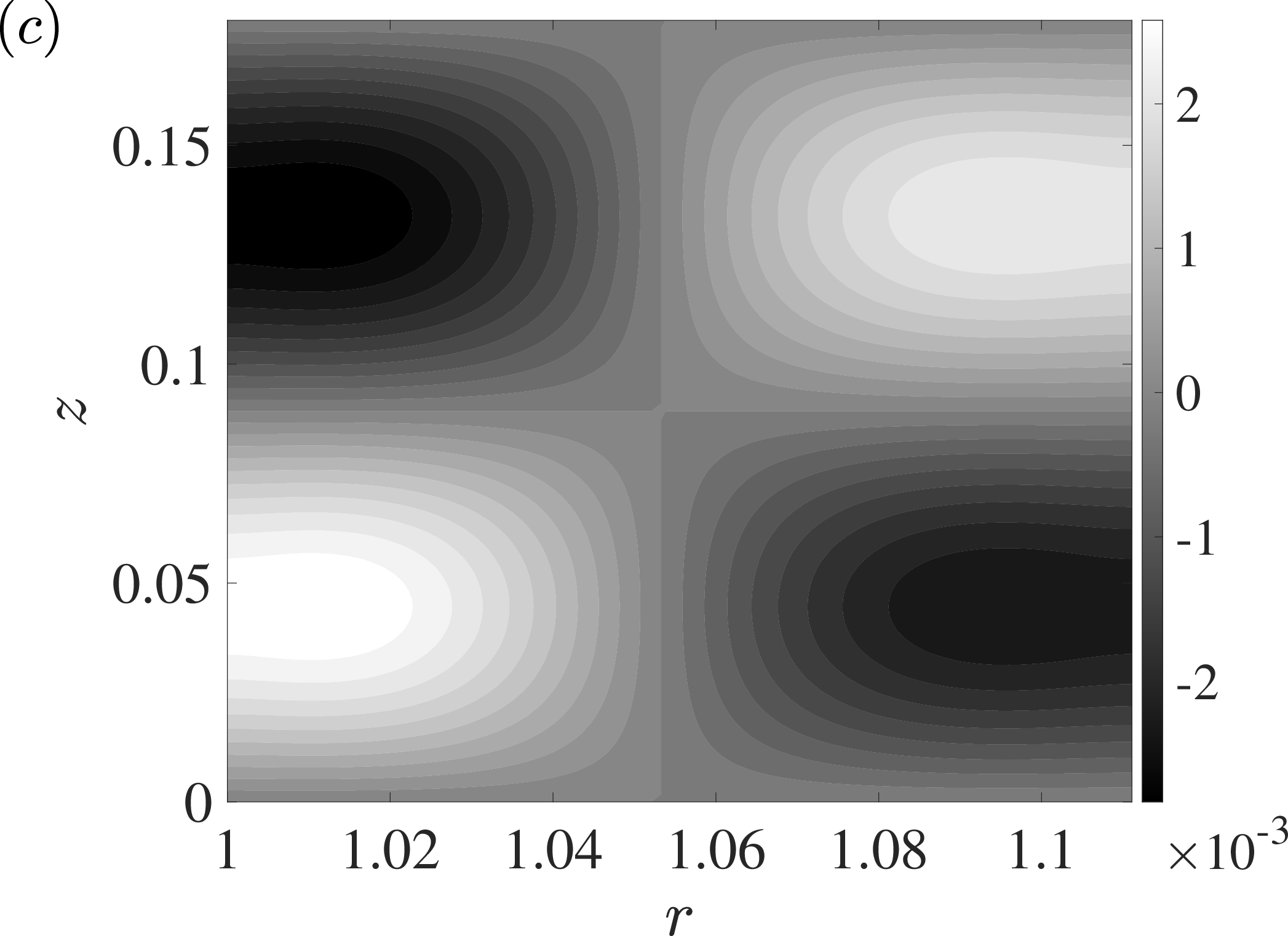}
  }
  \caption{
  (\textit{a},\textit{b}) Real (solid) and imaginary (dashed) parts of the mode shape $\hat{v}(r)$ and rescaled mode $\hat{T}(r)/Pr$ for $Pr=1$ (red), $Pr=10^{-2}$ (blue), and $Pr=10^{-4}$ (grey) at $\mu=0$, $\eta=0.9$, $Re_{i}=200$, $N=1$, $m=0$ and $k_{d}=3.91$.
  (\textit{c}) Perturbation temperature $T(r,z)$ reconstructed from $\hat{T}(r)$ for $Pr=10^{-2}$.    
  }
\label{fig:eigenfunction}
\end{figure}
Figure \ref{fig:eigenfunction} shows the real and imaginary parts of eigenmode shapes $\hat{v}(r)$ and $\hat{T}(r)$ for different $Pr$ for the wavenumber set $(m,k_{d})=(0,3.91)$.
The mode shapes are normalised by the maximum value of $\hat{v}$ and it is found that the mode shape $\hat{T}$ has a smaller amplitude than $\hat{v}$ for small $Pr$. 
As the Prandtl number $Pr$ decreases, the amplitude of $\hat{T}$ decreases and scales as $O(Pr)$. 
For a better comparison, Figure \ref{fig:eigenfunction}(\textit{b}) displays the mode shape $\hat{T}(r)$ divided by $Pr$ and we see that the rescaled $\hat{T}/Pr$ has the same mode shape for the cases with $Pr=10^{-2}$ and $Pr=10^{-4}$.
In Figure \ref{fig:eigenfunction}(\textit{c}), the perturbation temperature $T(r,z)=\hat{T}\exp(\mathrm{i}kz)+\hat{T}^{*}\exp(-\mathrm{i}kz)$ for $Pr=10^{-2}$ is plotted. 
Two in-phase waves, one near the inner cylinder and the other near the outer one, are clearly shown. 
The shape of this axisymmetric mode is different from that of a non-axisymmetric centrifugal instability mode, which is weakly sheared, or that of a strato-rotational instability mode that is out of phase \citep[][]{Park2017,Park2018}. 
The perturbation temperature has one node (i.e. the zero crossing) around $r\simeq1.053$ as it is the first mode which becomes the most unstable. 
Higher-order modes with more number of nodes are, on the other hand, found to be stable for the parameter set in Figure \ref{fig:eigenfunction} with $Pr=10^{-2}$. 

\subsection{Small-$Pr$ approximation}
The single dependence on $P_{N}$ for different $N$ and $Pr$ in the limit $Pr\ll1$ can be understood by taking the small-$Pr$ approximation.  
Consider the Taylor expansions in terms of small $Pr$ as $\hat{\textbf{u}}=\left(\hat{u},\hat{v},\hat{w}\right)^{\mathrm{T}}=\hat{\textbf{u}}^{(0)}+Pr\hat{\textbf{u}}^{(1)}+O\left(Pr^{2}\right)$ and similarly for $\hat{T}$ and $\hat{p}$ as $\hat{T}=\hat{T}^{(0)}+Pr\hat{T}^{(1)}+\cdots$ and $\hat{p}=\hat{p}^{(0)}+Pr\hat{p}^{(1)}+\cdots$.
For the P\'eclet number $Pe$ being small as $Pe=RePr\ll1$ (i.e. $Re\sim O(1)$ and $Pr\ll1$), we obtain the leading-order equation for $\hat{T}^{(0)}$ as 
\begin{equation}
\frac{1}{RePr}\hat{\nabla}^{2}\hat{T}^{(0)}=0,
\end{equation}
which leads to the analytic solution $\hat{T}^{(0)}(r)=a_{1}\mathrm{I}_{m}(kr)+a_{2}\mathrm{K}_{m}(kr)$ where $a_{1}$ and $a_{2}$ are constants and $\mathrm{I}_{m}$ and $\mathrm{K}_{m}$ are the modified Bessel functions of the first and second kinds, respectively \citep[][]{AbramowitzStegun1970}.  
The constants $a_{1}$ and $a_{2}$ become zero if the no-flux conditions $\mathrm{d}\hat{T}^{(0)}/\mathrm{d}r=0$ are imposed at both cylinders, thus the leading-order solution simply becomes zero (i.e. $\hat{T}^{(0)}(r)=0$).
At the next order, we obtain the following equations:
\begin{equation}
\label{eq:continuity_sPr}
\frac{\mathrm{d} \hat{u}^{(0)}}{\mathrm{d} r}+\frac{\hat{u}^{(0)}}{r}+\frac{\mathrm{i}m\hat{v}^{(0)}}{r}+\mathrm{i}k\hat{w}^{(0)}=0,
\end{equation}
\begin{equation}
\label{eq:mom_r_sPr}
-\mathrm{i}\omega\hat{u}^{(0)}+\mathrm{i}m\Omega\hat{u}^{(0)}-2\Omega \hat{v}^{(0)}=-\frac{\mathrm{d} \hat{p}^{(0)}}{\mathrm{d} r}+\frac{1}{Re}\left(\hat{\nabla}^{2}\hat{u}^{(0)}-\frac{\hat{u}^{(0)}}{r^{2}}-\frac{2\mathrm{i}m\hat{v}^{(0)}}{r^{2}}\right),
\end{equation}
\begin{equation}
\label{eq:mom_th_sPr}
-\mathrm{i}\omega\hat{v}^{(0)}+\mathrm{i}m\Omega\hat{v}^{(0)}+Z\hat{u}^{(0)}=-\frac{\mathrm{i}m\hat{p}^{(0)}}{r}+\frac{1}{Re}\left(\hat{\nabla}^{2}\hat{v}^{(0)}-\frac{\hat{v}^{(0)}}{r^{2}}+\frac{2\mathrm{i}m\hat{u}^{(0)}}{r^{2}}\right),
\end{equation}
\begin{equation}
\label{eq:mom_z_sPr}
-\mathrm{i}\omega\hat{w}^{(0)}+\mathrm{i}m\Omega\hat{w}^{(0)}=-\mathrm{i}k\hat{p}^{(0)}+P_{N}\hat{T}^{(1)}+\frac{1}{Re}\hat{\nabla}^{2}\hat{w}^{(0)},
\end{equation}
\begin{equation}
\label{eq:energy_sPr}
\hat{w}^{(0)}=\frac{1}{Re}\hat{\nabla}^{2}\hat{T}^{(1)},
\end{equation}
where $P_{N}=N^{2}Pr$.
We see that the two parameters $N$ and $Pr$ representing thermal stratification and diffusion are simplified into a single parameter $P_{N}$, as observed in other studies under the small-P\'eclet-number approximation \citep[e.g.][]{Lignieres1999SPA,Park2020AA}. 
By considering the continuity equation and eliminating the pressure, the equations (\ref{eq:continuity_sPr})-(\ref{eq:energy_sPr}) can further be simplified into the following eigenvalue problem
\begin{equation}
\label{eq:evp_spa}
-\mathrm{i}\omega\mathcal{A}^{(0)}\hat{\boldsymbol{q}}^{(0)}=\mathcal{B}^{(0)}\hat{\boldsymbol{q}}^{(0)},
\end{equation}
where $\hat{\boldsymbol{q}}^{(0)}$ is the mode shape vector and $\mathcal{A}^{(0)}$ and $\mathcal{B}^{(0)}$ are the operator matrices detailed in Appendix \ref{app:matrices}.
In Figure \ref{fig:nsc}(\textit{d}), it is clearly shown that a neutral stability curve computed from (\ref{eq:evp_spa}) overlaps with other neutral stability curves when they are plotted over the rescaled Prandtl number $P_{N}=N^{2}Pr$ as the abscissa.  

\subsection{Perturbation energy analysis}
The instability characteristics can also be understood by examining the evolution of perturbation energy. 
We first define the total perturbation energy $E(t)$ as
\begin{equation}
\label{eq:total_energy_ptb}
E(t)=\frac{1}{2}\int_{0}^{L_{z}}\int_{0}^{2\pi}\int_{1}^{1/\eta}\left(u_{r}^{2}+u_{\theta}^{2}+u_{z}^{2}+N^{2}T^{2}\right)r\mathrm{d}r\mathrm{d}\theta\mathrm{d}z=\frac{1}{2}\left<\textbf{q}_{E}~\textbf{:}~\textbf{q}_{E}\right>,
\end{equation}
where $L_{z}$ is a vertical length of the domain assumed to be periodic (e.g. $L_{z}=2\pi/k$ if we consider one periodic length of a normal mode with the axial wavenumber $k$), the angle brackets denote the volume integral defined as $\left<\textbf{X}\right>=\int_{0}^{L_{z}}\int_{0}^{2\pi}\int_{1}^{1/\eta}\textbf{X} ~r\mathrm{d}r\mathrm{d}\theta\mathrm{d}z$, $\textbf{q}_{E}=\left(u_{r},u_{\theta},u_{z},NT\right)^{\mathrm{T}}$, and $\textbf{:}$ denotes the Frobenius product \citep[for more details, see also][]{Park2017}.  
On the momentum and energy equations (\ref{eq:mom_r_ptb})-(\ref{eq:energy_ptb}), we multiply both sides by $\left(u_{r}, u_{\theta}, u_{z}, N^{2}T\right)^{\mathrm{T}}$ and, after some manipulations, we obtain the following equation for perturbation energy:
\begin{equation}
\label{eq:energy_evolution_ptb}
\frac{\partial E}{\partial t}=\left<-\frac{\mathrm{d}\Omega}{\mathrm{d}r}\left(ru_{r}u_{\theta}\right)-\frac{1}{Re}\left[\nabla\textbf{u}~\textbf{:}~\nabla\textbf{u}+\frac{N^{2}}{Pr}\left(\nabla T~\textbf{:}~\nabla T\right)\right]\right>,
\end{equation}
where the first term on the right-hand side within the angle brackets denotes the contribution from the mean angular shear $\mathrm{d}\Omega/\mathrm{d}r$ and the second term corresponds to the kinetic and potential energy dissipation.
The contribution from the mean angular shear is a main source of energy production if the momentum transfer term $u_{r}u_{\theta}$ is anti-correlated with the mean angular shear $\mathrm{d}\Omega/\mathrm{d}r$, a well-known mechanism for instability called the Orr mechanism \citep[][]{Orr1907}.
The momentum and thermal dissipation terms are always negative and thus stabilise the perturbation energy. 
We note that this mechanism is valid for both linear and non-linear cases as the evolution equation (\ref{eq:energy_evolution_ptb}) is derived from the non-linear perturbation equations (\ref{eq:mom_r_ptb})-(\ref{eq:energy_ptb}) and the non-linear terms are cancelled out in the derivation process in which the continuity is taken into account.
If we apply the normal mode (\ref{eq:normal_mode}) into the evolution equation (\ref{eq:energy_evolution_ptb}), we obtain the following expression for the growth rate
\begin{equation}
\label{eq:growth_rate_energy_relation}
\omega_{i}=\frac{1}{\hat{E}}\left<-\frac{\mathrm{d}\Omega}{\mathrm{d}r}\frac{r\left(\hat{u}^{*}\hat{v}+\hat{u}\hat{v}^{*}\right)}{2}-\frac{1}{Re}\left[\hat{\nabla}\hat{\textbf{u}}~\textbf{:}~\hat{\nabla}\hat{\textbf{u}}+\frac{N^{2}}{Pr}\left(\hat{\nabla}\hat{T}~\textbf{:}~\hat{\nabla}\hat{T}\right)\right]\right>_{r},
\end{equation}
where $*$ denotes the complex conjugate, $<>_{r}$ denotes the line integral over the radial coordinate $r$ as $\left<\hat{\textbf{X}}\right>_{r}=\int_{1}^{1/\eta}\hat{\textbf{X}} ~r\mathrm{d}r$ and $\hat{E}=\left<|\hat{u}|^{2}+|\hat{v}|^{2}+|\hat{w}|^{2}+|N\hat{T}|^{2}\right>_{r}$.

\begin{table}
  \begin{center}
\def~{\hphantom{0}}
  \begin{tabular}{ccccclccccc}
      $N$ & $Pr$ & $m$ & & $k_{d,\max}$ & & $\omega_{i,\max}$  & $\mathcal{P}_{\Omega}/\hat{E}$   &   $\epsilon_{k}/\hat{E}$ & $\epsilon_{p}/\hat{E}$ & $\left(\mathcal{P}_{\Omega}+\epsilon_{k}+\epsilon_{p}\right)/\hat{E}$ \\[3pt]
       0 & 1 & 0 && 3.74 &  &  0.667 & 1.813 & -1.146 & 0 & 0.667\\
       1 & 1 & 0 && 3.91 &  &  0.574 & 1.769 & -1.185 & -0.010 & 0.574\\
       1 & 1 & 1 && 3.90 &  &  0.564 & 1.763 & -1.188 & -0.011 & 0.564\\
       1 & $10^{-2}$ & 0 && 3.74 &  &  0.666 & 1.808 & -1.142 & -0.0003 & 0.666\\
       10 & $10^{-2}$ & 0 && 4.03 &  &  0.535 & 1.791 & -1.236 & -0.020 & 0.535\\
  \end{tabular}
  \caption{Values of the maximum growth rates $\omega_{i,\max}$ with the corresponding wavenumber $k_{d,\max}$, production and dissipation terms for various $N$, $Pr$ and $m$ at $\mu=0$, $\eta=0.9$ and $Re_{i}=200$.}
  \label{tab:growth_rate}
  \end{center}
\end{table}
Table \ref{tab:growth_rate} shows examples of the maximum growth rate $\omega_{i,\max}$ and corresponding wavenumber $k_{d,\max}$ for various parameter sets of $(N,Pr,m)$ at $\mu=0$, $\eta=0.9$ and $Re_{i}=200$.
The table also details the production and dissipation terms $\mathcal{P}_{\Omega}$, $\epsilon_{k}$ and $\epsilon_{p}$, all of which are defined as
\begin{equation}
\label{eq:prod_diss}
\mathcal{P}_{\Omega}=\left<-\frac{\mathrm{d}\Omega}{\mathrm{d}r}\frac{r\left(\hat{u}^{*}\hat{v}+\hat{u}\hat{v}^{*}\right)}{2}\right>_{r},~
\epsilon_{k}=-\frac{1}{Re}\left<\hat{\nabla}\hat{\textbf{u}}:\hat{\nabla}\hat{\textbf{u}}\right>_{r},~
\epsilon_{p}=-\frac{N^{2}}{RePr}\left<\hat{\nabla}\hat{T}~\textbf{:}~\hat{\nabla}\hat{T}\right>_{r}.
\end{equation}
The production and dissipation terms are calculated by implementing the eigenfunction into the expressions (\ref{eq:prod_diss}) and their sum shows a good agreement with the eigenvalue $\omega_{i,\max}$. 
For every case, the contribution from the thermal dissipation $\epsilon_{p}$ to the growth rate is small as $Pr\leq1$ and the growth rate $\omega_{i}$ is mainly determined by a difference between the production $\mathcal{P}_{\Omega}$ and the viscous dissipation $\epsilon_{k}$. 
For the cases with $Pr\ll1$, if we apply the small-$Pr$ approximation, we can re-express the thermal dissipation $\epsilon_{p}$ as
\begin{equation}
\label{eq:growth_rate_dissipation_thermal}
\epsilon_{p}\simeq-\frac{P_{N}}{Re}\left<\hat{\nabla}\hat{T}_{1}~\textbf{:}~\hat{\nabla}\hat{T}_{1}\right>_{r}\sim O\left(\frac{P_{N}}{Re}\right),
\end{equation}
where the dependence on the two parameters $N$ and $Pr$ is now expressed by the single parameter $P_{N}$ under the small-$Pr$ approximation. 
The scaling for thermal dissipation $\epsilon_{p}\sim O(P_{N}/Re)$ is based on the assumption $\hat{T}_{1}\sim O(1)$ under the small-$Pr$ approximation.

\subsection{Parametric investigations}
\begin{figure}
  \centerline{
  \includegraphics[height=3.5cm]{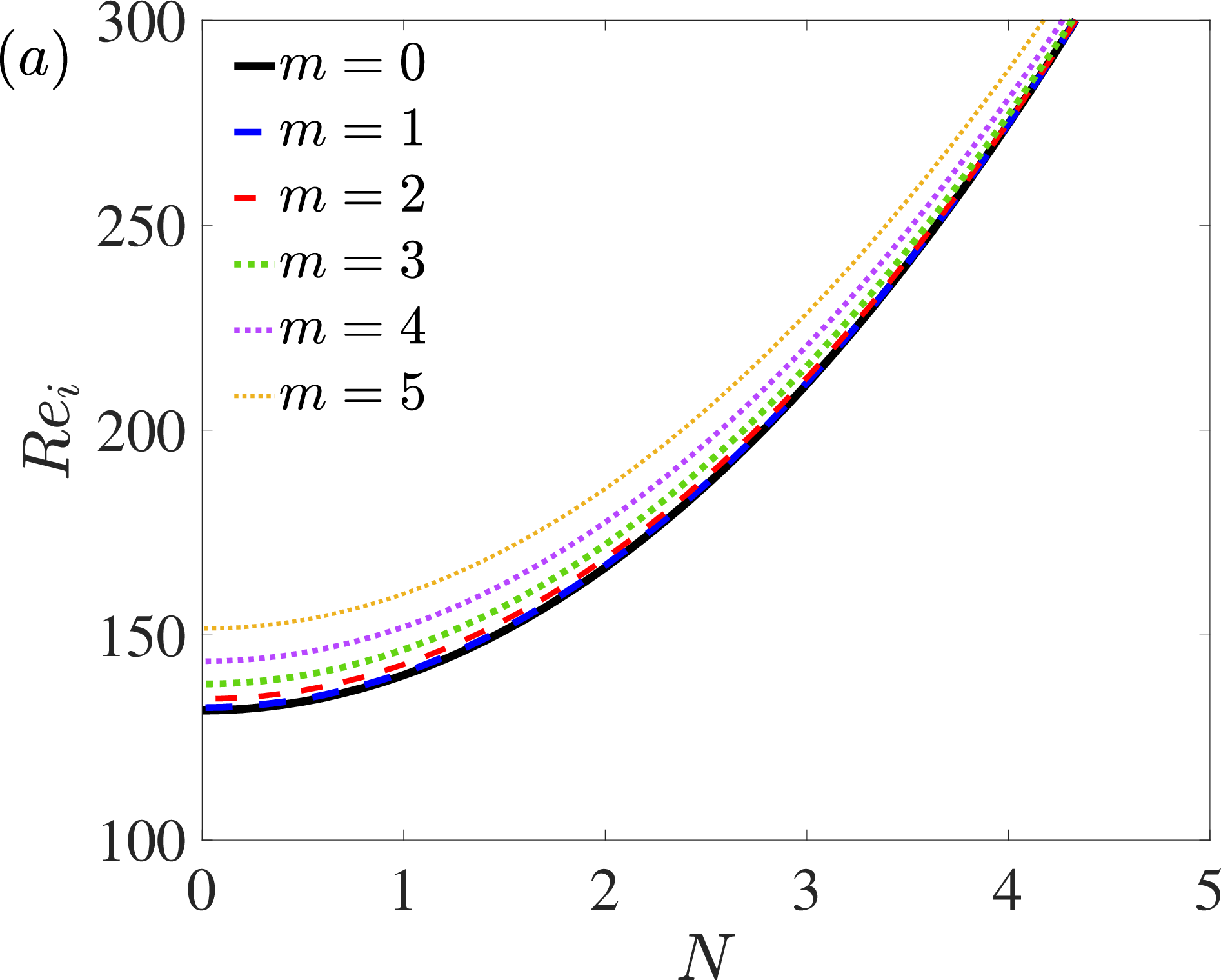}
    \includegraphics[height=3.5cm]{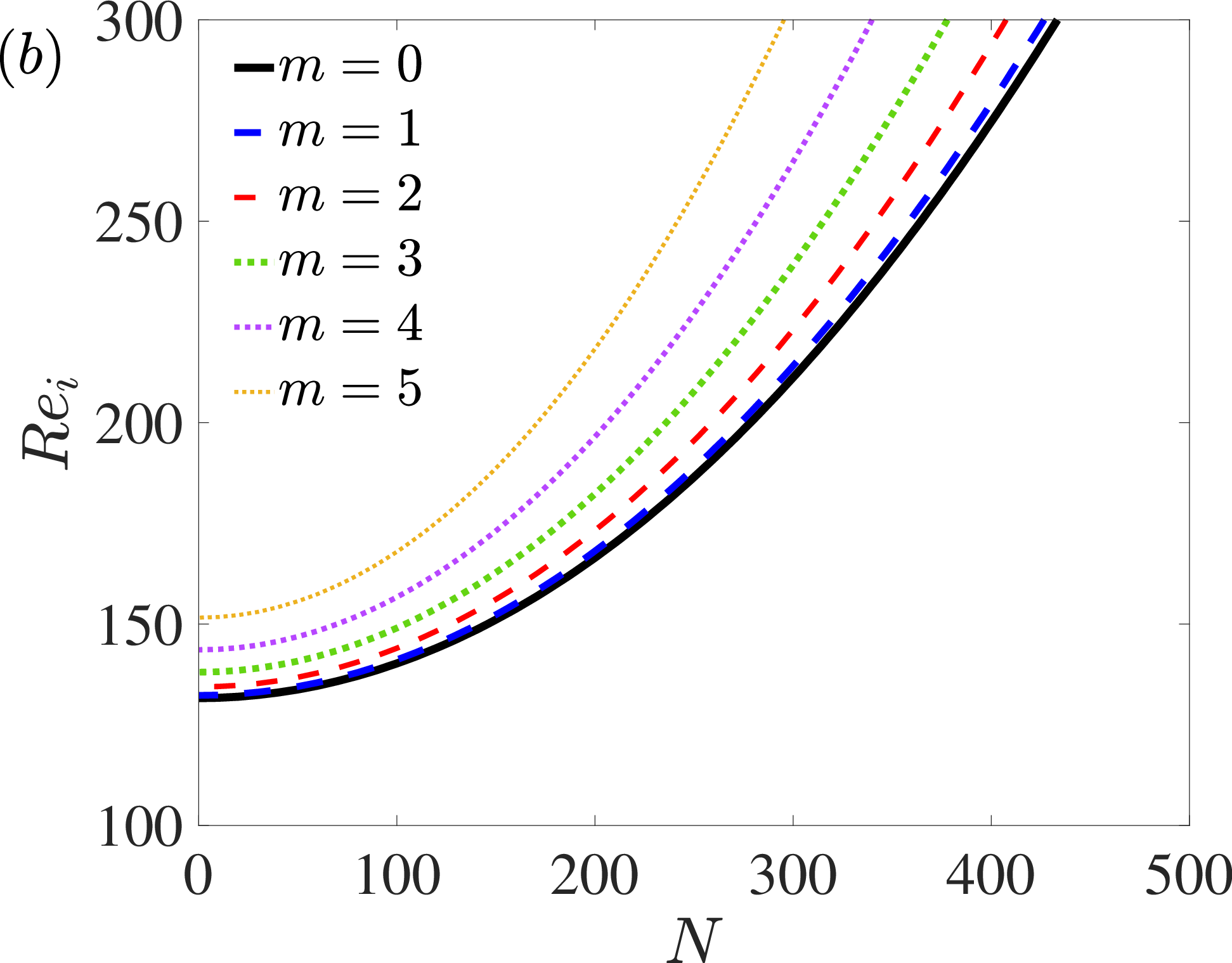}
      \includegraphics[height=3.5cm]{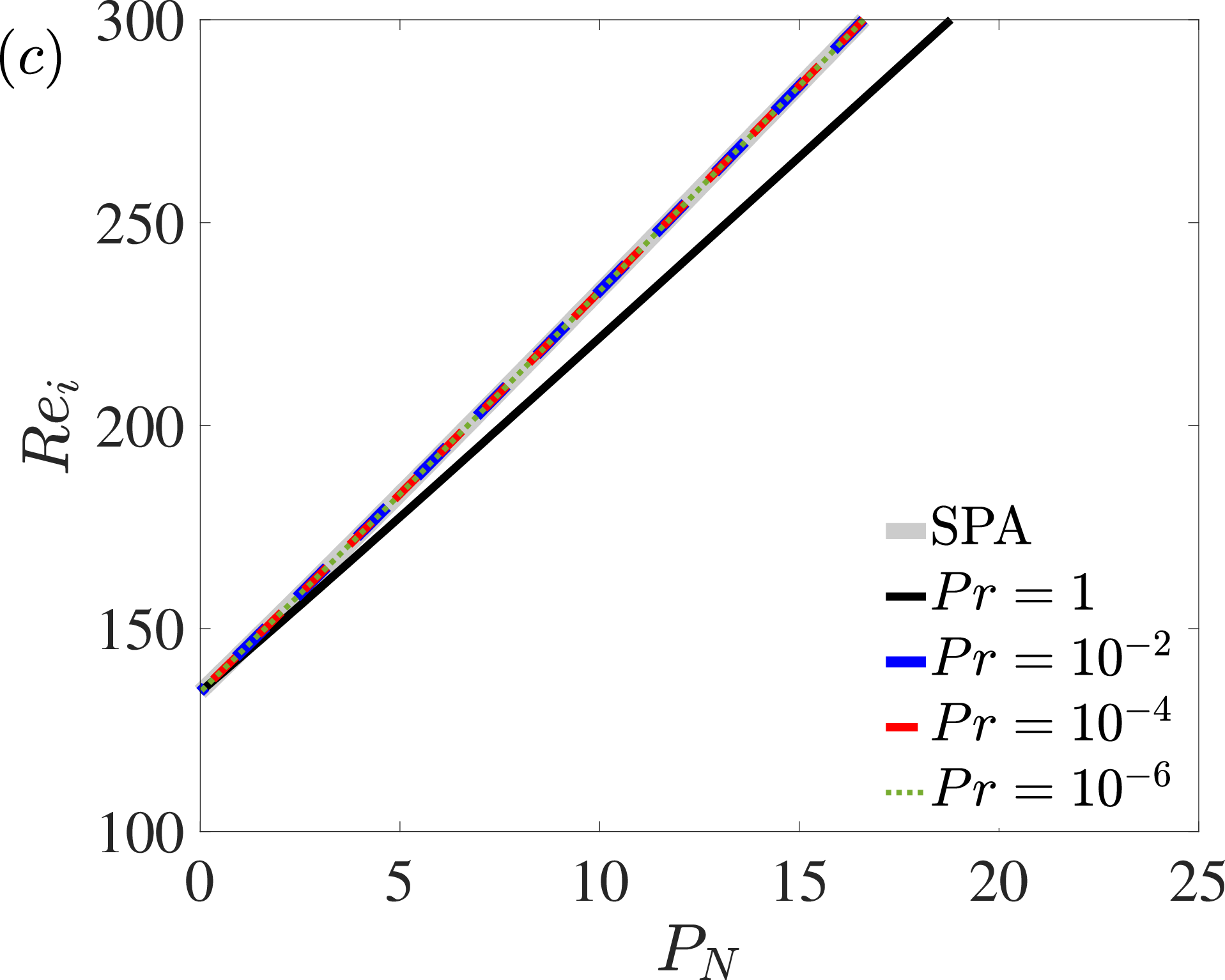}
  }
  \caption{
  (\textit{a},\textit{b}) Neutral stability curves for different $m$ for (\textit{a}) $Pr=1$ and (\textit{b}) $Pr=10^{-4}$ at $\mu=0$ and $\eta=0.9$.
  (\textit{c}) Neutral stability curves for different $Pr$ over the rescaled parameter $P_{N}$ at $\mu=0$, $\eta=0.9$ and $m=2$. A thick grey line denotes the neutral stability curve from the small-$Pr$ approximation. 
  }
\label{fig:lsa_non_axis}
\end{figure}
Figure \ref{fig:lsa_non_axis}(\textit{a}) and (\textit{b}) show neutral stability curves for axisymmetric and non-axisymmetric cases in the parameter space $(N,Re_{i})$ at $\mu=0$ and $\eta=0.9$ for $Pr=1$ and $10^{-4}$, respectively.
For both $Pr$, all the neutral stability curves increasing with $N$ ascend as $m$ increases, thus the lowest curves correspond the axisymmetric case with $m=0$.
A difference between (\textit{a}) and (\textit{b}) is that, for $Pr=1$, neutral stability curves for non-axisymmetric cases ($m>0$) stay closer to the curve for the axisymmetric case ($m=0$) as $N$ increases, while for $Pr=10^{-4}$, the curves for $m>0$ stay further to the curve for $m=0$ as $N$ increases. 
This implies that for strongly stratified fluids with $N\gg0$ for $Pr=1$, the axisymmetric mode will appear at the instability onset $Re_{i}=Re_{i,c}$ but then non-axisymmetric perturbations will also become unstable immediately after the onset, thus competition between the axisymmetric and non-axisymmetric modes will occur right above $Re_{i}>Re_{i,c}$. 
On the other hand, for low $Pr$ and above the instability onset $Re_{i}>Re_{i,c}$, the axisymmetric mode will become more dominant over non-axisymmetric modes as $N$ increases. 
Figure \ref{fig:lsa_non_axis}(\textit{c}) displays neutral stability curves over the rescaled $P_{N}$ for different $Pr$ at $m=2$.
As similarly observed for the axisymmetric case $m=0$ in Figure \ref{fig:nsc}(\textit{c}), the curves for $Pr\leq10^{-2}$ overlap and agree with the prediction from the small-$Pr$ approximation (SPA). 
It is found that the line from the SPA for $m=2$ scales as $Re_{i,c}= 134.4+9.971P_{N}$ where $Re_{c}=134.4$ as the $Re_{i}$-axis intercept at $P_{N}=0$ corresponds to the critical Reynolds number for unstratified case $N=0$.
The slope 9.971 for $m=2$ is higher than the slope 8.965 of the $m=0$ case.
It is verified that the critical Reynolds number and the slope increase with $m$ and thus they are at the lowest for $m=0$.
This implies that for low $Pr$, the axisymmetric mode with $m=0$ is expected to be the most unstable one for $P_{N}=N^{2}Pr\geq0$.

\begin{figure}
  \centerline{
  \includegraphics[height=5cm]{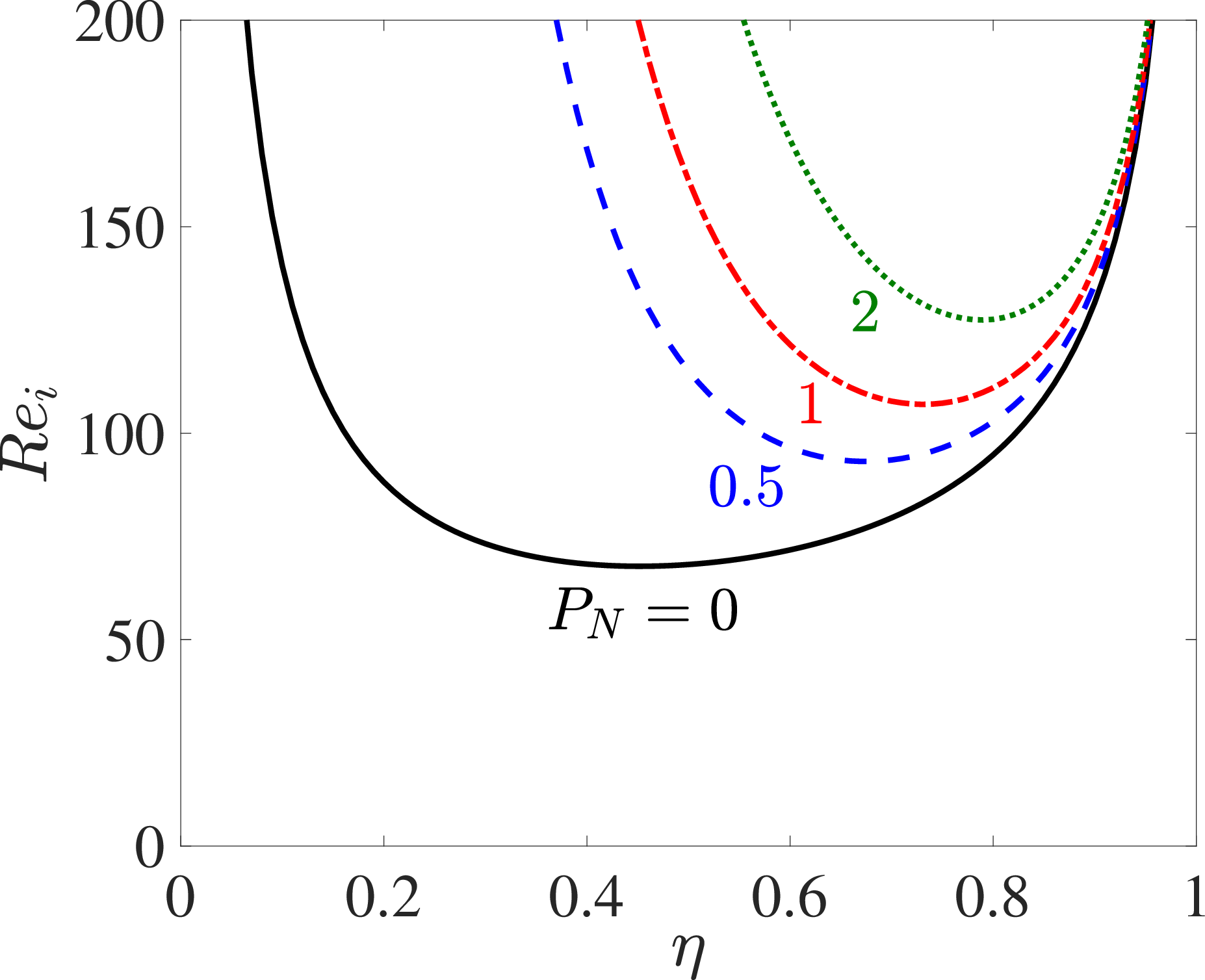}
  }
  \caption{Neutral stability curves obtained from the small-$Pr$ approximation at $\mu=0$. 
  }
\label{fig:nsc_eta_Re}
\end{figure}
Figure \ref{fig:nsc_eta_Re} shows neutral stability curves obtained from the small-$Pr$ approximation in the wide parameter space $(\eta,Re_{i})$ for different values of $P_{N}$ at $\mu=0$.
The curve for $P_{N}=0$ agrees with the result in \citet{DiPrima1984}. 
The azimuthal wavenumber of the curves corresponds to $m=0$ as the axisymmetric perturbation is found to be more unstable than non-axisymmetric perturbations. 
An advantage of using the small-$Pr$ approximation is that, for $Pr\ll1$, the stability curves depend solely on a single parameter $P_{N}=N^{2}Pr$ instead of two parameters $N$ and $Pr$, thus parametric investigations become simplified. 
As $P_{N}$ increases, the curves ascend and, in particular, the wide-gap Taylor-Couette flow with small $\eta$ is strongly stabilised.
It is also shown that the curves are less sensitive to the change in $P_{N}$ when $\eta$ is close to 1 (i.e. the Taylor-Couette flow with a small gap).  

Section \ref{sec:LSA} describes how linear centrifugal instability of stratified Taylor-Couette flow is affected by strong thermal diffusion. 
It is shown that the stratification acting a stabilising role on the instability is suppressed by strong thermal diffusion. 
For the cases with $Pr\ll1$, the dependence on $N$ and $Pr$ is simplified further by a single rescaled parameter $P_{N}=N^{2}Pr$ as derived by the small-$Pr$ approximation.
In the following Section \ref{sec:NCI}, we will study how the instability modes develop nonlinearly in stratified and diffusive fluids.  
Non-linear dynamical behaviours such as non-linear saturation, secondary instability of the saturated state or transition to chaotic states will be examined.

\section{Non-linear development of centrifugal instability}
\label{sec:NCI}
\begin{table}
  \begin{center}
\def~{\hphantom{0}}
  \begin{tabular}{cccccccccccc}
      Case & $Re_i$ & $N$ & $Pr$ & $N_{r}$ & $N_{\theta}$ & $N_{z}$ & $k$ & $k_{d}$ & $L_{z}$   &  $\mathcal{R}_{c}$  \\[3pt]
       1 & 145 & 1 & 1 & 60 & 33 & 33 & 30.6 & 3.40 & $2\pi/k$ & 1.035\\
       2 & 200 & 1 & 0.01 & 120 & 65 & 65 & 28.2 & 3.13 & $2\pi/k$ &   1.519\\
       3 & 200 & 1 & 1 & 120 & 65 & 65 & 30.6 & 3.40 & $2\pi/k$ & 1.427 \\
  \end{tabular}
  \caption{Physical and numerical parameters for representative 3D DNS cases.}
  \label{tab:DNS}
  \end{center}
\end{table}
In this section, we investigate via direct numerical simulation (DNS) how the centrifugal instability develops nonlinearly.
Table \ref{tab:DNS} provides details of physical and numerical parameters used for the main 3D DNS cases that are thoroughly analysed.
There are more 3D and 2D DNS results with similar parameters in the paper; however, we focus on the analysis of these three main cases featuring non-linear saturation, secondary instability, and transition to chaotic states. 
These non-linear features are thoroughly discussed in the following subsections.  
In the Table, parameters to be noticed are the axial wavenumber $k$ and the domain length $L_{z}=2\pi/k$ (i.e. one periodic length).
The wavenumber $k$ in Table \ref{tab:DNS} is chosen as $k=k_{c}$, which is the critical wavenumber at the critical Reynolds number $Re_{i}=Re_{i,c}$ at the onset of primary instability.
This wavenumber choice is coherent with previous studies that investigated axisymmetric Taylor vortices for $Re_{i}>Re_{i,c}$ by considering the characteristic wavenumber $k$ close to $k_{c}$ as $k\simeq k_{c}$, although a better agreement with experiments can be met for the torque of Taylor-vortex flow and wavy-vortex flow at high $Re_{i}\gg R_{i,c}$ if a suitable variation of the wavenumber is taken into account \citep[see e.g.][]{Meyer1966,Davey1968,DiPrima1984}.
Our study considers the Reynolds number $Re_{i}$ not too far from $Re_{i,c}$ (i.e. the Reynolds-number ratio $\mathcal{R}_{c}=Re_{i}/Re_{i,c}<2$), thus the flow is not turbulent. 
In this case, the choices of the wavenumber $k=k_{c}$ and the corresponding domain length $L_{z}=2\pi/k$ are seemingly appropriate for comparison with other studies.
We support this presumption by providing in Appendix \ref{sec:App_validation} the validation against experiments and numerical verification on the domain length dependence. 
The principal azimuthal wavenumber $m$ in DNS is $m=1$ so that the entire angle $\theta=[0,2\pi]$ with its azimuthal periodicity is considered as the azimuthal domain. 

If the Fourier representation (\ref{eq:Fourier_mode}) is taken into account, the perturbation energy $E(t)$ can be expressed as the sum: $E(t)=\sum_{j=-M}^{M}\sum_{l=-K}^{K}\tilde{E}_{jl}(t)$ where $\tilde{E}_{jl}$ is the modal energy defined as
\begin{equation}
\label{eq:ptb_energy_modal}
\tilde{E}_{jl}=\pi L_{z}\int_{1}^{1/\eta}\left(|\tilde{u}_{jl}|^{2}+|\tilde{v}_{jl}|^{2}+|\tilde{w}_{jl}|^{2}+N^{2}|\tilde{T}_{jl}|^{2}\right)r\mathrm{d}r.
\end{equation}
For 3D DNS conducted in this study, we consider a controlled initial condition as the combination of axisymmetric modes with $(j,l)=(0,\pm1)$ and $\tilde{E}_{jl}=5\times10^{-7}$ for each mode, which is most unstable, and other unstable non-axisymmetric modes with $(j,l)=(\pm1,\pm1)$ at a smaller amplitude with $\tilde{E}_{jl}=5\times10^{-9}$ for each mode.
These modes are computed from the 1D local LSA and the non-axisymmetric modes are normalised to be out of phase, the case in which secondary instability can be promoted at the equilibrium state (i.e. axisymmetric Taylor vortices) reached by nonlinear saturation of the axisymmetric mode \citep[see also,][]{Davey1968,Eagles1974}. 
Consideration of the non-axisymmetric modes with $j=\pm1$ (i.e. $m=\pm1$) is essential to allow the nonlinear energy transfer to modes with higher azimuthal wavenumber $|m|>1$ and axial wavenumber $jk$ with $|j|\geq1$.
For the analysis of secondary instability with the 2D bi-global LSA presented in the following subsections, 2D DNS are also conducted by considering the axisymmetric modes only (i.e. $M=0$ and $N_{\theta}=1$).  

\subsection{Non-linear saturation to axisymmetric Taylor vortices}
\begin{figure}
  \centerline{
      \includegraphics[height=4.3cm]{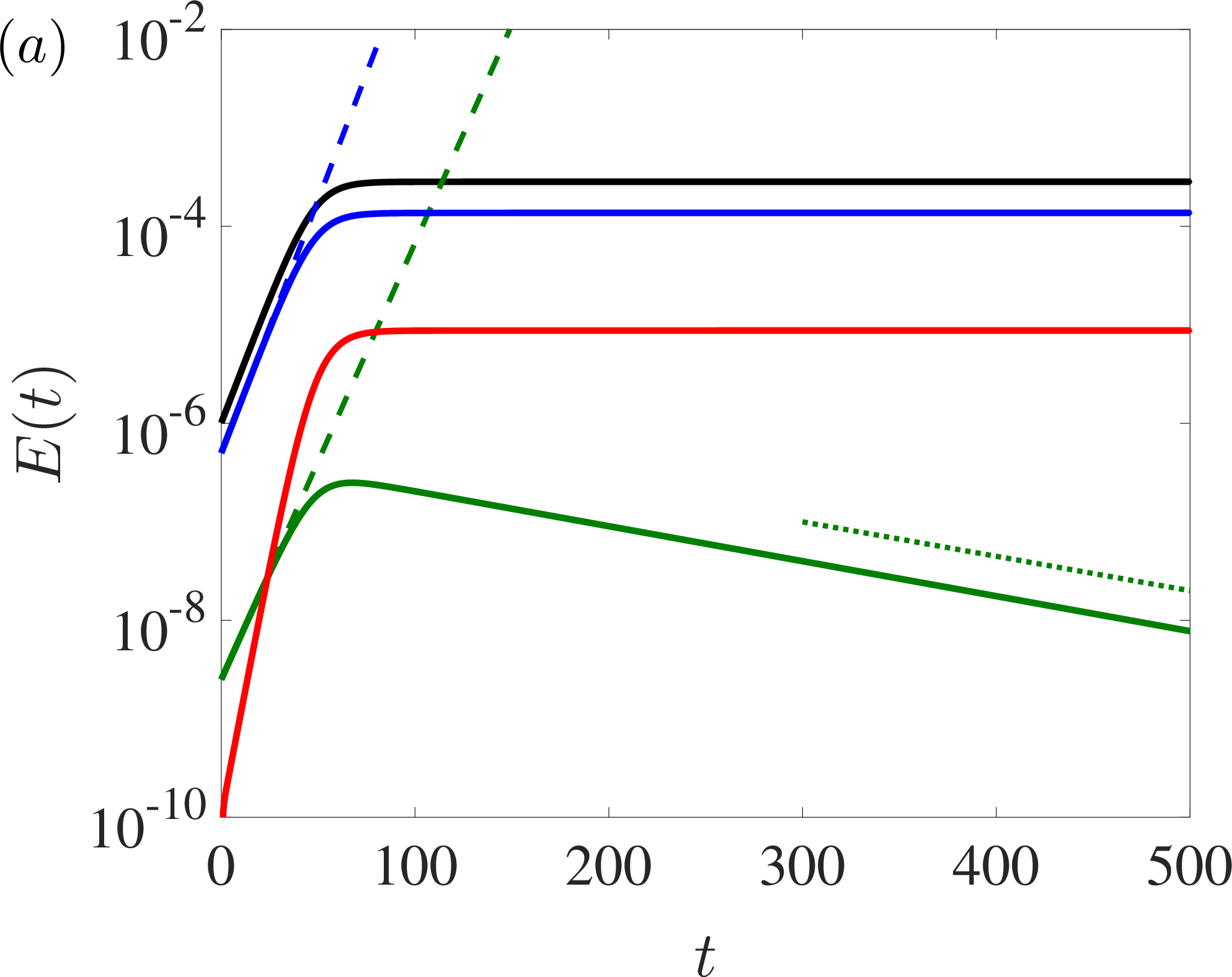}
      \includegraphics[height=4.3cm]{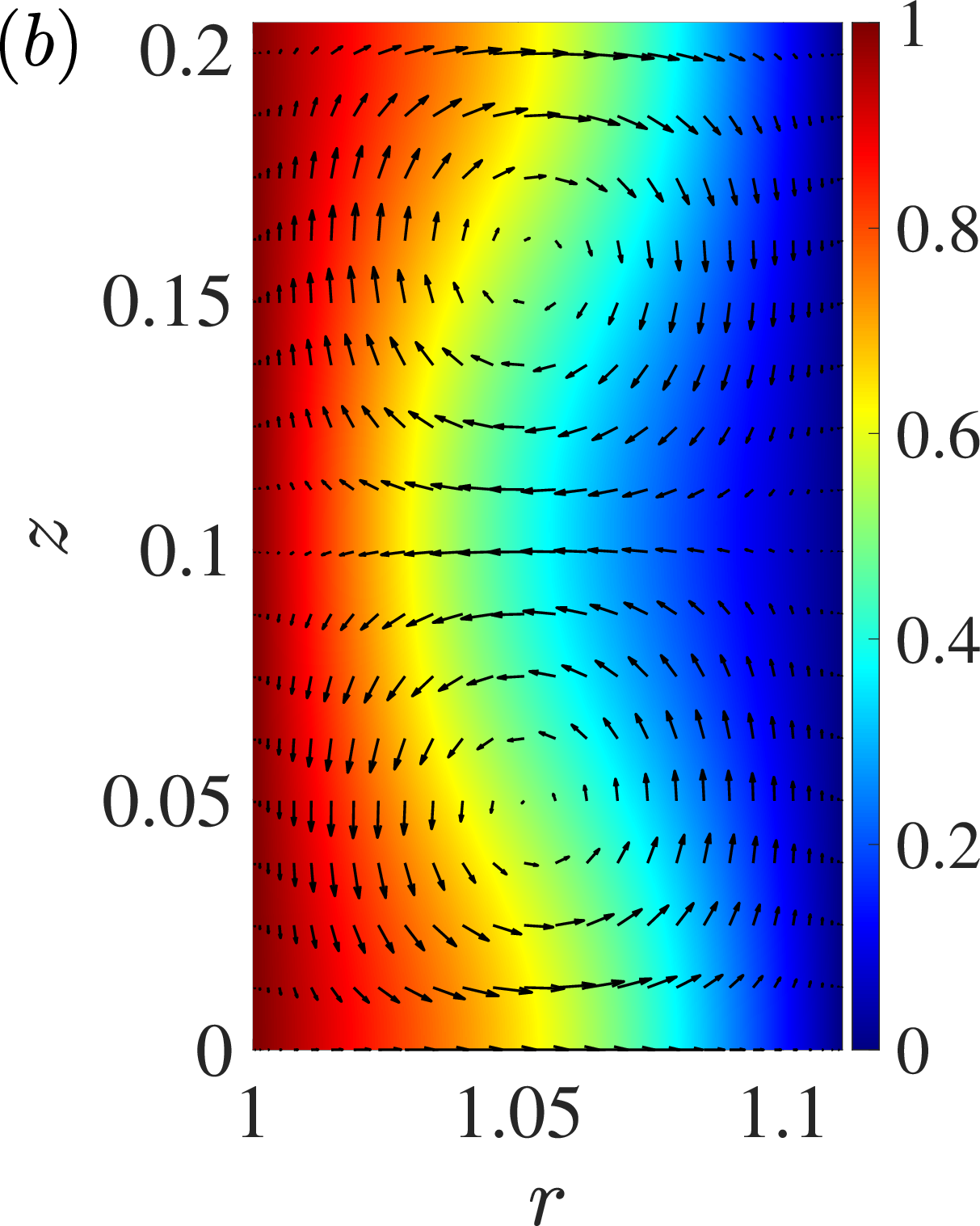}
      \includegraphics[height=4.3cm]{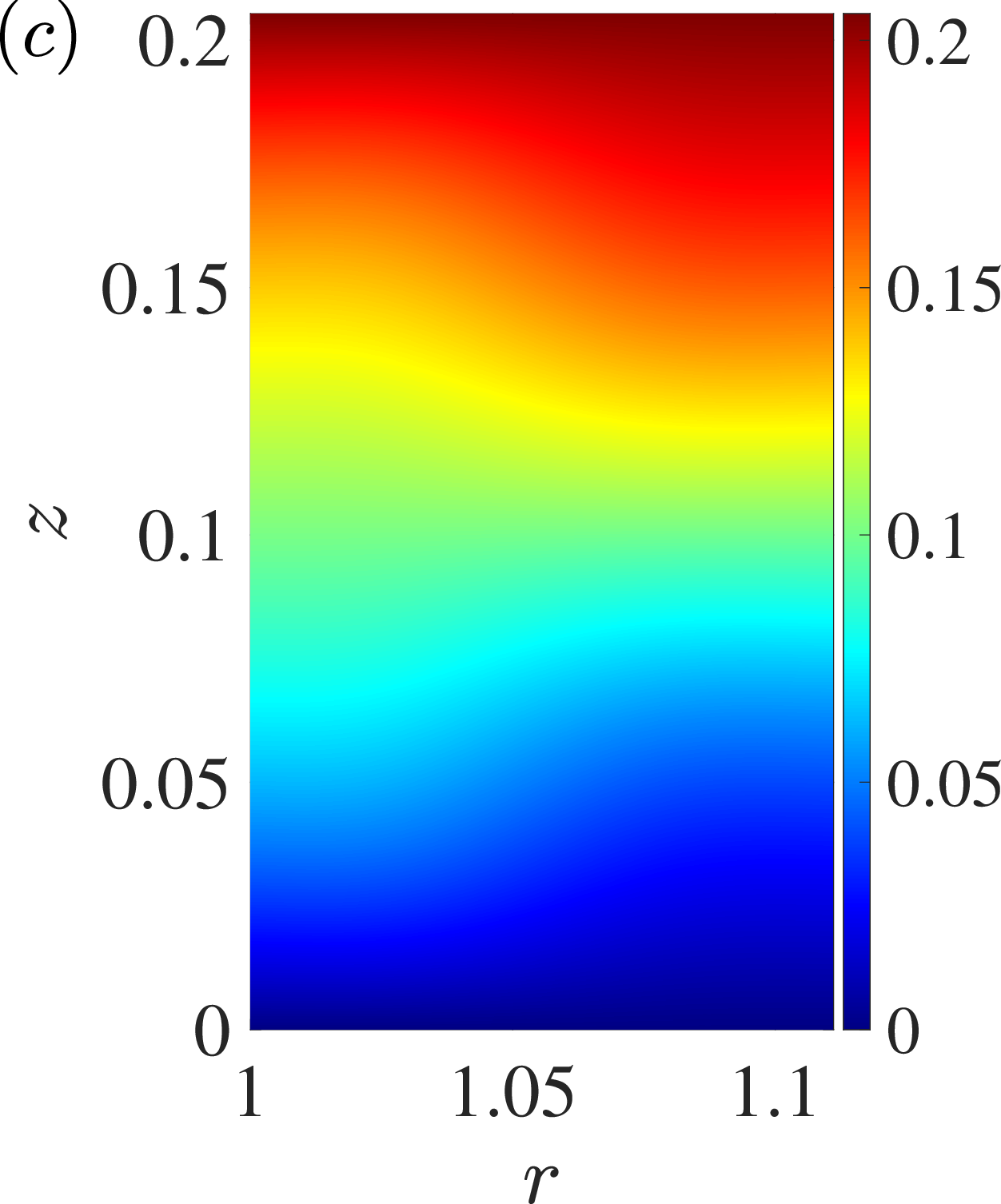}
  }
  \caption{
  (\textit{a}) Time evolution of the total energy $E(t)$ (black) and representative modal energy components: $\tilde{E}_{01}$ (blue), $\tilde{E}_{11}$ (green), and $\tilde{E}_{00}$ (red) for Case 1. 
  The dashed lines denote 1D LSA predictions on the growth of the modes with ($m=0$) (blue) and $m=1$ (green). 
  The green dotted line denotes a 2D bi-global LSA prediction on the decay of the non-axisymmetric mode ($m=1$). 
  (\textit{b}) Velocity field on the plane $(r,z)$ at $\theta=0$ and $t=500$ with contours denoting the total azimuthal velocity $U_{\theta}$ and vector plot denoting the transverse velocity field $(u_{r},u_{z})$.
  (\textit{c}) The total temperature profile $\Uptheta(r,z)$ at $t=500$. 
  }
\label{fig:saturation}
\end{figure}
Figure \ref{fig:saturation}(\textit{a}) shows an example of the time evolution of the total energy $E(t)$ for Case 1 and a few representative examples of the modal energy $\hat{E}_{jl}$ for the axisymmetric mode with $(j,l)=(0,1)$, the non-axisymmetric mode with $(j,l)=(1,1)$, and the mean-flow distortion with $(j,l)=(0,0)$.
It is reminded that the indices $j$ and $l$ are from the azimuthal wavenumber $m_{j}=jm$ and axial wavenumber $k_{l}=lk$.
At the initial stage, the perturbation energy is small enough and the total energy grows exponentially as $E(t)\sim\exp(2\Imag(\omega_{01})t)$ where $\omega_{01}$ is the growth rate of the most unstable axisymmetric mode (i.e. $j=0$ and $l=1$). 
A good agreement between the LSA prediction and DNS is clearly shown in Figure \ref{fig:saturation}$(a)$ for the growth of the $m=0$ and $m=1$ modes at early stage.  
Once the energy of the axisymmetric mode increases and saturates, the flow reaches an equilibrium state featuring axisymmetric Taylor vortices as shown in Figure \ref{fig:saturation}($b$).
In this saturation process, the mean-flow distortion also grows and its energy $\hat{E}_{00}$ saturates. 
While the energies of the axisymmetric mode and mean-flow distortion remain constant at large $t$, the energy of the non-axisymmetric mode with $(j,l)=(1,1)$ decays as the Taylor vortices appear.
It is not shown here but other modes with higher $m$ or $k$ also decay as the Taylor vortices appear. 
Using the Taylor vortices at saturation as a new base state, the bi-global LSA can predict the decay of non-axisymmetric modes with the index $l=1$ as we see a good agreement on the decay rate between the DNS and the bi-global LSA predictions.
In Figure \ref{fig:saturation}($c$), the total temperature $\Uptheta(r,z)=T_{B}+T$ at the equilibrium state is displayed.
For Case 1, thermal diffusion is moderate with $Pr=1$ and thus the temperature perturbation $T$ can grow and have an amplitude comparable to $T_{B}$ at the equilibrium. 
In this case, the total temperature $\Uptheta$ no longer varies linearly but is affected and mixed by the Taylor vortices; i.e. the temperature $\Uptheta$ increases (decreases) in the region where the direction of the velocity is upwards (downwards).   
The Taylor vortices in stratified fluids lead to the baroclinicity with a radial temperature gradient; however, for Case 1 with moderate $N=1$ and $Re_{i}=145$, the overturning of the temperature does not occur and it does not become secondarily unstable.
Movie 1 is also available in Supplementary Material to demonstrate the non-linear development of the Taylor vortices and perturbation through the saturation process.
It is not shown here but with the same physical parameters of Case 1, different sets of numerical parameters are tested (i.e. a higher resolution case with $(N_{r},N_{\theta},N_{z})=(120,65,65)$ and another case with a longer domain $L_{z}=8\pi/k$ with the resolution $(N_{r},N_{\theta},N_{z})=(60,33,129)$). 
The results are found to be the same qualitatively with the appearance of axisymmetric Taylor vortices without secondary instability and quantitatively with insignificant differences in the energy growth curves.  

\subsection{Bi-global linear stability analysis and secondary instability}
\begin{figure}
  \centerline{
  \includegraphics[width=5.6cm]{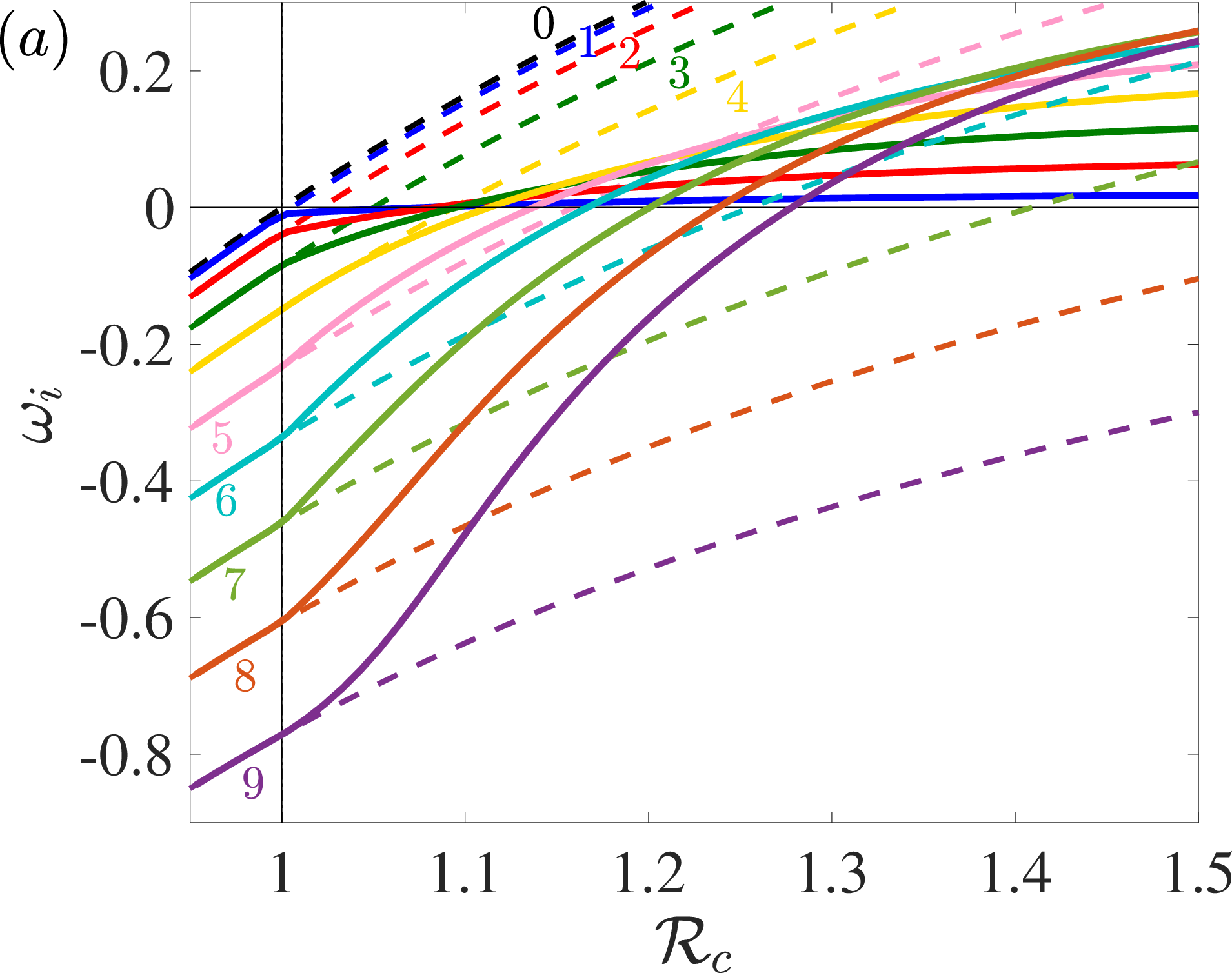}
    \includegraphics[width=5.6cm]{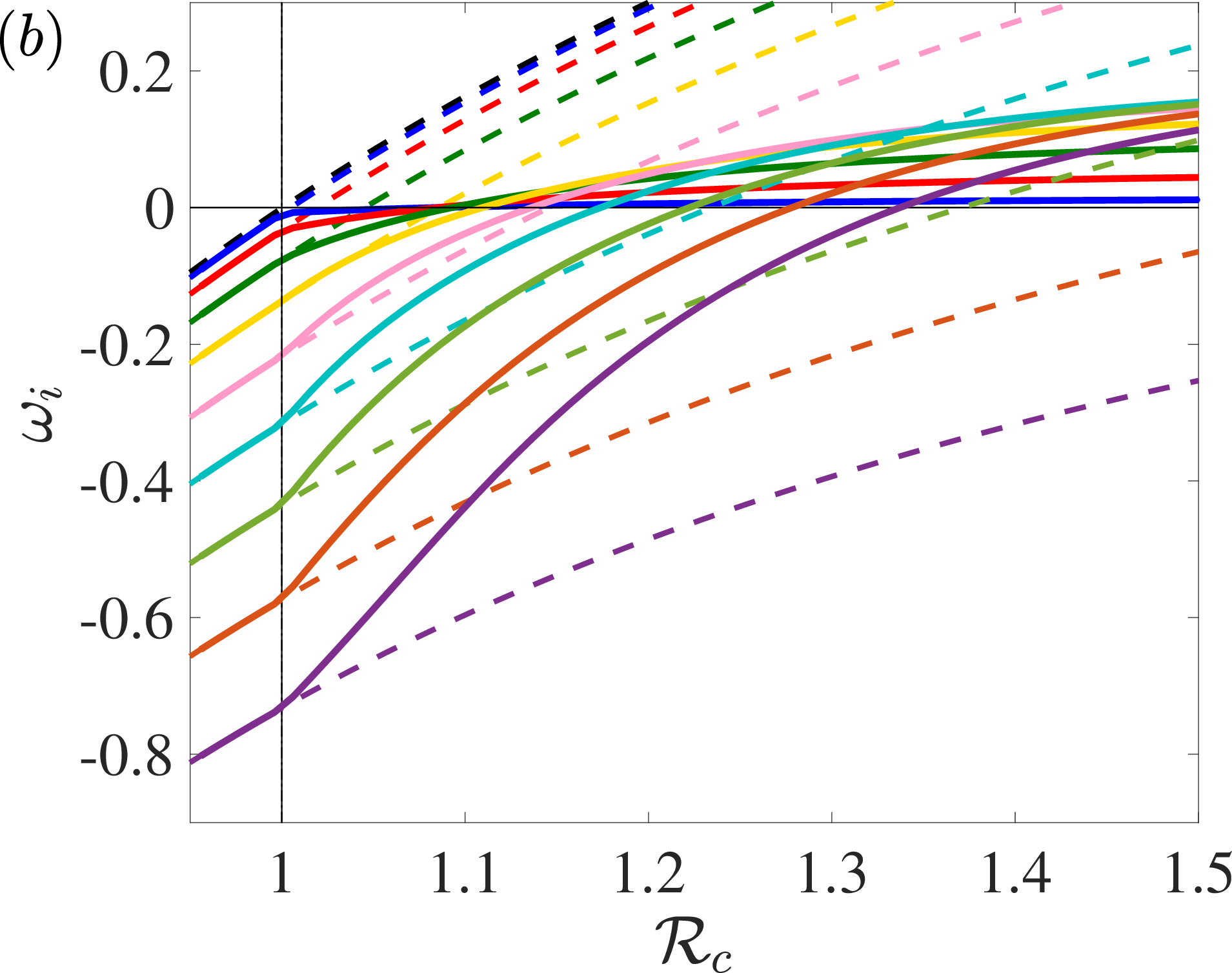}
    }
    \centerline{
        \includegraphics[width=5.6cm]{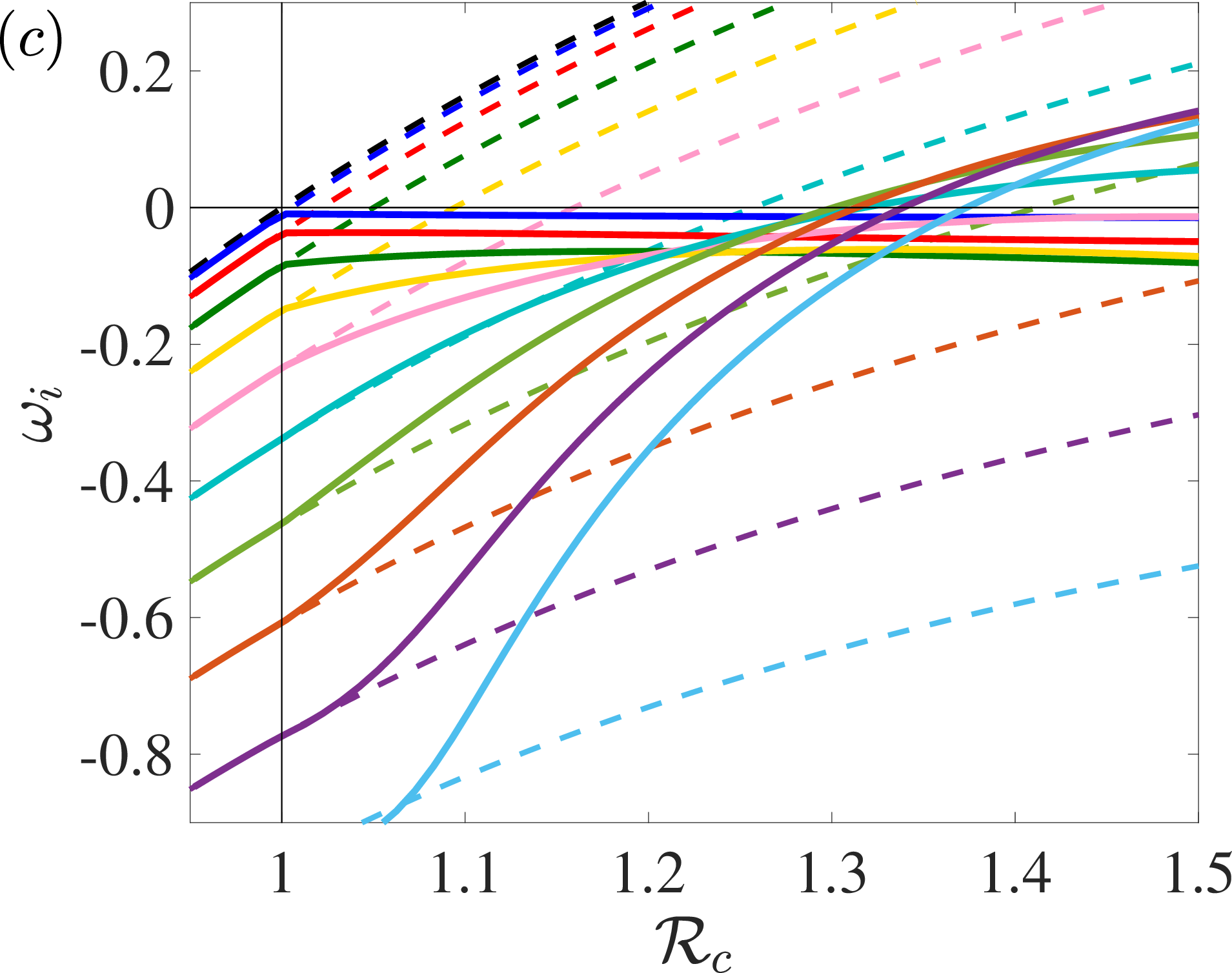}
            \includegraphics[width=5.6cm]{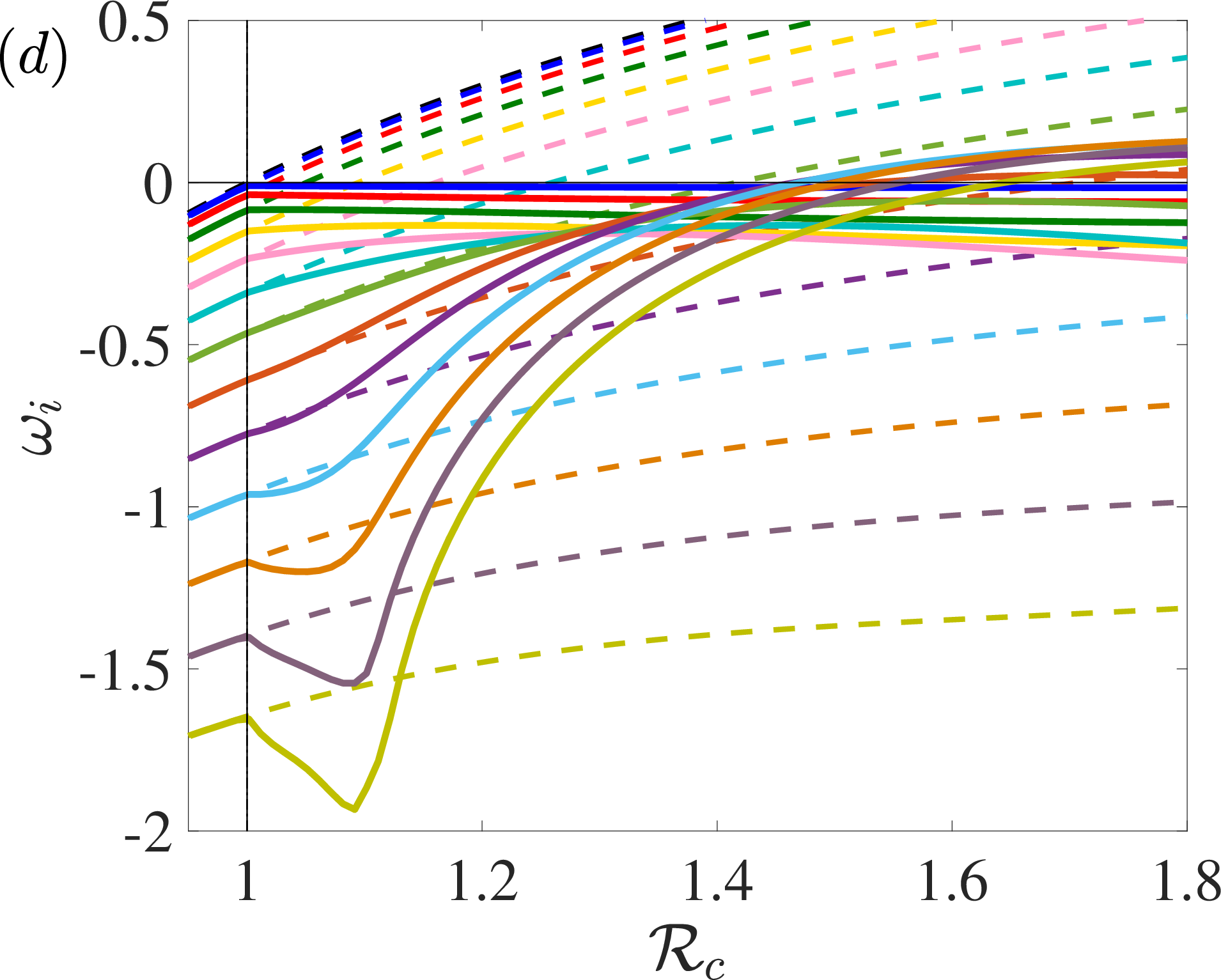}
  }
  \caption{
  Growth rate curves from 1D LSA (dashed lines) and 2D bi-global LSA (solid lines) for $(a)$ unstratified case $N=0$ with numbers denoting $m$, $(b)$ $(N,Pr)=(1,1)$, $(c)$ $(N,Pr)=(1,0.01)$ and $(d)$ $(N,Pr)=(1.5,0.01)$.
  Black dashed lines denote the growth rate of the axisymmetric mode $m=0$ and other coloured curves in the descending order (for $\mathcal{R}_{c}<1$) denote the growth rate for $m=1$ and higher $m$.    
  }
\label{fig:global_sa}
\end{figure}
For the Taylor vortex flow, which is axisymmetric, we can explore its secondary instability by performing a 2D bi-global LSA.  
To obtain the axisymmetric base state $\bar{\boldsymbol{Q}}(r,z)=(\bar{\boldsymbol{U}}(r,z),\bar{T}(r,z))$ for the bi-global LSA, 2D DNS with $N_{\theta}=1$ (i.e., $M=0$) are conducted instead of 3D DNS for computational efficiency.
Another reason for 2D DNS is that the base state $\bar{\boldsymbol{Q}}$ does not become secondarily unstable but remains saturated, thus we can use this steady and axially-periodic base state in the bi-global LSA \citep[see also,][]{Park2011}. 
Figure \ref{fig:global_sa} shows examples of the growth rates of the most unstable modes versus the Reynolds-number ratio $\mathcal{R}_{c}=Re_{i}/Re_{i,c}$. 
Results are computed from the 1D local LSA (dashed lines) and 2D bi-global LSA (solid lines). 
Highly non-axisymmetric modes not appeared in Figure \ref{fig:global_sa} are more stable than the modes shown in Figure \ref{fig:global_sa}.
For every 1D case in Figure \ref{fig:global_sa}, the axisymmetric mode with $m=0$ becomes primarily unstable, as shown by black dashed lines. 
In both 1D and 2D LSA, the axial wavenumber $k=k_{c}$ of the axisymmetric mode at the onset of instability $\mathcal{R}_{c}=1$.
In the bi-global LSA, we use the Taylor vortices with the axial wavenumber $k=k_{c}$, the value at the onset of instability $Re_{i}=Re_{i,c}$, and compute growth rates of non-axisymmetric modes with $m>0$ by increasing $\mathcal{R}_{c}$.    
For $\mathcal{R}_{c}<1$, growth rates of non-axisymmetric modes computed from the bi-global LSA are the same as those from the 1D LSA since the axisymmetric Taylor vortices are not developed and the base state obtained from 2D DNS is essentially cylindrical Couette flow (\ref{eq:base}). 
For the unstratified case with $N=0$ in Figure \ref{fig:global_sa}(\textit{a}), the axisymmetric mode becomes unstable at $Re_{i,c}\simeq131.6$. 
As the axisymmetric Taylor vortices appear for $Re_{i}>Re_{i,c}$, the growth rates of non-axisymmetric modes for $m\geq1$ are attenuated by this new base state as shown by coloured solid lines in Figure \ref{fig:global_sa}(\textit{a}).
In the presence of the Taylor vortices, growth rates of weakly non-axisymmetric modes for $1\leq m\leq 4$ increase slowly with $\mathcal{R}_{c}$ and the non-axisymmetric mode with $m=1$ becomes unstable at $\mathcal{R}_{c}=1.075$ (i.e. $Re_{i}=Re_{i,2}\simeq141.5$, where $Re_{i,2}$ denotes the secondary critical Reynolds number at which a non-axisymmetric mode becomes secondarily unstable). 
When $\mathcal{R}_{c}=1.075$, the growth rates of highly non-axisymmetric modes increase faster with $\mathcal{R}_{c}$ for the base flow case with Taylor vortices than the case with cylindrical Couette flow.
This implies that highly non-axisymmetric modes are strongly promoted by the axisymmetric Taylor vortices for large $\mathcal{R}_{c}$. 
However, the $m=1$ mode is the second unstable mode and it is difficult to predict in advance which non-axisymmetric mode becomes the next dominant mode for $Re_{i}>Re_{i,2}$ as non-linear interactions involving the growth of the $m=1$ mode will lead to a new non-axisymmetric base state.

Characteristics of secondary instability are similar for a stratified case with $(N,Pr)=(1,1)$ in Figure \ref{fig:global_sa}(\textit{b}), the case where the axisymmetric mode becomes primarily unstable at a higher $Re_{i,c}\simeq140.10$ and the non-axisymmetric mode with $m=1$ becomes secondarily unstable at a higher Reynolds-number ratio $\mathcal{R}_{c}=1.085$ (i.e. $Re_{i,2}\simeq152.1$) than the unstratified case. 
For a highly-diffusive case with $Pr=0.01$ in Figure \ref{fig:global_sa}(\textit{c}), characteristics of secondary instability change as weakly non-axisymmetric modes with $m=1$ and $2$ are stabilised by the Taylor vortices while a highly non-axisymmetric mode with $m=7$ becomes secondarily unstable at $\mathcal{R}_{c}=1.3$ (i.e, $Re_{i,2}=171.31$).
This implies that the onset of secondary instability is delayed by strong thermal diffusion at $Pr=0.01$ and highly non-axisymmetric modes become dominant while weakly non-axisymmetric modes are suppressed. 
The onset of secondary instability is further delayed as stratification becomes stronger with $N=1.5$ as shown in Figure \ref{fig:global_sa}(\textit{d}), the case where the Taylor vortices become unstable by the $m=9$ mode at a higher ratio with $\mathcal{R}_{c}=1.466$ (i.e. $Re_{i,2}=193.25$).  

The prominence of highly non-axisymmetric modes, which become dominant over weakly non-axisymmetric modes with low $m$, can be understood by conducting the energetics analysis.  
Similar to the equation (\ref{eq:energy_evolution_ptb}), one can obtain an equation for the temporal evolution of perturbation energy from the linearised perturbation equations (\ref{eq:lpe_2D_continuity})-(\ref{eq:lpe_2D_energy}) as
\begin{equation}
\label{eq:energy_evolution_ptb_2D}
\frac{\partial \bar{E}}{\partial t}=\mathcal{P}_{\bar{U}}+\mathcal{P}_{\bar{V}}+\mathcal{P}_{\bar{W}}+\mathcal{P}_{\bar{\mathcal{T}}}-\mathcal{D}_{k}-\mathcal{D}_{p},
\end{equation}
where $\mathcal{P}_{\bar{U},\bar{V},\bar{W},\bar{\mathcal{T}}}$ are the production terms induced by axisymmetric Taylor vortices with $(\bar{U},\bar{V},\bar{W},\bar{\mathcal{T}})$ and $\mathcal{D}_{K,P}$ are the viscous and thermal dissipations terms, respectively, all of which are defined as follows:
\begin{eqnarray}
&&\mathcal{P}_{\bar{U}}=-\left<\frac{\partial\bar{U}}{\partial r}\bar{u}^{2}+\frac{\bar{U}}{r}\bar{v}^{2}+\frac{\partial\bar{U}}{\partial z}\bar{u}\bar{w}\right>,~
\mathcal{P}_{\bar{V}}=-\left<\left(\frac{\partial\bar{V}}{\partial r}-\frac{\bar{V}}{r}\right)\bar{u}\bar{v}+\frac{\partial\bar{V}}{\partial z}\bar{v}\bar{w}\right>,\nonumber\\
&&
\mathcal{P}_{\bar{W}}=-\left<\frac{\partial\bar{W}}{\partial r}\bar{u}\bar{w}+\frac{\partial\bar{W}}{\partial z}\bar{w}^{2}\right>,~
\mathcal{P}_{\bar{\mathcal{T}}}=-N^{2}\left<\frac{\partial\bar{\mathcal{T}}}{\partial r}\bar{u}\bar{T}+\frac{\partial\bar{\mathcal{T}}}{\partial z}\bar{w}\bar{T}\right>,\nonumber\\
&&\mathcal{D}_{K}=\frac{1}{Re}\left<{\nabla}\bar{\textbf{u}}~\textbf{:}~{\nabla}\bar{\textbf{u}}\right>,~
\mathcal{D}_{P}=\frac{N^{2}}{RePr}\left<{\nabla}\bar{T}~\textbf{:}~{\nabla}\bar{T}\right>.
\end{eqnarray}
We note that the dissipation terms $\mathcal{D}_{K,P}$ are always positive and play a stabilising role while the production terms $\mathcal{P}_{\bar{U},\bar{V},\bar{W},\bar{\mathcal{T}}}$ are not necessarily positive as they depend on the correlation between the base state $(\bar{U},\bar{V},\bar{W},\bar{\mathcal{T}})$ and perturbation variables $(\bar{u},\bar{v},\bar{w},\bar{T})$.
Applying the normal mode (\ref{eq:normal_mode_global}) to the above equation (\ref{eq:energy_evolution_ptb_2D}) and considering the normalisation $\hat{E}_{m}=1$ where $\hat{E}_{m}$ is the modal energy defined as
\begin{equation}
\hat{E}_{m}=\left<|\hat{u}_{m}|^{2}+|\hat{b}_{m}|^{2}+|\hat{w}_{m}|^{2}+N^{2}|\hat{T}_{m}|^{2}\right>_{rz},~\left<\hat{\textbf{X}}_{m}\right>_{rz}=\int_{0}^{L_{z}}\int_{1}^{1/\eta}\hat{\textbf{X}}_{m} ~r\mathrm{d}r\mathrm{d}z,
\end{equation}
we obtain the following relation between the growth rate $\omega_{m,i}$ and the modal contribution terms
\begin{equation}
\label{eq:growth_rate_balance_2D}
\omega_{m,i}=\hat{\mathcal{P}}_{m,\bar{U}}+\hat{\mathcal{P}}_{m,\bar{V}}+\hat{\mathcal{P}}_{m,\bar{W}}+\hat{\mathcal{P}}_{m,\bar{\mathcal{T}}}-\hat{\mathcal{D}}_{m,K}-\hat{\mathcal{D}}_{m,P},
\end{equation}
where
\begin{eqnarray}
\label{eq:contribution_terms}
&&\hat{\mathcal{P}}_{m,\bar{U}}=-\left<\frac{\partial\bar{U}}{\partial r}|\hat{u}_{m}|^{2}+\frac{\bar{U}}{r}|\hat{v}_{m}|^{2}+\frac{\partial\bar{U}}{\partial z}\left(\frac{\hat{u}_{m}^{*}\hat{w}_{m}+\hat{u}_{m}\hat{w}^{*}_{m}}{2}\right)\right>_{rz},\nonumber\\
&&\hat{\mathcal{P}}_{m,\bar{V}}=-\left<\left(\frac{\partial\bar{V}}{\partial r}-\frac{\bar{V}}{r}\right)\left(\frac{\hat{u}_{m}^{*}\hat{v}_{m}+\hat{u}_{m}\hat{v}^{*}_{m}}{2}\right)+\frac{\partial\bar{V}}{\partial z}\left(\frac{\hat{v}_{m}^{*}\hat{w}_{m}+\hat{v}_{m}\hat{w}^{*}_{m}}{2}\right)\right>_{rz},\nonumber\\
&&
\hat{\mathcal{P}}_{m,\bar{W}}=-\left<\frac{\partial\bar{W}}{\partial r}\left(\frac{\hat{u}_{m}^{*}\hat{w}_{m}+\hat{u}_{m}\hat{w}^{*}_{m}}{2}\right)+\frac{\partial\bar{W}}{\partial z}|\hat{w}_{m}|^{2}\right>_{rz},\nonumber\\
&&
\hat{\mathcal{P}}_{m,\bar{\mathcal{T}}}=-N^{2}\left<\frac{\partial\bar{\mathcal{T}}}{\partial r}\left(\frac{\hat{u}_{m}^{*}\hat{T}_{m}+\hat{u}_{m}\hat{T}^{*}_{m}}{2}\right)+\frac{\partial\bar{\mathcal{T}}}{\partial z}\left(\frac{\hat{w}_{m}^{*}\hat{T}_{m}+\hat{w}_{m}\hat{T}^{*}_{m}}{2}\right)\right>_{rz},\nonumber\\
&&\hat{\mathcal{D}}_{m,K}=\frac{1}{Re}\left<\hat{\nabla}_{m}\hat{\textbf{u}}_{m}~\textbf{:}~\hat{\nabla}_{m}\hat{\textbf{u}}_{m}\right>_{rz},~
\hat{\mathcal{D}}_{m,P}=\frac{N^{2}}{RePr}\left<\hat{\nabla}_{m}\hat{T}_{m}~\textbf{:}~\hat{\nabla}_{m}\hat{T}_{m}\right>_{rz}.
\end{eqnarray}

\begin{table}
  \begin{center}
\def~{\hphantom{0}}
  \begin{tabular}{ccccccccccccc}
  \hline
      $Re_{i}$ & $N$ & $Pr$ & $m$ & Case & $\omega_{m,i}$  & $\hat{\mathcal{P}}_{m,\bar{U}}$   &  $\hat{\mathcal{P}}_{m,\bar{V}}$ & $\hat{\mathcal{P}}_{m,\bar{W}}$ & $\hat{\mathcal{P}}_{m,\bar{T}}$ & $\hat{\mathcal{D}}_{m,K}$ & $\hat{\mathcal{D}}_{m,P}$  & The sum (\ref{eq:growth_rate_balance_2D})\\
      \hline
       200 & 1 & 0.01 & 0 & 1D LSA  &  0.6274 & 0 & 1.5854 & 0 & 0 & 0.9577 & 0.0003 & 0.6273 \\
       200 & 1 & 0.01 & 1 & 1D LSA  &  0.6165 & 0 & 1.5821 & 0 & 0 & 0.9654 & 0.0004 & 0.6164\\
       200 & 1 & 0.01 & 1 & 2D LSA  &  -0.0147 & -0.0005 & 1.1594 & 0.0030 & 0.0000 & 1.1764 & 0.0005 & -0.0150\\
       200 & 1 & 0.01 & 5 & 2D LSA  &  -0.0134 & 0.0112 & 1.2137 & -0.0219 & 0.0000 & 1.2121 & 0.0046& -0.0136 \\
       200 & 1 & 0.01 & 9 & 2D LSA  &  0.1525 & 0.0335 & 1.4639 & -0.0641 & 0.0000 & 1.2700 & 0.0110 & 0.1522 \\
       200 & 1 & 0.01 & 13 & 2D LSA  &  -0.0303 & 0.0371 & 1.4953 & -0.1295 & 0.0000 & 1.4202 & 0.0133 & -0.0306 \\
  \hline
       200 & 1 & 1 & 0 & 1D LSA  &  0.5441 & 0 & 1.5699 & 0 & 0 & 1.0143 & 0.0115& 0.5441\\
       200 & 1 & 1 & 1 & 1D LSA  &  0.5346 & 0 & 1.5677 & 0 & 0 & 1.0213 & 0.0119 & 0.5345\\
       200 & 1 & 1 & 1 & 2D LSA  &  0.0105 & 0.0008 & 1.1757 & 0.0018 & 0.0269 & 1.1470 & 0.0479 & 0.0103\\
       200 & 1 & 1 & 5 & 2D LSA  &  0.1336 & 0.0147 & 1.3567 & -0.0086 & 0.0192 & 1.2128 & 0.0360 & 0.1334 \\
       200 & 1 & 1 & 9 & 2D LSA  &  0.0736 & 0.0259 & 1.4480 & -0.0430 & 0.0112 & 1.3309 & 0.0377 & 0.0734 \\
       200 & 1 & 1 & 13 & 2D LSA  &  -0.1690 & 0.0267 & 1.4451 & -0.0981 & 0.0054 & 1.5029 & 0.0453 & -0.1692 \\
       \hline
  \end{tabular}
  \caption{Growth rates $\omega_{m,i}$ and contribution terms computed from 1D local LSA and 2D bi-global LSA for $(Re_{i},N,Pr)=(200,1,0.01)$ and $(Re_{i},N,Pr)=(200,1,1)$.}
  \label{tab:2D_growth_rate}
  \end{center}
\end{table}
Examples of the growth rate $\omega_{i}$ and instability contribution terms (\ref{eq:contribution_terms}) computed from the eigenfunction $\hat{\textbf{q}}_{m}$ are presented in Table \ref{tab:2D_growth_rate}.
In the Table, the growth rate $\omega_{m,i}$ computed from the eigenvalue problems (\ref{eq:evp}) and (\ref{eq:evp_biglobal}) and the sum on the right hand side of (\ref{eq:growth_rate_balance_2D}), which is obtained by integrating the eigenfunction computed from either (\ref{eq:evp}) for 1D LSA cases or (\ref{eq:evp_biglobal}) for 2D LSA cases, are in good agreement with a very small difference of the order $O(10^{-4})$.
For 1D LSA cases in which the base flow is cylindrical Couette flow, the axisymmetric mode with $m=0$ is most unstable due to the largest production $\hat{\mathcal{P}}_{m,\bar{V}}$ and the least total dissipation $\hat{\mathcal{D}}_{m}=\hat{\mathcal{D}}_{m,K}+\hat{\mathcal{D}}_{m,P}$.
This applies to both cases with $Pr=0.01$ and $Pr=1$ although the thermal dissipation $\hat{\mathcal{D}}_{m,P}$ is small when $Pr$ is small. 
As $m$ increases, the production $\hat{\mathcal{P}}_{m,\bar{V}}$ decreases while the dissipation $\hat{\mathcal{D}}_{m}$ increases, which leads to the decrease of the growth rate. 
For $m=1$, we compare the 1D LSA cases against the 2D LSA cases where the base flow is 2D axisymmetric Taylor vortices. 
It is clearly shown that the growth rate decreases as the flow becomes two-dimensional with decrease in the total production term $\hat{\mathcal{P}}_{m}=\hat{\mathcal{P}}_{m,\bar{U}}+\hat{\mathcal{P}}_{m,\bar{V}}+\hat{\mathcal{P}}_{m,\bar{W}}+\hat{\mathcal{P}}_{m,\bar{T}}$ and increase in the total dissipation $\hat{\mathcal{D}}_{m}$.  
While other production terms such as $\hat{\mathcal{P}}_{m,\bar{U}}$, $\hat{\mathcal{P}}_{m,\bar{W}}$ and $\hat{\mathcal{P}}_{m,\bar{T}}$ appear for two-dimensional cases due to the velocity and temperature gradients of the Taylor vortices in the radial and axial directions (see also, Figure \ref{fig:saturation}\textit{b,c}), their contribution to the growth rate is smaller than the contribution from the azimuthal velocity $\bar{V}$ (i.e. $\hat{\mathcal{P}}_{m,\bar{V}}$).  
What is noteworthy in secondary instability of axisymmetric Taylor vortices is that the total dissipation $\hat{\mathcal{D}}_{m}$ increases monotonically with $m$ while the total production $\hat{\mathcal{P}}_{m}$ increases and then decreases as $m$ increases. 

\begin{figure}
  \centerline{
  \includegraphics[height=3.6cm]{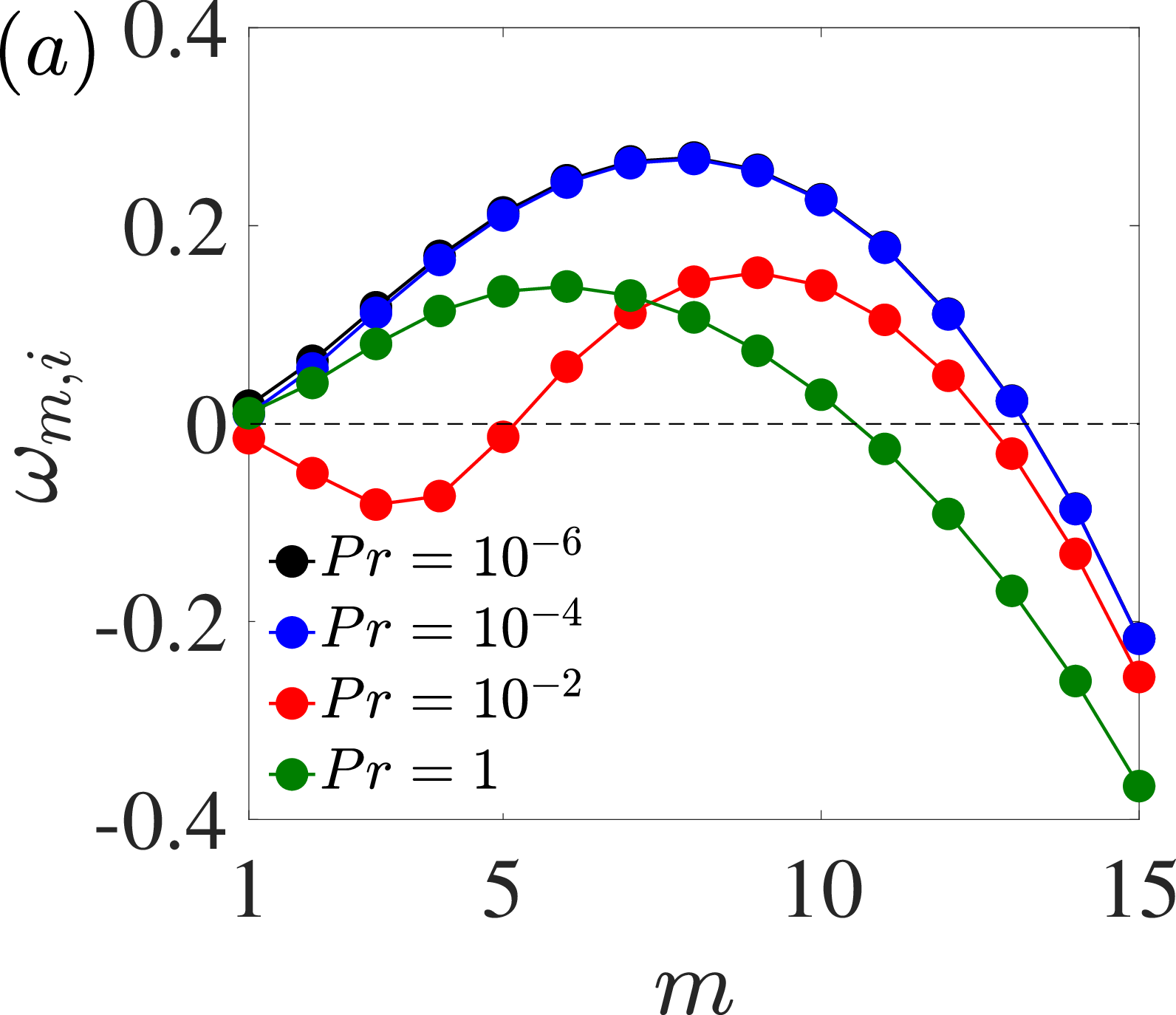}
    \includegraphics[height=3.6cm]{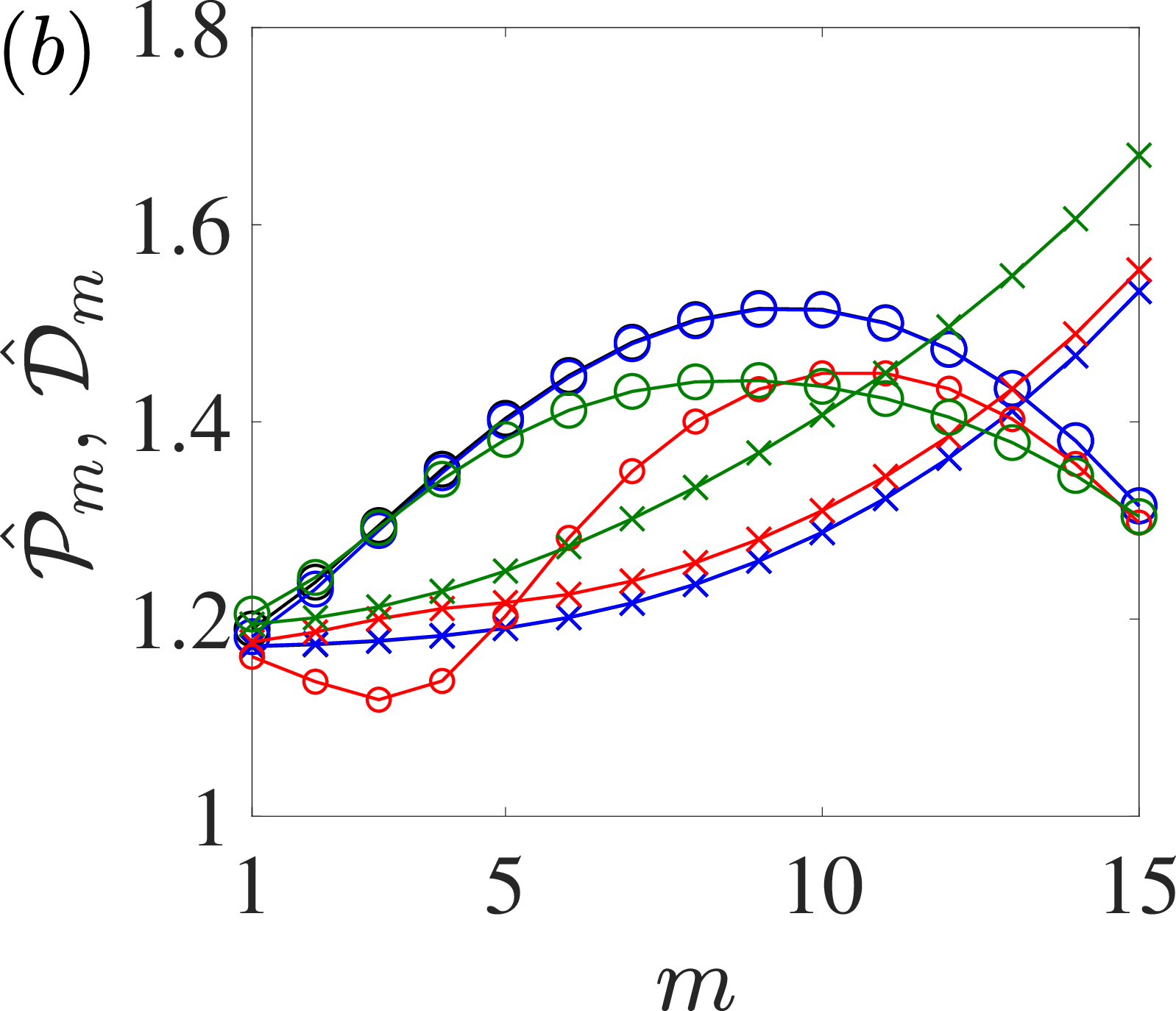}
      \includegraphics[height=3.6cm]{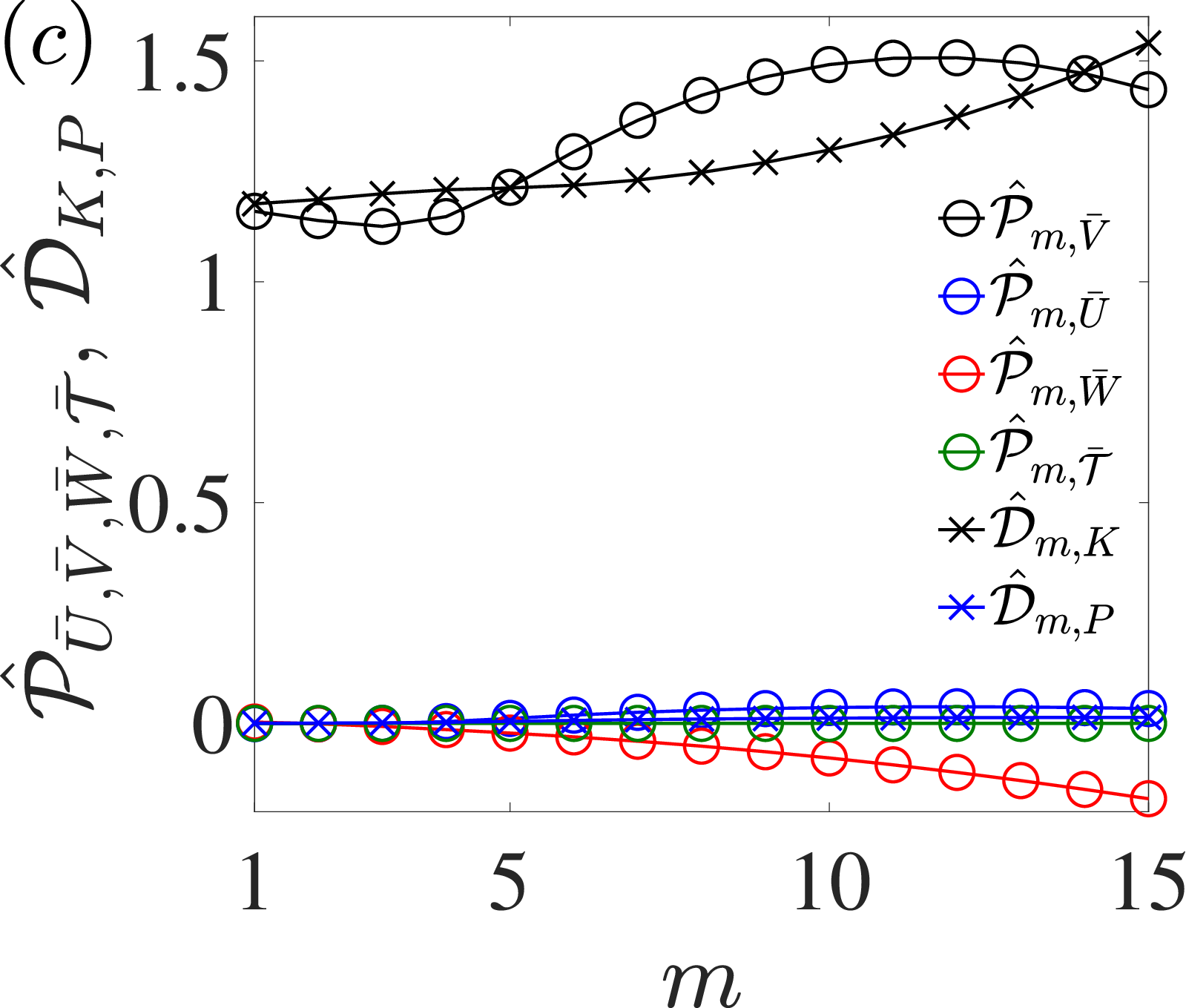}
  }
  \caption{
  (\textit{a},\textit{b}) Growth rate $\omega_{m,i}$ (filled circles in \textit{a}), production $\hat{\mathcal{P}}_{m}$ (empty circles in \textit{b}) and dissipation $\hat{\mathcal{D}}_{m}$ (crosses in \textit{b}) versus azimuthal wavenumber $m$ for different $Pr$: $Pr=10^{-6}$ (black), $10^{-4}$ (blue), $0.01$ (red) and $1$ (green) at $(Re_{i},N)=(200,1)$. 
  (\textit{c}) Production and dissipation terms versus $m$ for $(Re_{i},N,Pr)=(200,1,0.01)$. 
  }
\label{fig:PD_2DLSA}
\end{figure}
Figure \ref{fig:PD_2DLSA}(\textit{a,b}) demonstrate behaviours of the growth rate $\omega_{m,i}$, production $\hat{\mathcal{P}}_{m}$ and dissipation $\hat{\mathcal{D}}_{m}$ with varying $m$ for different $Pr$ at $(Re_{i},N)=(200,1)$. 
Except for $Pr=0.01$, the growth rate $\omega_{m,i}$ increases first and then decreases as $m$ increases (Figure \ref{fig:PD_2DLSA}\textit{a}).
This is due to the fact that the production $\hat{\mathcal{P}}_{m}$ increases only for small $m$ and decreases as $m$ increases further while the dissipation $\hat{\mathcal{D}}_{m}$ increases exponentially with $m$ (Figure \ref{fig:PD_2DLSA}\textit{b}).
This results in the growth rate $\omega_{m,i}$ being positive only in a finite range of $m$.
The growth rates for $Pr=10^{-4}$ and $Pr=10^{-6}$ are very similar to each other and are higher than $\omega_{m,i}$ for $Pr=1$. 
The growth rate behaviour for $Pr=0.01$ is more complicated as $\omega_{m,i}$ is negative for $m\leq5$ before it shows a similar increasing/decreasing trend.
Only highly non-axisymmetric modes in the range $6\leq m\leq 12$ are secondarily unstable for $Pr=0.01$ while non-axisymmetric modes for other $Pr$ cases are unstable in broader ranges of $m$ such as $1\leq m\leq 10$ for $Pr=1$ and $1\leq m\leq 13$ for $Pr=10^{-4}$ and $10^{-6}$. 
This implies that a highly non-axisymmetric flow pattern, as further discussed in the next subsection, can appear for $Pr=0.01$ by secondary instability of Taylor vortices. 
Figure \ref{fig:PD_2DLSA}(\textit{c}) describes variations of all the production and dissipation terms with $m$ at $(Re_{i},N,Pr)=(200,1,0.01)$. 
The dominant production and dissipation terms are $\hat{\mathcal{P}}_{\bar{V}}$ and $\hat{\mathcal{D}}_{K}$ while other terms are small.
The production term $\hat{\mathcal{P}}_{\bar{W}}$ becomes negative as $m>1$ increases implying the stabilizing role of the axial velocity $\bar{W}$.  

\begin{figure}
  \centerline{
  \includegraphics[height=4.4cm]{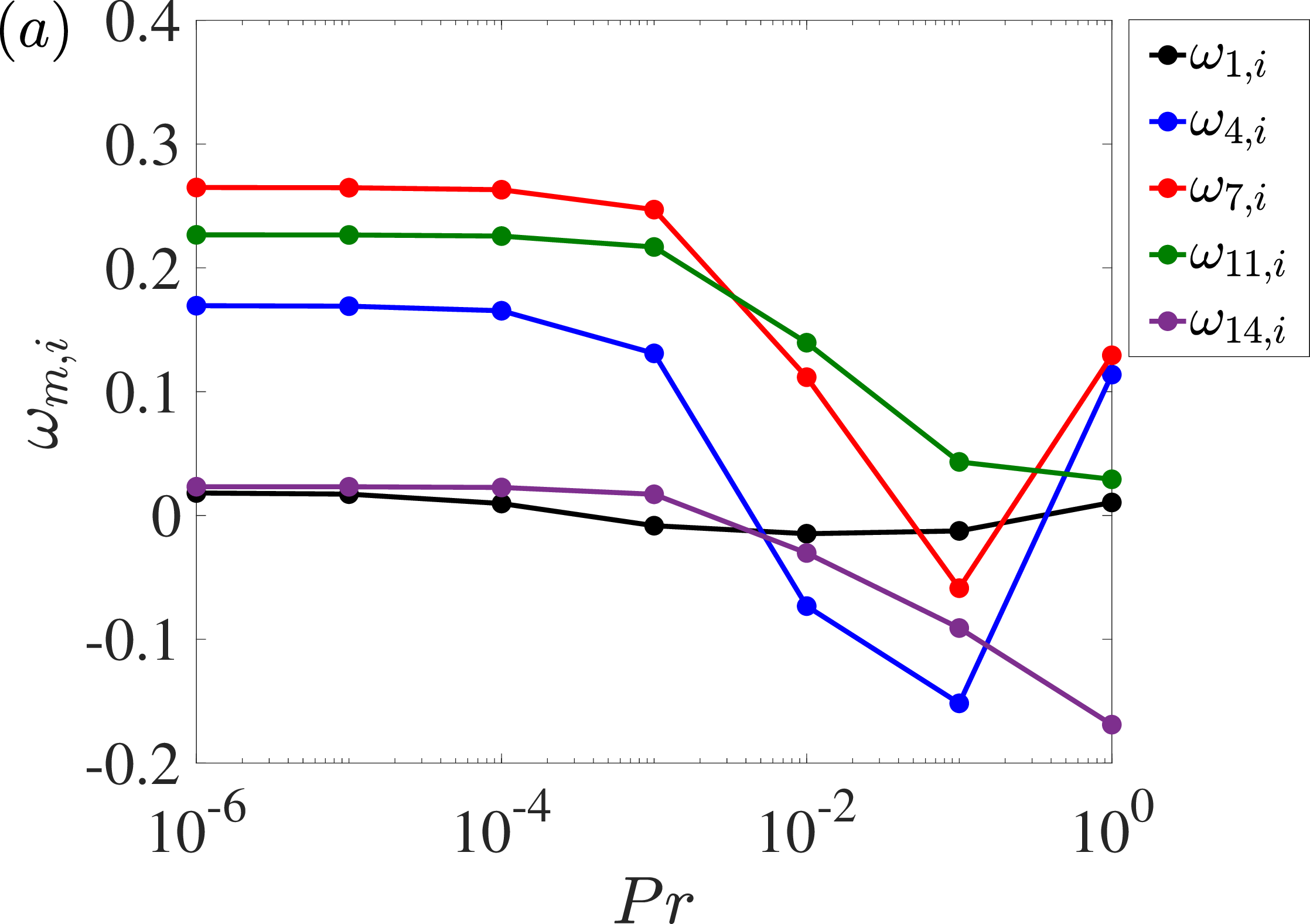}
  \includegraphics[height=4.4cm]{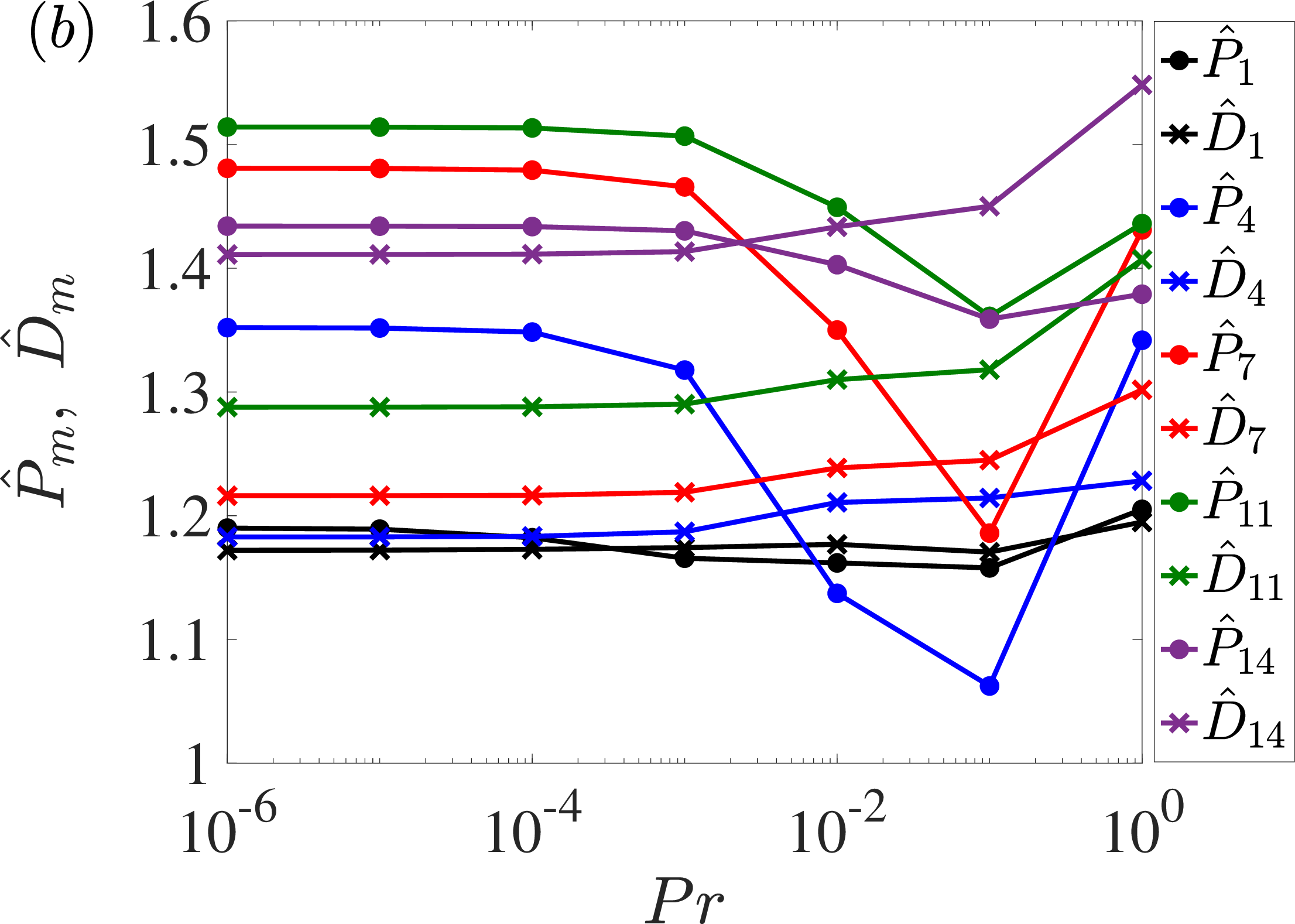}
    }
    \centerline{
      \includegraphics[height=4.4cm]{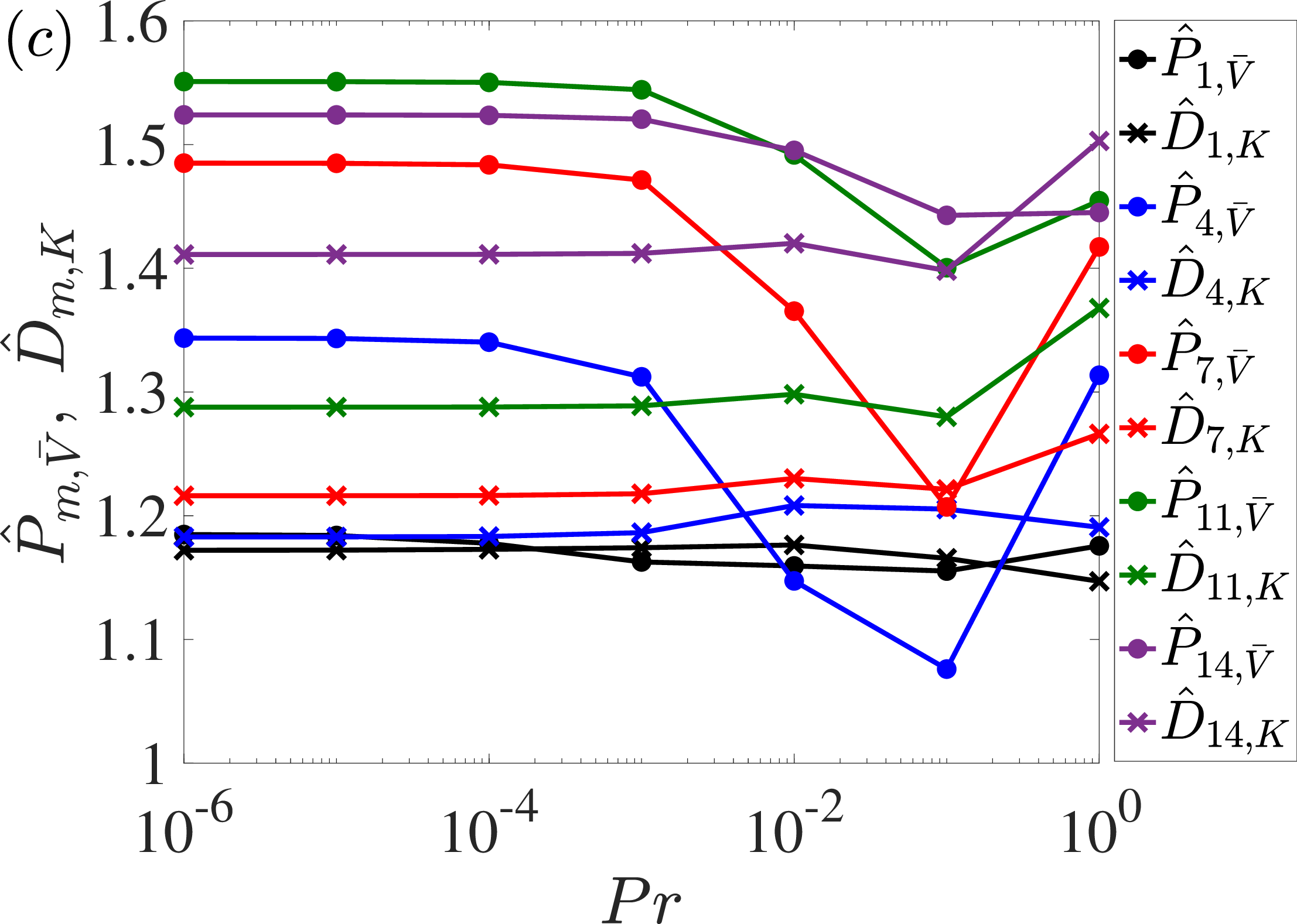}
        \includegraphics[height=4.4cm]{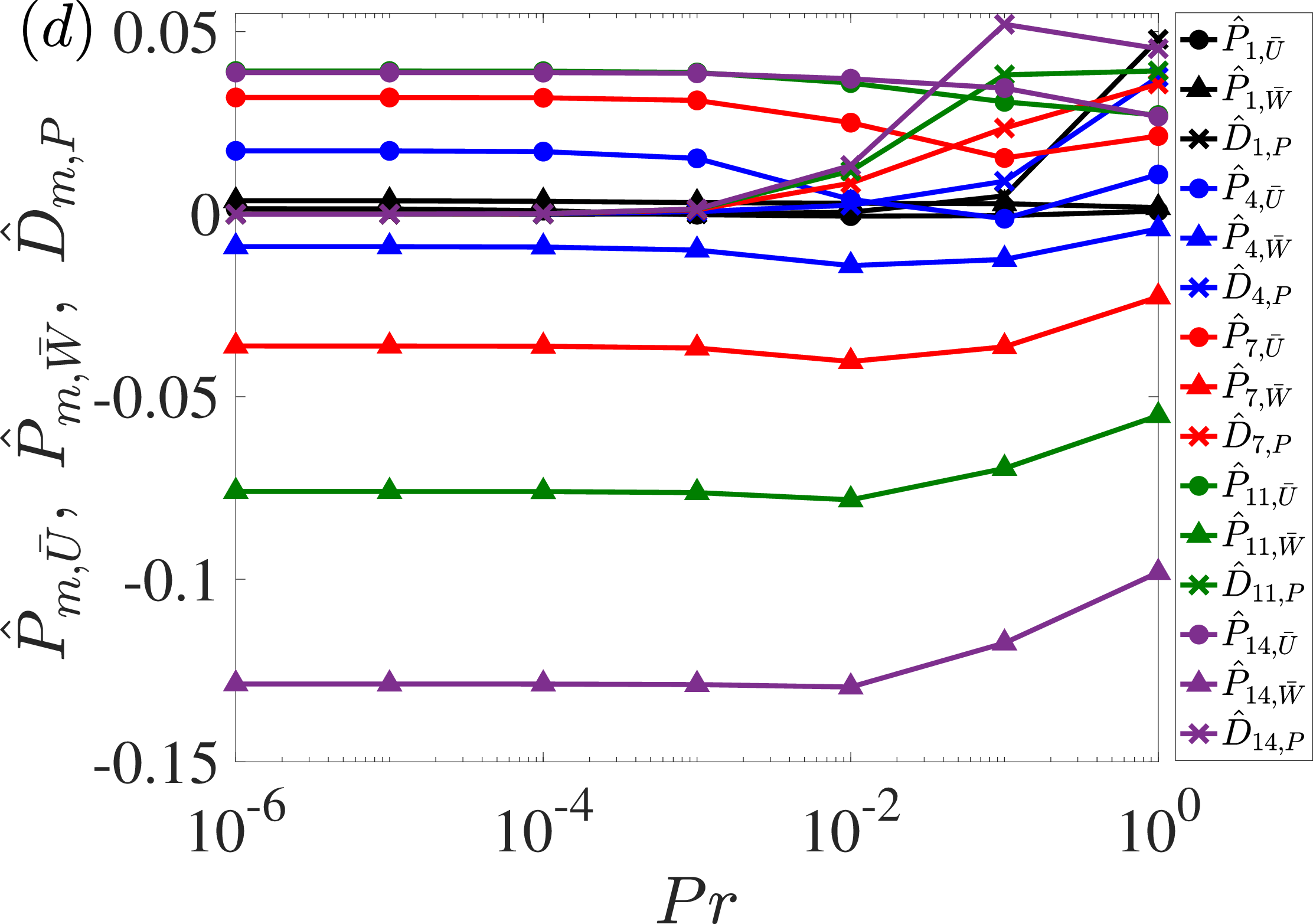}
  }
  \caption{
Variations with the Prandtl number $Pr$ for $(a)$ the growth rate $\omega_{m,i}$, $(b)$ the total production $\hat{P}_{m}$ and dissipation $\hat{D}_{m}$, $(c)$ the dominant production term $\hat{P}_{m,\bar{V}}$ and dissipation term $\hat{D}_{m,K}$ and $(d)$ other production terms $\hat{P}_{m,\bar{U}}$, $\hat{P}_{m,\bar{W}}$ and dissipation term $\hat{D}_{m,P}$ for different $m=1,4,7,11$ and 14 at $Re_{i}=200$ and $N=1$. 
  }
\label{fig:production_dissipation_Pr}
\end{figure}
Figure \ref{fig:production_dissipation_Pr} details how the growth rate, production and dissipation vary with $Pr$ for various non-axisymmetric modes at $Re_{i}=200$ and $N=1$. 
In Figure \ref{fig:production_dissipation_Pr}$(a)$, we see that, for low $m\leq7$, the growth rate $\omega_{m,i}$ decreases as $Pr$ decreases from $Pr=1$ and then it increases as $Pr$ further decrases from $Pr=10^{-2}\sim10^{-1}$.
For high $m\geq8$, the growth rate increases monotonically as $Pr$ decreases.    
Such behaviours are similar for the total production term $\hat{P}_{m}$ while the dissipation term $\hat{D}_{m}$ decreases overall monotically as $Pr$ decreases, as shown in Figure \ref{fig:production_dissipation_Pr}$(b)$.  
As revealed in Figure \ref{fig:PD_2DLSA}$(c)$, the dominant contributions to the production and dissipation come from $\hat{P}_{m,\bar{V}}$ and $\hat{D}_{m,K}$, respectively, which have the magnitude of $O(1)$ (Figure \ref{fig:production_dissipation_Pr}$c$).
The $Pr$-behaviour of $\hat{P}_{m,\bar{V}}$ is similar to that of the total production $\hat{P}_{m}$, which drives the growth rate. 
Although it is not straightfoward to delineate the $Pr$-tendency on which non-axisymmetric mode becomes most unstable, one can infer from Figure \ref{fig:production_dissipation_Pr}$(c)$ that highly non-axisymmetric modes overall contribute to the secondary instability via its interaction with the azimuthal velocity $\bar{V}$ when $Pr$ is sufficiently low as $Pr<10^{-2}$, while the behaviour is more complicated in the range $10^{-2}<Pr<1$.
In Figure \ref{fig:production_dissipation_Pr}($d$), we describe other contribution terms $\hat{P}_{m,\bar{U}}$, $\hat{P}_{m,\bar{W}}$ and $\hat{D}_{m,P}$, which are minor with the magnitude of $O(0.1)$ or less. 
The production $\hat{P}_{m,\bar{U}}$ is overall positive except for $m=1$ around $Pr=10^{-2}$ and behaves similar to $\hat{P}_{m,\bar{V}}$ while the contribution $\hat{P}_{m,\bar{W}}$ is overall negative except for the case with $m=1$, which is positive, and overall decreases as $Pr$ decreases.
As can be inferred from the expression in (\ref{eq:contribution_terms}), thermal dissipation $\hat{D}_{m,P}$ overall decreases to zero as $Pr$ decreases.   

\begin{figure}
  \centerline{
  \includegraphics[height=5cm]{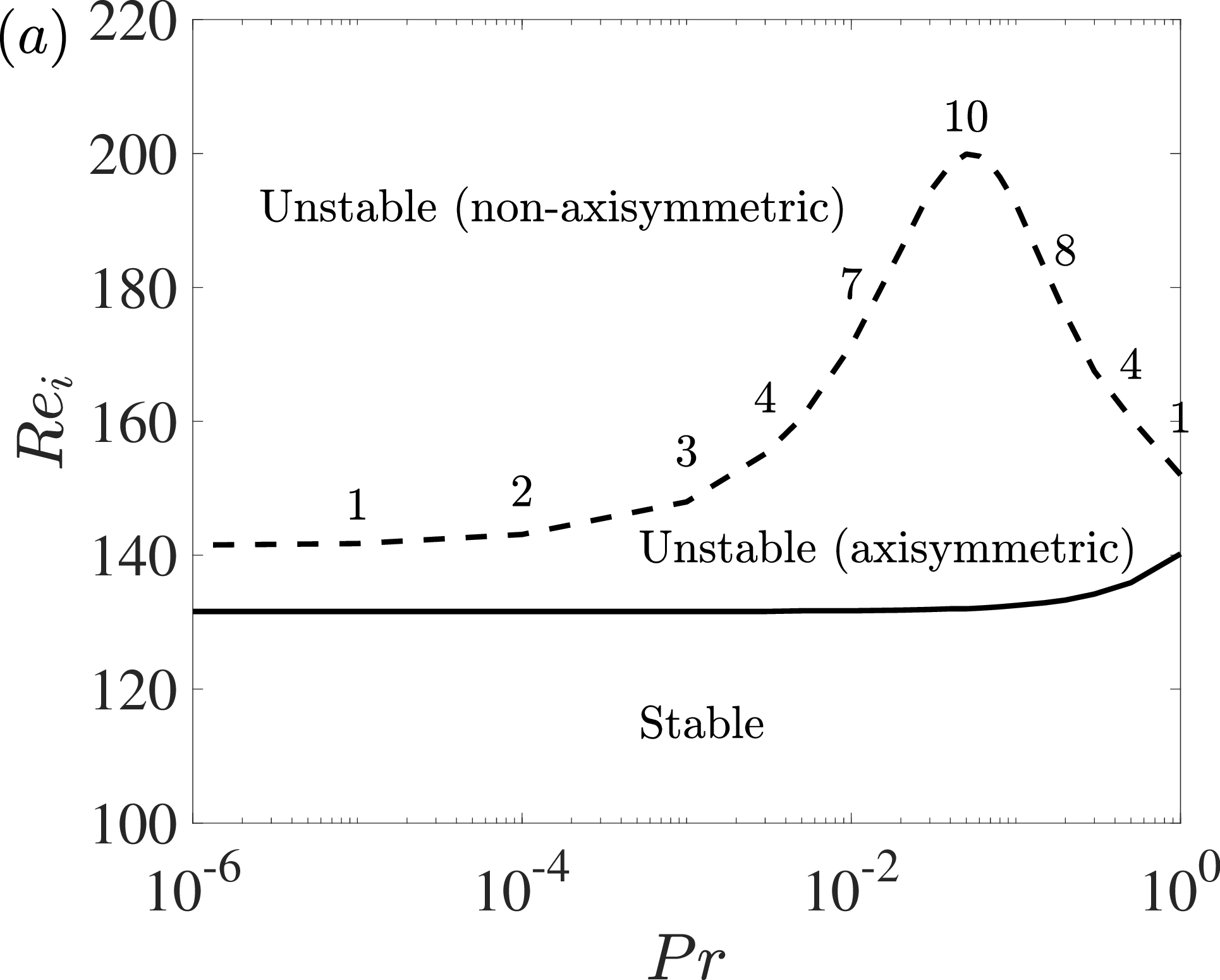}
    \includegraphics[height=5cm]{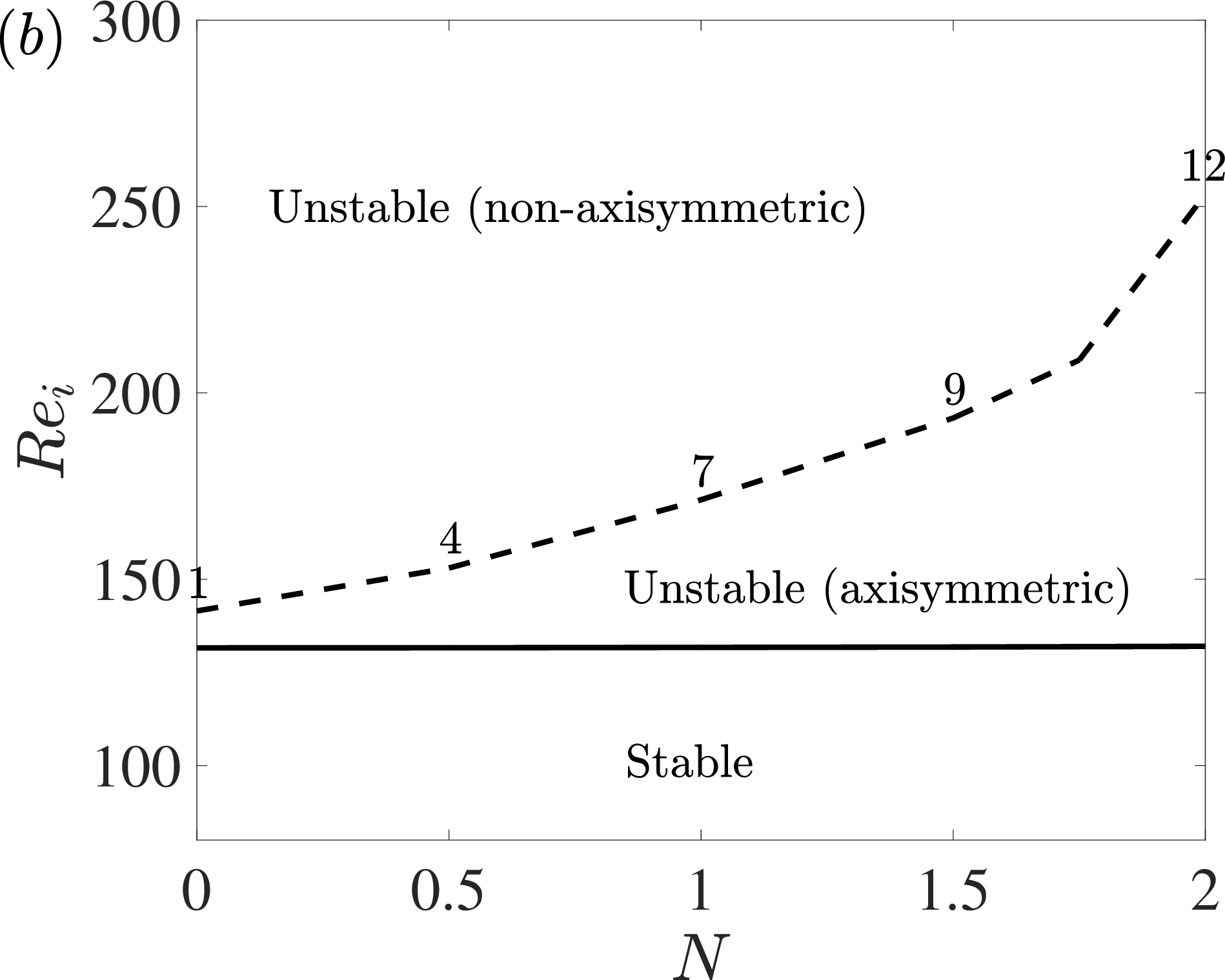}
  }
  \caption{Neutral stability curves from the 1D LSA on the axisymmetric mode $m=0$ (solid lines denoting $Re_{i,c}$) and bi-global LSA on the non-axisymmetric modes (dashed lines denoting $Re_{i,2}$) for (a) $N=1$ and (b) $Pr=0.01$.     
  The numbers above the dashed lines indicate the aziumthal wavenumbers of the non-axisymmetric mode which becomes secondarily unstable. 
  }
\label{fig:nsc_global}
\end{figure}
Figure \ref{fig:nsc_global} displays different neutral stability curves obtained from the 1D local LSA (solid lines) and 2D bi-global LSA (dashed lines) to show how the critical Reynolds numbers $Re_{i,c}$ and $Re_{i,2}$ vary with the Prandtl number $Pr$ or the Brunt-V\"ais\"al\"a frequency $N$. 
For $(N,Pr)=(1,1)$ in Figure \ref{fig:nsc_global}($a$), primary centrifugal instability by an axisymmetric mode occurs at $Re_{i,c}=140.2$ while secondary instability occurs at $Re_{i,2}=152.0$ by a non-axisymmetric mode with $m=1$. 
At $N=1$, the critical Reynolds number $Re_{i,c}$ decreases monotonically as $Pr$ decreases while the secondary critical Reynolds number $Re_{i,2}$ increases as $Pr$ decreases from $Pr=1$ to $Pr=0.06$ and then $Re_{i,2}$ decreases monotonically from the peak at $Pr=0.06$ as $Pr$ further decreases. 
It is noteworthy that, in the range $10^{-3}<Pr<1$, the secondary instability is triggered by highly non-axisymmetric modes with $m\geq4$, thus for $Re_{i}>Re_{i,2}$, we expect to observe highly non-axisymmetric flow patterns in this range of $Pr$.
This highly non-axisymmetric pattern is further discussed in the following subsection. 
In the range $Pr\leq10^{-3}$, the critical Reynolds numbers $Re_{i,c}$ and $Re_{i,2}$ approaches those of unstratified case at $N=0$ (i.e. $Re_{i,c}=131.6$ and $Re_{i,2}=141.5$), the case with secondary instability triggered by a non-axisymmetric mode with $m=1$. 
At $Pr=0.01$, it is shown in Figure \ref{fig:nsc_global}$(b)$ that the critical Reynolds number $Re_{i,c}$ does not change significantly in the range $0\leq N\leq2$ (e.g. $Re_{i,c}=131.6$ at $N=0$ and $Re_{i,c}=132.0$ at $N=2$) while the secondary critical Reynolds number $Re_{i,2}$ increases monotonically with $N$. 
The corresponding azimuthal wavenumber of the non-axisymmetric mode, which triggers secondary instability, also increases with $N$. 

Neutral stability curves in Figure \ref{fig:nsc_global} delineate the regimes of axisymmetric and non-axisymmetric flow patterns in the parameter space $(Re_{i},Pr)$ or $(Re_{i},N)$.  
The results are obtained by 1D and 2D linear stability analyses and we discuss in the next subsection about non-linear simulation results on how flow patterns change from axisymmetric Taylor vortices to non-axisymmetric wavy vortices via secondary instability or how the flow transitions to a chaotic and irregular state, a precursor stage prior to turbulence.

\subsection{Transition of flow patterns}
\begin{figure}
  \centerline{
  \includegraphics[height=3.9cm]{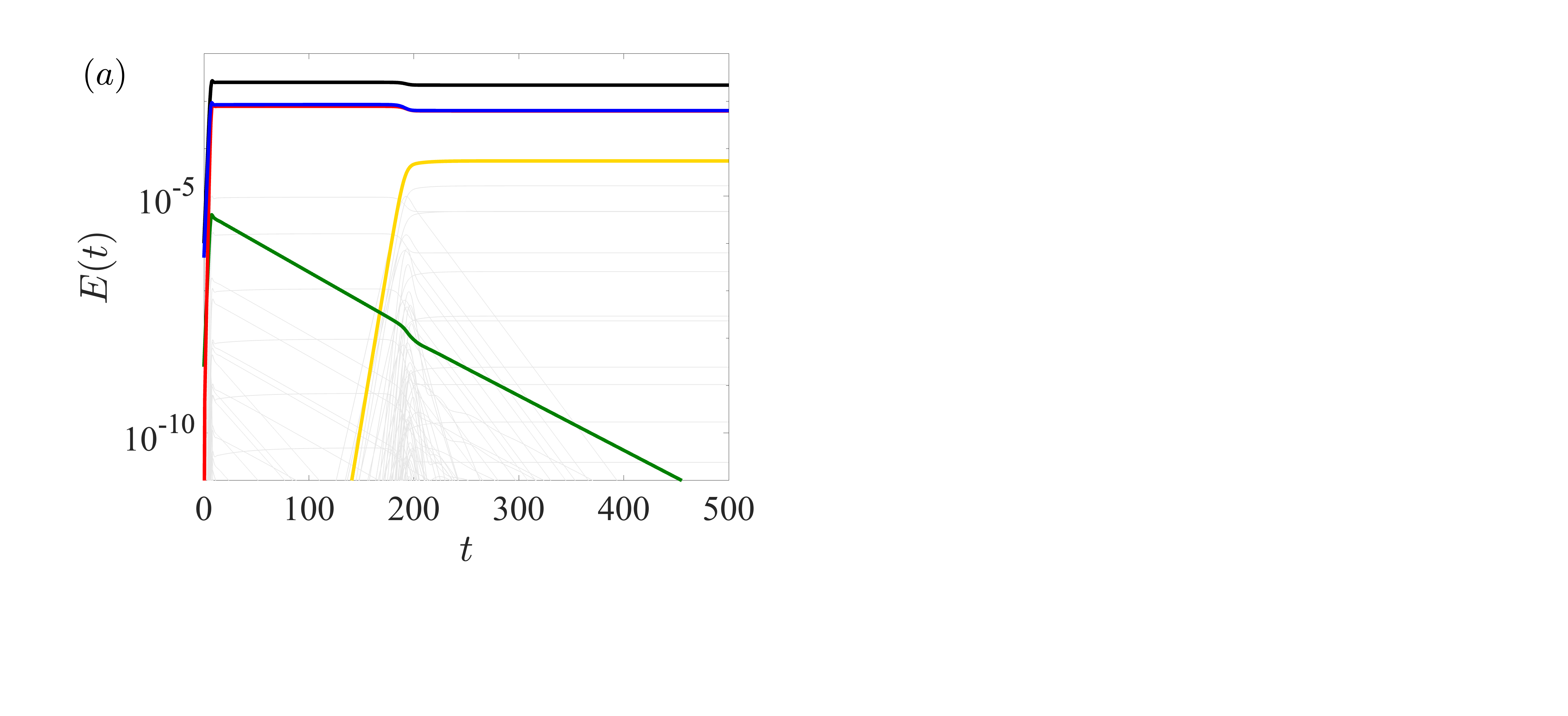}
        \includegraphics[height=4cm]{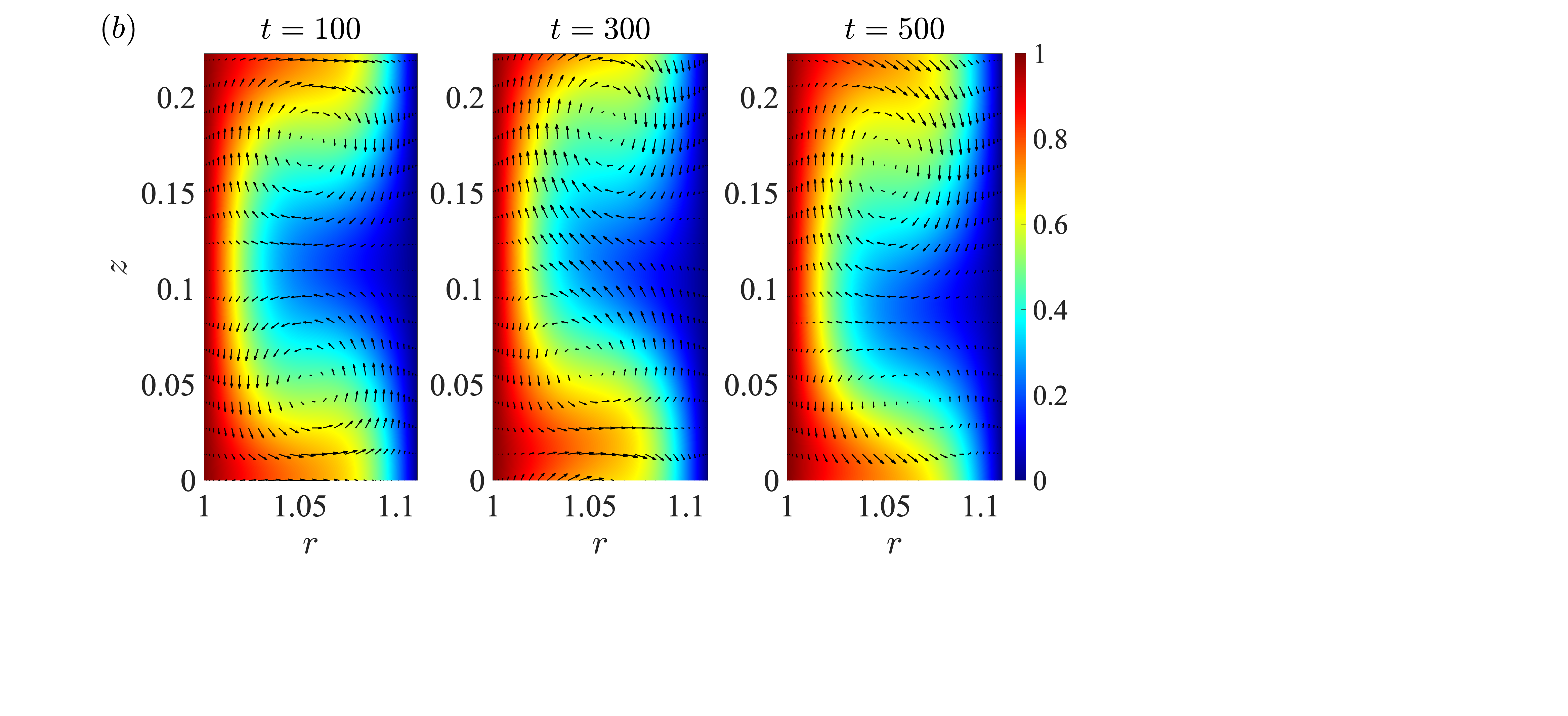}
  }
  \caption{(a) Temporal evolution of the total energy $E(t)$ (black) and modal energy components $\tilde{E}_{jl}$ for Case 2: $\tilde{E}_{00}$ (red), $\tilde{E}_{01}$ (blue), $\tilde{E}_{11}$ (green), $\tilde{E}_{91}$ (yellow), and other energy components dnoted by grey lines. 
    (b) The corresponding instantaneous velocity profiles at different times on the plane $(r,z)$ at $\theta=0$. 
    The contours denote the azimuthal velocity $U_{\theta}(r,z)$ and the vector plot denotes the transverse velocity field $(U_{r},U_{z})$. 
  }
\label{fig:ptb_energy_case2}
\end{figure}
\begin{figure}
  \center
  \includegraphics[height=4.5cm]{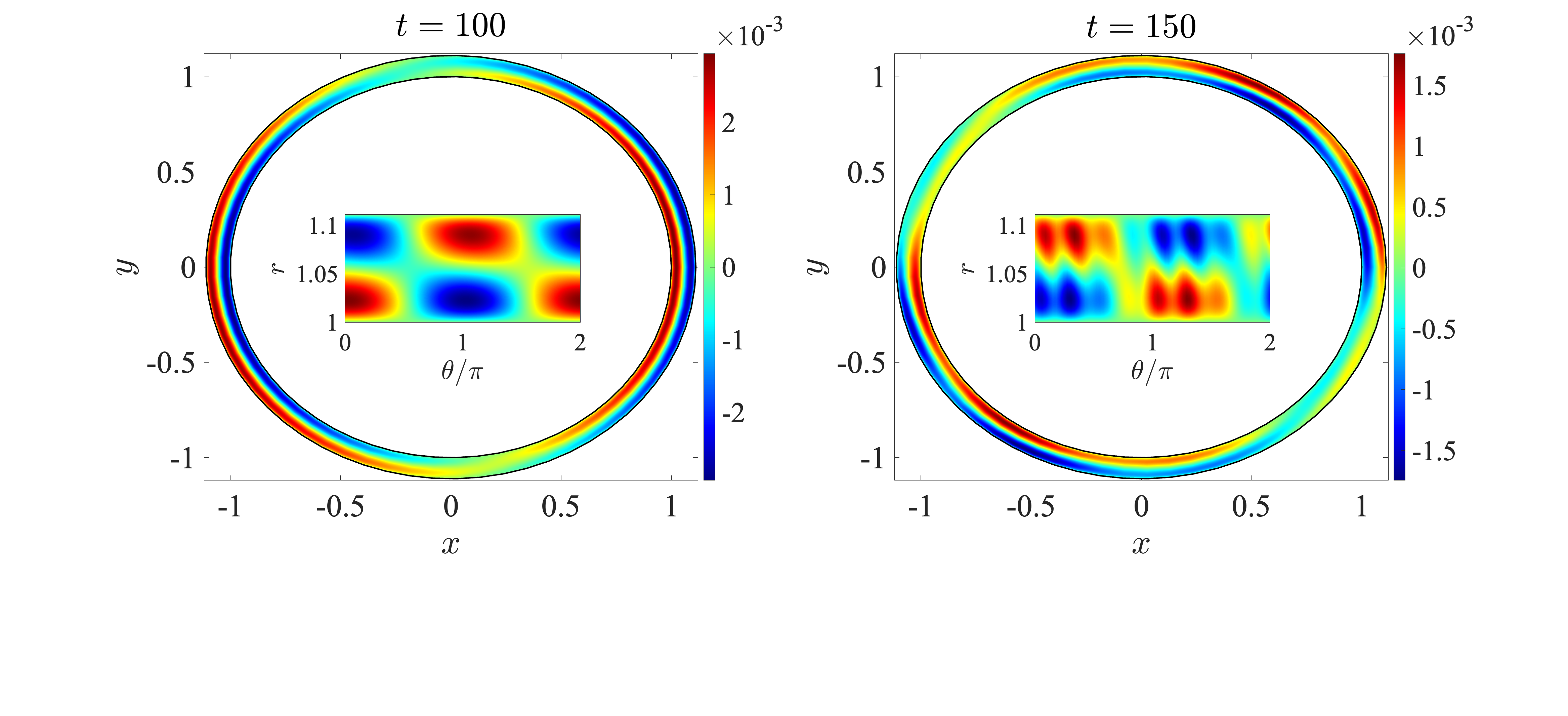}
    \includegraphics[height=4.5cm]{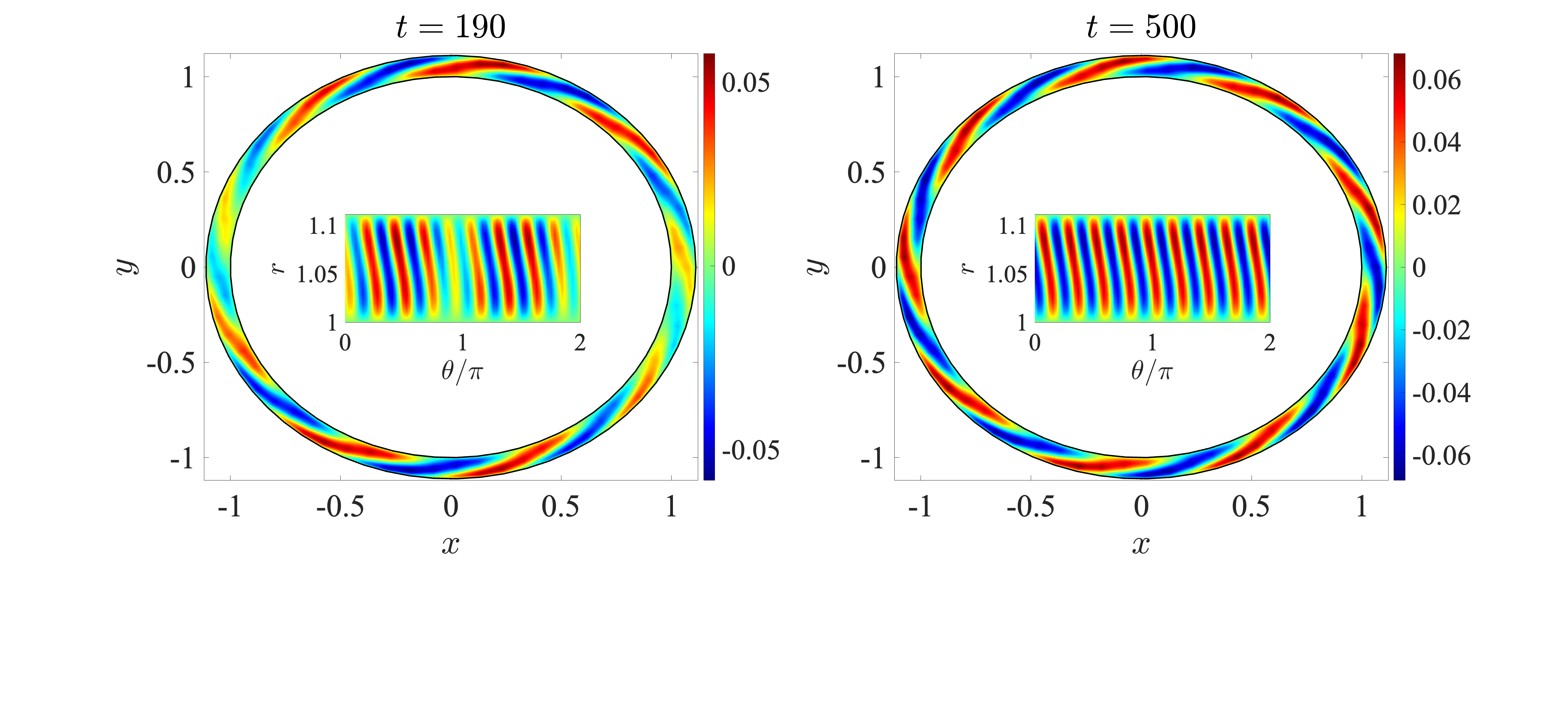}
  \caption{Time evolution of the vertical velocity $u_{z}$ in the $(x,y)$ plane at $z=0$ for Case 2.       
  }
\label{fig:ptb_velocity_uz_case2}
\end{figure}
\begin{figure}
  \centerline{
  \includegraphics[height=3.6cm]{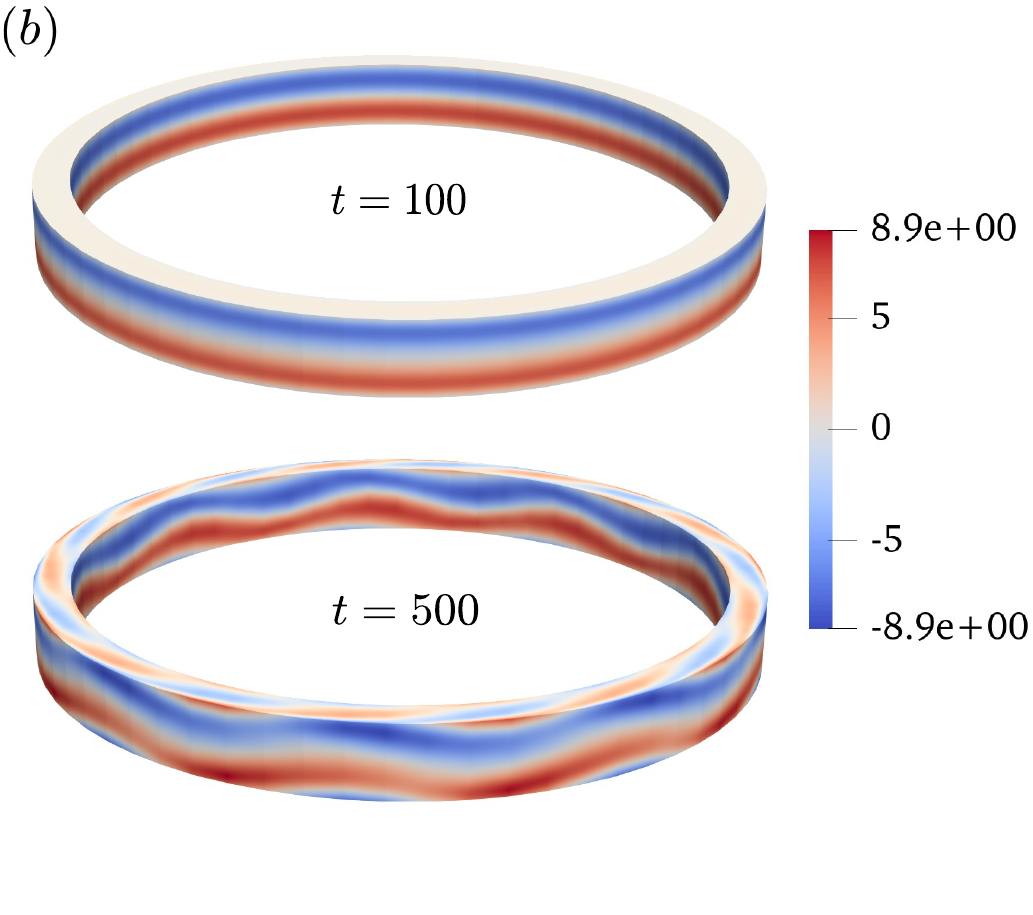}
  \includegraphics[height=3.6cm]{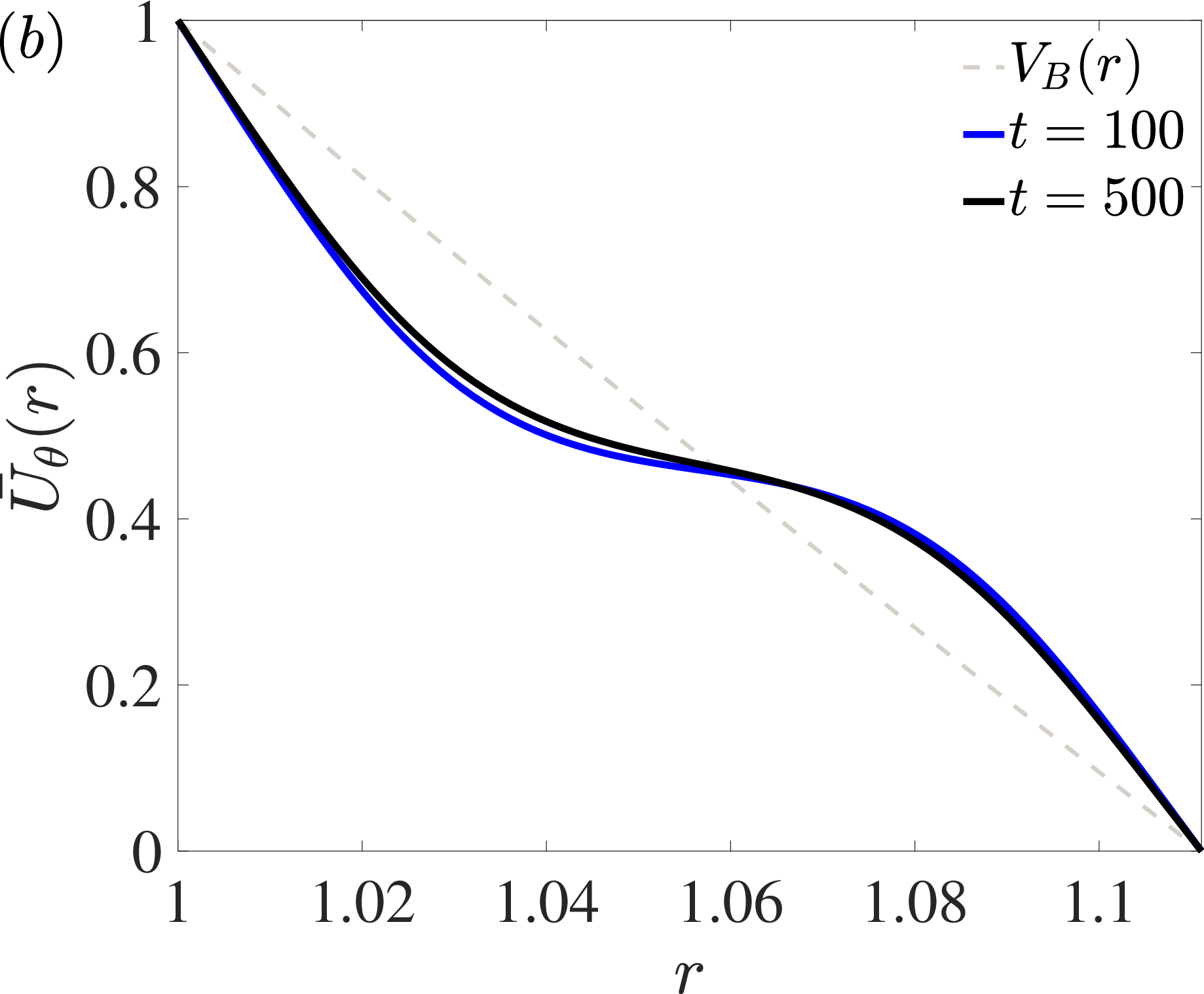}\hspace{0.1cm}
    \includegraphics[height=3.6cm]{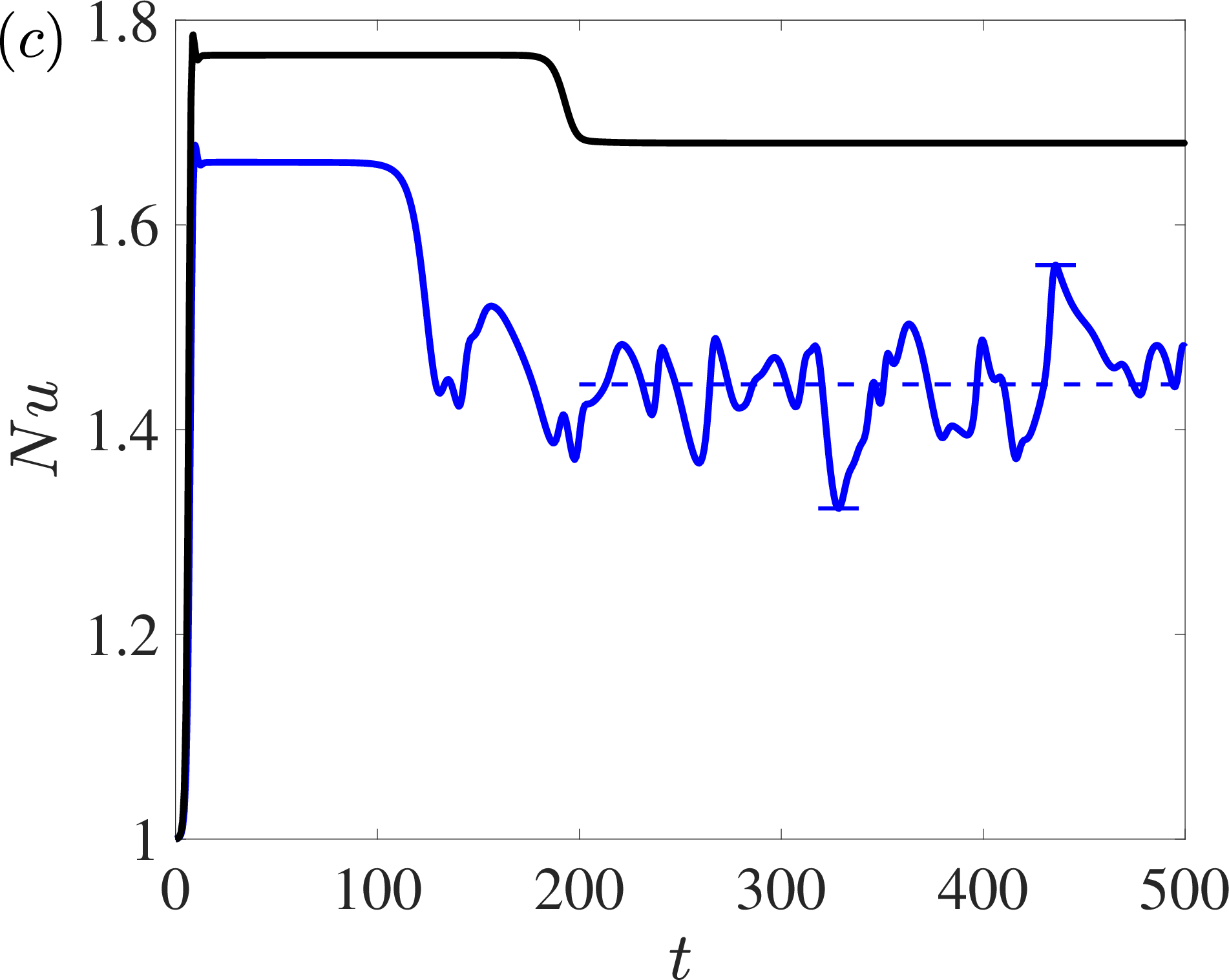}
  }
  \caption{
  $(a)$ Contours of the azimuthal vorticity $\omega_{\theta}$ at the boundary surfaces for Case 2.
  $(b)$ Profiles of the averaged velocity $\bar{U}_{\theta}(r)$ and cylidrical Couette flow $V_{B}(r)$ for Case 2.
   $(c)$ Nusselt number $Nu$ as a function of time $t$ for Case 2 (black) and Case 3 (blue). 
  }
\label{fig:ptb_others_case2}
\end{figure}

We now consider Case 2 at $(Re_{i},N,Pr)=(200,1,0.01)$, a case where the axisymmetric Taylor vortices become secondarily unstable by highly non-axisymmetric modes (see also, Figure \ref{fig:global_sa}$c$ and Figure \ref{fig:nsc_global}$a$).
The perturbation energy plot in Figure \ref{fig:ptb_energy_case2}$(a)$ shows that the axisymmetric mode becomes unstable and saturates quickly to the state of axisymmetric Taylor vortices (e.g. Figure \ref{fig:ptb_energy_case2}$b$ at $t=100$). 
At this first saturation state, the energies of weakly non-axisymmetric modes like the $m=1$ mode (green line in $a$) decay exponentially.
On the contrary, highly non-axisymmetric modes like the $m=9$ mode (yellow line in $a$) become unstable as the axisymmetric Taylor vortices sustain and their amplitudes become comparable to that of the axisymmetric mode as $t>200$.
The flow reaches a second saturation state in which the Taylor vortices start to oscillate vertically (Figure \ref{fig:ptb_energy_case2}$b$ and Movie 2) and become non-axisymmetric.
The transition on which non-axisymmetric modes become dominant is clearly illustrated by the vertical velocity $u_{z}(x,y)$ plot for different time $t$ (Figure \ref{fig:ptb_velocity_uz_case2} and Movie 3). 
The most energetic non-axisymmetric mode with $m=1$ loses its energy at the first transition stage (e.g., $u_{z}$ at $t=100$) while other non-axisymmetric modes with higher $m$ gain their energies and become comparable in the transition stage $150<t<200$. 
Weakly and highly non-axisymmetric modes compete during the transition (e.g., $u_{z}$ at $t=150$ and $190$) and after $t>200$, the non-axisymmetric mode with $m=9$ becomes the second dominant mode after the axisymmetric mode $m=0$ forming the wavy Taylor vortices. 
This non-axisymmetric pattern in $u_{z}$ has the maximum amplitude around the centreline between the two cylinders and is different from the feature of strato-rotational instability, which is triggered by two out-of-phase inertia-gravity waves confined near the two cylinders and has the maxima near the cylinders \citep[][]{Park2013JFM,Park2017}.
Contours of the azimuthal vorticity $\omega_{\theta}=\partial u_{r}/\partial z-\partial u_{z}/\partial r$ in Figure \ref{fig:ptb_others_case2}$(a)$ clearly show the axisymmetric Taylor vortices at $t=100$ as the first saturation state and non-axisymmetric wavy Taylor vortices as the second saturation state.
Various modal energies of the perturbation are expected to either saturate or decay after $t>500$ and thus non-axisymmetric Taylor vortices sustain for large $t$. 

Figure \ref{fig:ptb_others_case2}$(b)$ shows profiles of the total averaged azimuthal velocity $\bar{U}_{\theta}(r,t)=V_{B}(r)+\tilde{v}_{00}(r,t)$, the latter which is the azimuthal velocity of the mean-flow distortion denoting the azimuthal perturbation velocity averaged in the directions $\theta$ and $z$. 
Compared to the profile of $\bar{U}_{\theta}$ at $t=100$ for the axisymmetric Taylor vortices at the first saturation, the distortion in the azimuthal velocity is reduced for non-axisymmetric wavy Taylor vortices at the second saturation. 
The reduced distortion in mean flow implies the reduction in the velocity gradient at the inner cylinder and the corresponding torque applied at the inner cylinder.  
Following \citet{Martinez2015}, we define an axially-averaged non-dimensional torque $G$ at the inner cylinder as
\begin{equation}
\label{eq:def_torque}
G=\left(\frac{\eta Re_{i}}{1-\eta}\right)\frac{1}{2\pi L_{z}}\int_{0}^{2\pi}\int_{0}^{L_{z}}\left.\left(\frac{U_{\theta}}{r}-\frac{\partial U_{\theta}}{\partial r}\right)\right|_{r=1}\mathrm{d}\theta\mathrm{d}z,~
\end{equation}
which gives the laminar torque $G_{\mathrm{lam}}$ for cylindrical Couette flow $U_{\theta}=V_{B}(r)$ as
\begin{equation}
\label{eq:def_G_lam}
G_{\mathrm{lam}}=\frac{2\eta(1-\mu)Re_{i}}{(1+\eta)(1-\eta)^{2}}.
\end{equation} 
We also define the Nusselt number $Nu$ as $Nu=G/G_{\mathrm{lam}}$, which is the ratio between the transverse convective transport of angular velocity to the molecular transport of angular velocity as a measure of the non-dimensional angular momentum transfer \citep[][]{Dubrulle2002,Eckhardt2007,Martinez2015}.  
Figure \ref{fig:ptb_others_case2}$(c)$ shows how the Nusselt number $Nu$ changes over time for Case 2 (black line). 
Compared to the axisymmetric Taylor vortices in the range $50\leq t\leq 150$ having a constant $Nu=1.77$ during the first saturation process, the wavy Taylor vortices at the second saturation in the range $t>200$ have a lower $Nu=1.68$. 
This implies that the non-axisymmetric Taylor vortices after secondary instability lead to reduced angular momentum transfer compared to the axisymmetric Taylor vortices after primary centrifugal instability. 

\begin{figure}
  \centerline{
    \includegraphics[height=3.4cm]{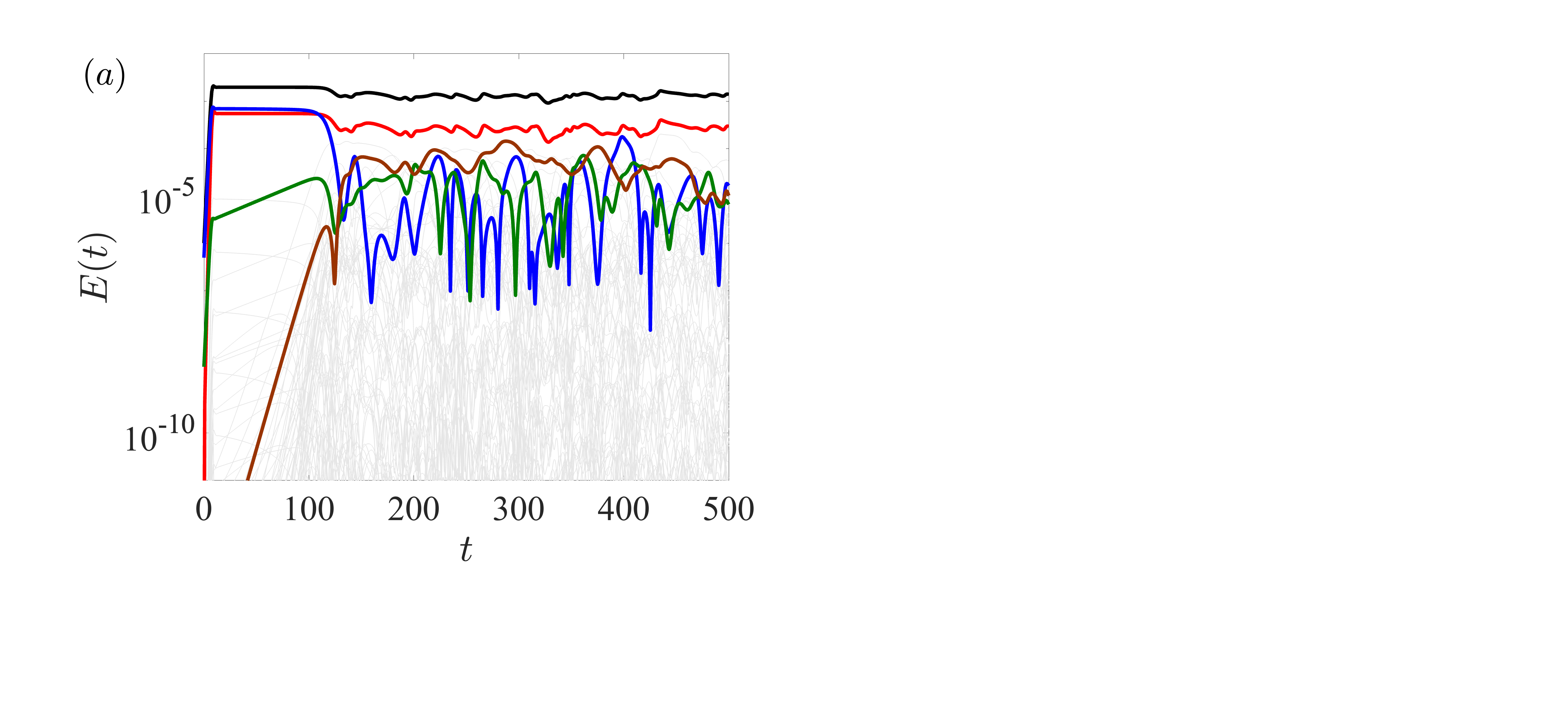}
        \includegraphics[height=3.4cm]{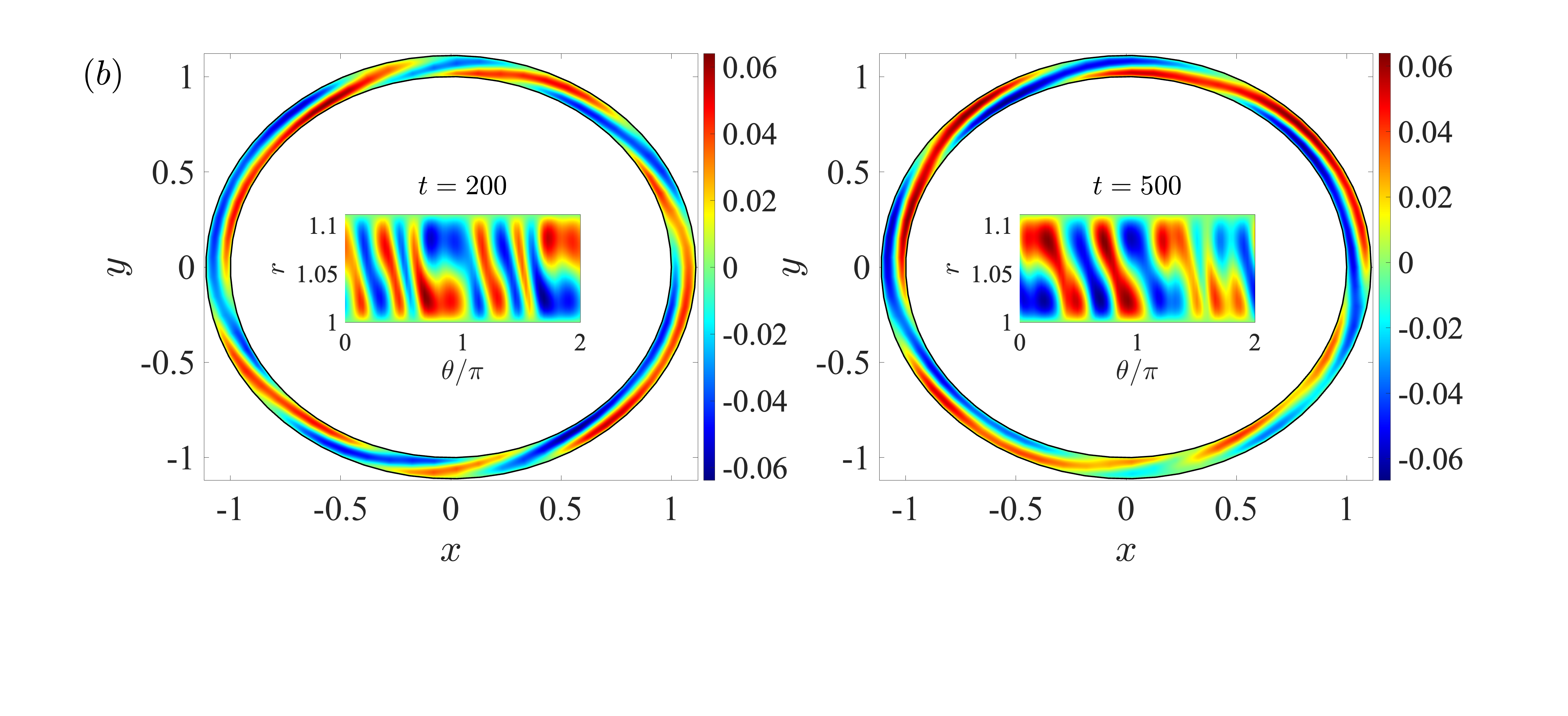}
  }
  \caption{(a) Temporal evolution of the total energy $E(t)$ (black) and modal energy components $\tilde{E}_{jl}$ for Case 3: $\tilde{E}_{00}$ (red), $\tilde{E}_{01}$ (blue), $\tilde{E}_{11}$ (green), $\tilde{E}_{41}$ (brown), and other energy components denoted by grey lines.
(b) Corresponding instantaneous velocity field $u_{z}(x,y)$ at $z=0$ at $t=200$ and $t=500$.
  }
\label{fig:ptb_energy_case3}
\end{figure}
\begin{figure}
  \centerline{
    \includegraphics[height=3.15cm]{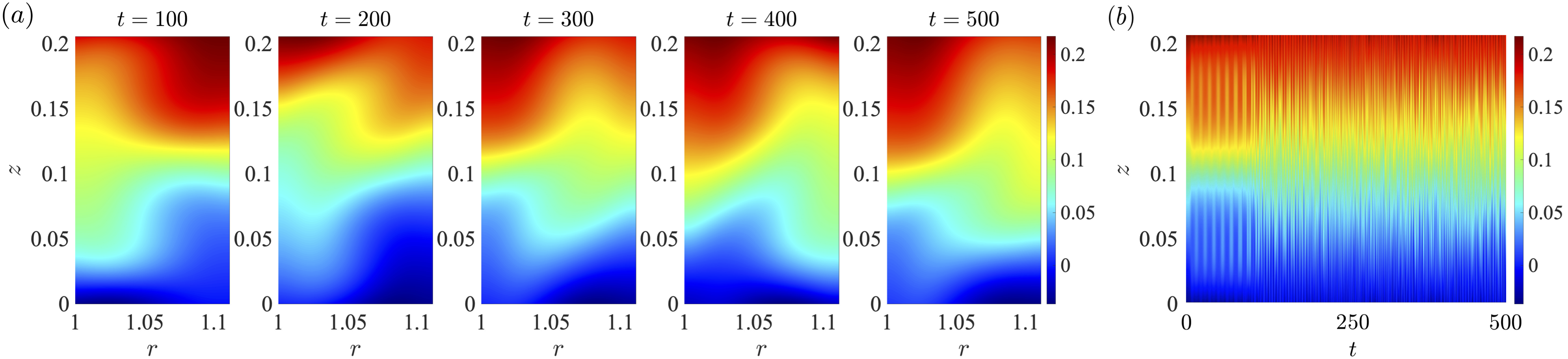}
  }
  \caption{(a) Evolution of the total temperature $\Uptheta(r,z)$ at $\theta=0$ for Case 3.\\
  (b) Corresponding spatio-temporal diagram of $\Uptheta(t,z)$ measured at $r=1.05$ and $\theta=0$.  
  }
\label{fig:ptb_temperature_case3}
\end{figure}
Case 3 with $(N,Pr)=(1,1)$ considers the Reynolds number $Re_{i}=200$ at the ratio $\mathcal{R}_{2}=Re_{i}/Re_{i,2}=1.316$, which is larger than the ratio $\mathcal{R}_{2}=1.168$ in Case 2. 
In this case, it is observed that the secondary instability bypasses the saturation but leads to an irregular fluctuation as shown by the temporal variation of the Nusselt number $Nu$ in Figure \ref{fig:ptb_others_case2}$(c)$ and modal energies of perturbation in Figure \ref{fig:ptb_energy_case3}$(a)$.
Primary centrifugal instability breaks down as $t>100$ where other non-axisymmetric modes including the mode with $m=1$ become dominant with strong non-linear modal interaction.
For Case 3, the most energetic mode varies over time and the flow exhibits a rather chaotic temporal fluctuation as shown by the energy evolution in Figure \ref{fig:ptb_energy_case3}$(a)$ and by the velocity and temperature fields in the $(r,z)$-plane in Movies 4 and 5. 
Although this irregular flow state does not show a small length-scale structure (see also, Figure \ref{fig:ptb_energy_case3}$b$ and Movie 6), the levels of the modal energies of large wavenumber modes are not negligible and turbulence is expected to occur for $(N,Pr)=(1,1)$ at high Reynolds number as $Re_{i}>200$, the range to be further explored in the future study. 
The Prandtl number $Pr=1$ is not small and the temperature perturbation $T$ is comparable to the base temperature $T_{B}$. 
Therefore, the fluctuation in the total temperature $\Uptheta=T_{B}+T$ induced by the velocity fluctuation is visible as shown in Figure \ref{fig:ptb_temperature_case3}$(a)$ although the overturning of temperature, which is a potential source of baroclinic instability, does not occur. 
Chaotic fluctuation of the temperature $\Uptheta$ after $t>100$ is also clearly shown in the spatio-temporal diagram measured near the mid-point at $r=1.05$ (Figure \ref{fig:ptb_temperature_case3}$b$).

\begin{figure}
  \centerline{
    \includegraphics[height=3.8cm]{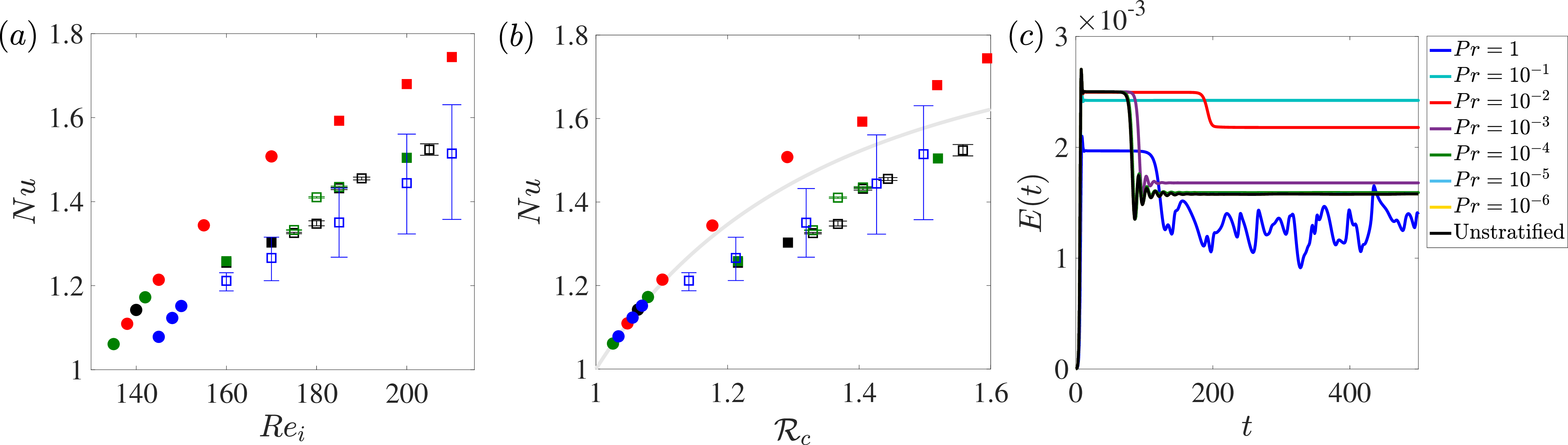}
  }
  \caption{$(a)$ Nusselt number $Nu$ versus Reynolds number $Re_{i}$  for $N=0$ (black), $(N,Pr)=(1,1)$ (blue), $(N,Pr)=(1,0.01)$ (red) and $(N,Pr)=(1,10^{-4})$ (green). 
  Filled circles and squares indicate $Nu$ for the axisymmetric Taylor vortices and non-axisymmetric wavy vortices at saturation, respectively. 
  Empty squares with error bars indicate statistically averaged $Nu$ and the minimum and maximum of $Nu$ in the time interval considered (see also, Figure \ref{fig:ptb_others_case2}$c$).    
 $(b)$ $Nu$ same as $(a)$ but over the rescaled Reynolds-number ratio $\mathcal{R}_{c}$.
  The grey line indicates the Nusselt number from (\ref{eq:Nu_DiPrima}).
  $(c)$ Temporal evolution of the total energy $E(t)$ for different $Pr$ at $Re_{i}=200$ and $N=1$. Black line denotes the unstratified case with $N=0$.
  }
\label{fig:torque_all}
\end{figure}
Figure \ref{fig:torque_all}$(a)$ shows how the Nusselt number $Nu$ changes with the Reynolds number $Re_{i}$ for different sets of $(N,Pr)$: $(1,1)$ (blue), $(1,0.01)$ (red), $(1,10^{-4})$ (green), and the unstratified case with $N=0$ (black). 
The Nusselt number $Nu$ is computed at the saturation state of the axisymmetric Taylor vortices (filled circles), which are computed through 3D DNS with a numerical resolution same as the one in Case 1, or for non-axisymmetric wavy vortices at the second saturation (filled squares) computed by 3D DNS with a numerical resolution same as the one in Case 2. 
For the flow in which its instability does not saturate and $Nu$ fluctuates like Case 3, empty squares with error bars are used to indicate the mean, minimum and maximum $Nu$ averaged in the time interval of fluctuation (see e.g. Figure \ref{fig:ptb_others_case2}$c$). 
In these cases with fluctuations, the same numerical resolution as Case 3 is used. 
It is clearly shown that $Nu$ for cases with the axisymmetric Taylor vortices increase fast with $Re_{i}$ and it increases slowly down as secondary instability occurs.
The increasing trend of the Nusselt number $Nu$ is similar for unstratified cases with $N=0$ and stratified cases with $(N,Pr)=(1,10^{-4})$. 
For both sets, secondary instability leading to a fluctuating flow state appears for $Re_{i}>160$ and there is a jump in $Nu$ around $Re_{i}=180$.
The Nusselt number $Nu$ continues to increase similarly for both as $Re_{i}$ increases further for $Re_{i}>180$.
For the cases with $(N,Pr)=(1,0.01)$,  the Nusselt number $Nu$ is higher than other cases as secondary instability occurs at a later stage as $Re_{i,2}=171.31$ for $Pr=0.01$.
Up to $Re_{i}=210$, the breakdown of secondary instability is not observed for $Pr=0.01$.   
Unlike other cases with the critical Reynolds numbers $Re_{i,c}=131.6\sim131.7$, the centrifugal instability for $(N,Pr)=(1,1)$ occurs at a higher $Re_{i,c}=140.2$ and thus the Nusselt number $Nu$ in this case is smaller than other cases at the same $Re_{i}$. 
Although the onset of primary centrifugal instability is delayed for $(N,Pr)=(1,1)$, it is found that the breakdown of instability occurs early and the fluctuation in the Nusselt number $Nu$ is larger than in other cases. 

For a better comparison among the cases, we display in Figure \ref{fig:torque_all}$(b)$ the same $Nu$ against the Reynolds-number ratio $\mathcal{R}_{c}=Re_{i}/Re_{i,c}$. 
For the axisymmetric Taylor vortices, \citet{DiPrima1984} proposed the following scaling law for $Nu$ as a function of the ratio $\mathcal{R}_{c}$ as follows: 
\begin{equation}
\label{eq:Nu_DiPrima}
Nu=1+A\left(1-\frac{1}{\mathcal{R}_{c}^{2}}\right)+B\left(1-\frac{1}{\mathcal{R}_{c}^{2}}\right)^{2},
\end{equation}
where $A=1.246$ and $B=-0.37$ are the constants for $\eta=0.9$ according to \citet{DiPrima1984}.
It is clearly shown in Figure \ref{fig:torque_all}$(b)$ that the Nusselt number $Nu$ for every case agrees well with the scaling law (\ref{eq:Nu_DiPrima}) in the range $\mathcal{R}_{c}\leq1.1$ where the axisymmetric Taylor vortices are present. 
This implies that the Nusselt number $Nu$ for the Taylor vortices, which are induced by primary centrifugal instability, is not strongly influenced by $Pr$.  
When the flow is secondarily unstable or chaotic, the cases with $(N,Pr)=(1,0.01)$ for which secondary instability appears late as $\mathcal{R}_{c}>1.3$ have higher $Nu$ than the scaling law (\ref{eq:Nu_DiPrima}) while other cases for which secondary instability appears early as $\mathcal{R}_{c}>1.1$ have lower $Nu$ than the proposed scaling (\ref{eq:Nu_DiPrima}).
This implies that the angular momentum transport represented by the Nusselt number $Nu$ depends on the Prandtl number $Pr$ once the flow becomes secondarily unstable or chaotic. 
However, as Figure \ref{fig:torque_all}$(a,b)$ only considers $N=1$, more detailed investigations are required to confirm the relation between $Nu$ and $Pr$ in a wider parameter space of $(N,Re_{i})$, an important topic to be further studied in the future to unravel the physics of turbulent angular momentum transport that depends on stratification and thermal diffusion.   

\begin{figure}
  \centerline{
    \includegraphics[height=3.4cm]{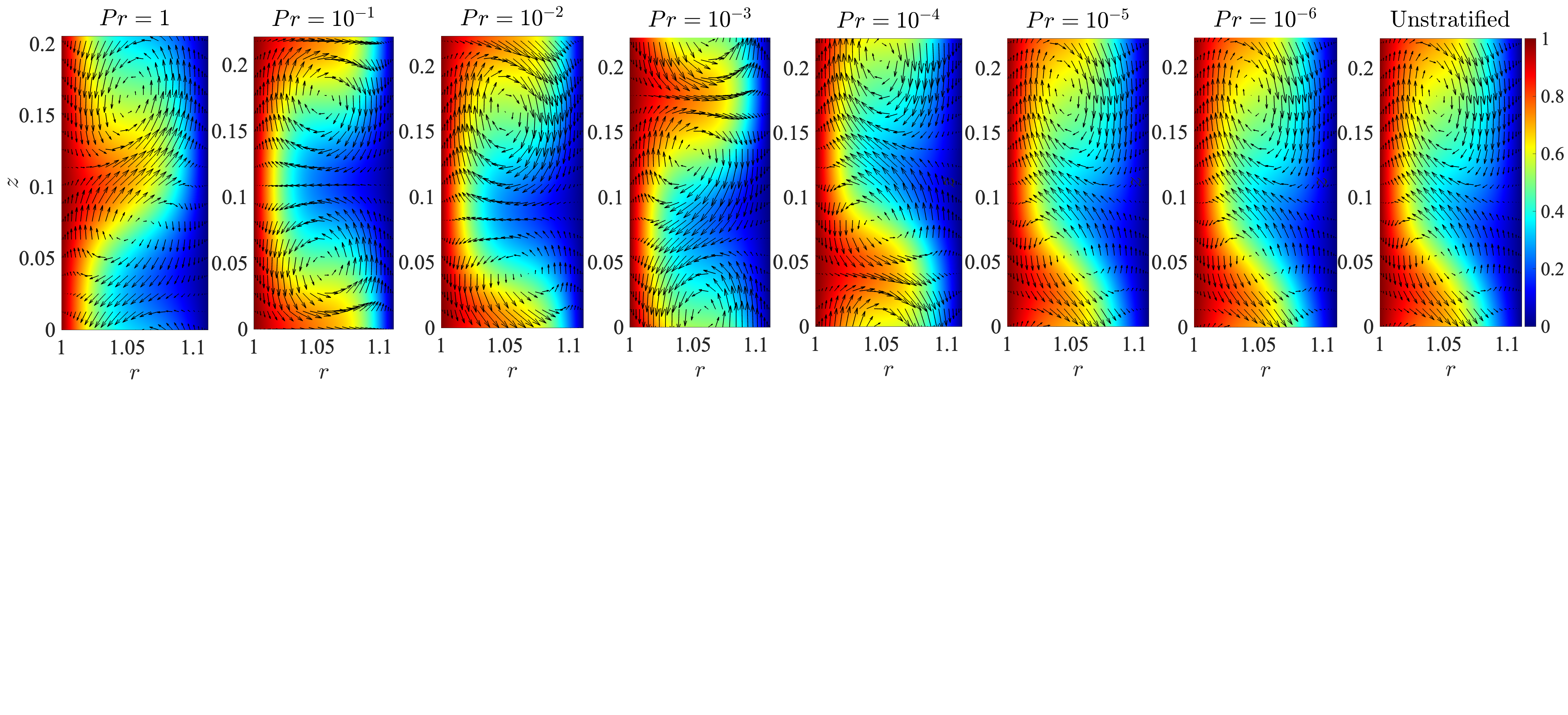}
    }
  \caption{Instantaneous velocity contours for $U_{\theta}$ and vector plots for the transverse velocity field $(U_{r},U_{z})$ on the plane $(r,z)$ for different $Pr$ at $t=500$, $\theta=0$, $Re_{i}=200$ and $N=1$. The rightmost figure corresponds to the unstratified case with $N=0$.
  }
\label{fig:Velocity_Pr}
\end{figure}
As a summary on the nonlinear dynamics with varying $Pr$, we display in Figure \ref{fig:torque_all}$(c)$ and Figure \ref{fig:Velocity_Pr} the temporal evolution of the total energy $E(t)$ and instantaneous velocity fields for different $Pr$ at $Re_{i}=200$ and $N=1$. 
We see in Figure \ref{fig:torque_all}$(c)$ that, as $Pr$ is small as $Pr<10^{-4}$, the energy curves collapse and become identical to the curve of the unstratified case with $N=0$. 
Instantaneous velocity fields obtained at $t=500$ from various DNS, which started from the same configuration of the initial condition described in the early part of Section \ref{sec:NCI}, are also identical bewteen the stratified cases with $Pr=10^{-5}$ and $10^{-6}$ and unstratified case with $N=0$.
These results support the explanation in the introduction and Figure \ref{fig:illust} that the stratification effect is suppressed by strong thermal diffusion and the flow behaves as the unstratifed flow in both linear and non-linear regimes.
Nonetheless, there still remains a question on whether our conclusion will remain valid for turbulent cases and if so, up to which Reynolds number $Re$.
Although DNS investigations like those by \citet{Prat2013} verified this question for $Re$ of the order $O(100)$, further investigations should be conducted at higher Reynolds numbers.

\section{Conclusion and discussion}
\label{sec:conclusion}
In this paper, linear and non-linear dynamics of centrifugal instability of stratified Taylor-Couette flow are studied for diffusive fluids in which the Prandtl number $Pr$ is low as $Pr\leq1$. 
The linear stability analysis (LSA) reveals that the stratification effect, which suppresses the centrifugal instability \citep[][]{Boubnov1995,Caton2000}, is suppressed by strong thermal diffusion.
This implies that the thermal diffusion promotes centrifugal instability of stratified TC flow, the situation which can be found similarly in other contexts like shear instabilities in stratified flows in the low-$Pr$ limit \citep[][]{Lignieres1999SPA,Garaud2020}.
When the thermal diffusion is sufficiently strong, we apply the small-$Pr$ approximation and find that the instability characteristics are self-similar with the new scaled parameter $P_{N}=N^{2}Pr$ \citep[see also,][]{Lignieres1999SPA,Park2020AA}.
The thermal diffusion effect as promoting centrifugal instability is found to be the same for both axisymmetric and non-axisymmetric perturbations. 
For the parameters considered here (i.e. $\mu=0$, $\eta=0.9$ and $Pr\leq1$), it is found that the axisymmetric perturbation with $m=0$ is most unstable compared to non-axisymmetric perturbations. 
By conducting direct numerical simulation (DNS) for the Reynolds number $Re_{i}$ above the critical one as $Re_{i}>Re_{i,c}$, we also study the nonlinear evolution of centrifugal instability. 
In the DNS, we use controlled initial conditions with both the axisymmetric perturbation and a smaller-amplitude non-axisymmetric perturbation to see how they grow and interact nonlinearly and to understand their non-linear modal interaction leading to different states.
At the initial stage, the axisymmetric perturbation grows at fastest and saturates along the non-linear interaction with base flow. 
Depending on $Re_{i}$, a new mean flow in the shape of axisymmetric Taylor vortices can become secondarily unstable by the growth of non-axisymmetric perturbation. 
Unlike the case of primary instability, which is promoted by strong thermal diffusion, the 2D bi-global LSA on the Taylor vortices reveals that thermal diffusion in the range $10^{-3}<Pr<1$ at $N=1$ delays the onset of secondary instability and potentially laminar-turbulent transition triggerred by highly non-axisymmetric perturbations. 
We note that the appearance of such a non-axisymmetric flow pattern by the onset of secondary instability in highly-diffusive and stratified flows has not been explored properly in the previous studies.
At $Pr=1$, we also examine irregular flow patterns leading to chaotic mixing of temperature, the state as a precursor to turbulence. 
Furthermore, we analyse how the Nusselt number $Nu$ as a measure of angular momentum transfer varies with the Reynolds number $Re_{i}$ for various sets of $(N,Pr)$. 
It is verified that the secondary instability slows down the increase of $Nu$, implying the momentum transfer is slowly enhanced after the onset of secondary instability. 

In recent years, stratified turbulence in the low-$Pr$ regime has been studied increasingly not only in fluid dynamics but also in astrophysics due to its relevance to the interior of stars and the Sun where the Prandtl number $Pr$ is low as $10^{-6}$ or below \citep[][]{Garaud2020,Dymott2023,Garaud2024}. 
Although the Reynolds numbers of astrophysical flows cannot be reached due to their immensely large length scales, simulations at lower Reynolds numbers and theories unveil new regimes of turbulence and give us a hint on how stratified turbulence behaves with strong thermal diffusion \citep[][]{Cope2020,Skoutnev2023,Garaud2024JFM}. 
What has not been explored yet though in these turbulence studies is how the turbulence is generated via laminar-turbulent transition processes. 
In this regard, the current study contributes to our knowledge of how the centrifugal instability develops nonlinearly towards turbulence in stratified and highly-diffusive fluids. 
Beyond this paper, we will aim to study first the characteristics of fully-developed stratified-rotating turbulence triggered by centrifugal instability in the low-$Pr$ regime. 
At low $Pr$, an unexpected highly non-axisymmetric pattern can generate turbulent flow with characteristics different from those of unstratified turbulence. 
Another topic to be investigated in the future is how the strato-rotational instability (SRI) will behave in the presence of strong thermal diffusion. 
From the perspective of linear instability, the thermal diffusion suppressing the stratification effect is expected to suppress the SRI as well.  
Linear and non-linear SRI need to be further explored in the low-$Pr$ regime. 
The non-linear development of the SRI is also of great interest in geophysical and astrophysical fluid dynamics, in particular for flows in the Keplerian or super-rotation regimes, the latter in which the outer part rotates faster than the inner region \citep[][]{Dubrulle2005,LeDizes2010,Park2013JFM}. 

\backsection[Supplementary data]{\label{SupMat}Supplementary movies are available.}

\backsection[Funding]{The author acknowledges support from the Engineering and Physical Sciences Research Council (EPSRC) through the EPSRC mathematical sciences small grant (EP/W019558/1).}

\backsection[Declaration of interests]
{The authors report no conflict of interest.}


\backsection[Author ORCIDs]{Junho Park, https://orcid.org/0000-0001-5947-0064}


\appendix
\section{Details of vectors and operator matrices}
\label{app:matrices}
\subsection{Vectors and operator matrices for the evolution equation (\ref{eq:evolution_slm}) and 1D linear stability analysis}
\label{app:matrices_1dlsa}
For the case with the index $l\neq0$, we consider in (\ref{eq:evolution_slm}) the following vectors and matrices
\begin{equation}
\tilde{\boldsymbol{q}}_{jl}=
\left(
\begin{array}{c}
\tilde{u}_{jl}\\
\tilde{v}_{jl}\\
\tilde{T}_{jl}
\end{array}
\right)
,~
\tilde{\boldsymbol{N}}_{jl}=
\left(
\begin{array}{c}
-\tilde{N}_{r,jl}-\frac{\mathrm{i}}{k_{l}}\frac{\partial\tilde{N}_{z,jl}}{\partial r}\\
-\tilde{N}_{\theta,jl}+\frac{m_{j}\tilde{N}_{z,jl}}{k_{l}r}\\
-\tilde{N}_{T,jl}
\end{array}
\right)
,
\end{equation}
\begin{equation}
\mathcal{A}_{jl}=
\left(
\begin{array}{ccc}
\mathcal{A}_{jl}^{11} & \mathcal{A}_{jl}^{12} & 0\\
\mathcal{A}_{jl}^{21} & \mathcal{A}_{jl}^{22} & 0\\
0 & 0 & 1\\
\end{array}
\right),~~
\mathcal{B}_{jl}=
\left(
\begin{array}{ccc}
\mathcal{B}_{jl}^{11} & \mathcal{B}_{jl}^{12} & \mathcal{B}_{jl}^{13}\\
\mathcal{B}_{jl}^{21} & \mathcal{B}_{jl}^{22} & \mathcal{B}_{jl}^{23}\\
\mathcal{B}_{jl}^{31} & \mathcal{B}_{jl}^{32} & \mathcal{B}_{jl}^{33}\\
\end{array}
\right),
\end{equation}
where
\begin{eqnarray}
\mathcal{A}_{jl}^{11}&=&1-\frac{1}{k_{l}^{2}}\frac{\partial}{\partial r}\left(\frac{\partial}{\partial r}+\frac{1}{r}\right),~~
\mathcal{A}_{jl}^{12}=-\frac{\mathrm{i}m_{j}}{k_{l}^{2}r}\left(\frac{\partial}{\partial r}-\frac{1}{r}\right),\nonumber\\
\mathcal{A}_{jl}^{21}&=&-\frac{\mathrm{i}m_{j}}{k_{l}^{2}r}\left(\frac{\partial}{\partial r}+\frac{1}{r}\right),~
\mathcal{A}_{jl}^{22}=1+\frac{m_{j}^{2}}{k_{l}^{2}r^{2}},
\end{eqnarray}
\begin{eqnarray}
\mathcal{B}_{jl}^{11}&=&-\mathrm{i}m_{j}\Omega+\frac{\mathrm{i}m_{j}}{k_{l}^{2}}\left(\Omega\frac{\partial}{\partial r}+\frac{\mathrm{d}\Omega}{\mathrm{d}r}\right)\left(\frac{\partial}{\partial r}+\frac{1}{r}\right)+\frac{1}{Re}\left[\tilde{\nabla}_{jl}^{2}-\frac{1}{r^{2}}-\frac{1}{k_{l}^{2}}\frac{\partial}{\partial r}\tilde{\nabla}_{jl}^{2}\left(\frac{\partial}{\partial r}+\frac{1}{r}\right)\right],\nonumber\\
\mathcal{B}_{jl}^{12}&=&2\Omega+\frac{\mathrm{i}m_{j}}{k_{l}^{2}}\left(\Omega\frac{\partial}{\partial r}+\frac{\mathrm{d}\Omega}{\mathrm{d}r}\right)\left(\frac{\mathrm{i}m_{j}}{r}\right)
+\frac{1}{Re}\left[-\frac{2\mathrm{i}m_{j}}{r^{2}}-\frac{1}{k_{l}^{2}}\frac{\partial}{\partial r}\tilde{\nabla}_{jl}^{2}\left(\frac{\mathrm{i}m_{j}}{r}\right)\right],~
\mathcal{B}_{jl}^{13}=\frac{\mathrm{i}N^{2}}{k_{l}}\frac{\partial}{\partial r},\nonumber\\
\mathcal{B}_{jl}^{21}&=&-Z-\frac{m_{j}^{2}\Omega}{k_{l}^{2}r}\left(\frac{\partial}{\partial r}+\frac{1}{r}\right)+\frac{1}{Re}\left[\frac{2\mathrm{i}m_{j}}{r^{2}}-\frac{\mathrm{i}m_{j}}{k_{l}^{2}r}\tilde{\nabla}_{jl}^{2}\left(\frac{\partial}{\partial r}+\frac{1}{r}\right)\right],\nonumber\\
\mathcal{B}_{jl}^{22}&=&-\mathrm{i}m_{j}\Omega-\frac{\mathrm{i}m_{j}^{3}\Omega}{k_{l}^{2}r^{2}}+\frac{1}{Re}\left[\tilde{\nabla}_{jl}^{2}-\frac{1}{r^{2}}+\frac{m_{j}^{2}}{k_{l}^{2}r}\tilde{\nabla}_{jl}^{2}\left(\frac{1}{r}\right)\right],~
\mathcal{B}_{jl}^{23}=-\frac{m_{j}N^{2}}{k_{l}r},\nonumber\\
\mathcal{B}_{jl}^{31}&=&-\frac{\mathrm{i}}{k}\left(\frac{\partial}{\partial r}+\frac{1}{r}\right),~
\mathcal{B}_{jl}^{32}=\frac{m_{j}}{k_{l}r},~
\mathcal{B}_{jl}^{33}=-\mathrm{i}m_{j}\Omega+\frac{1}{RePr}\tilde{\nabla}_{jl}^{2}.
\end{eqnarray}
For the case with $l=0$ and $j\neq0$, we consider in (\ref{eq:evolution_slm}) the following vectors and matrices
\begin{equation}
\tilde{\boldsymbol{q}}_{j0}=
\left(
\begin{array}{c}
\tilde{u}_{j0}\\
\tilde{w}_{j0}\\
\tilde{T}_{j0}
\end{array}
\right)
,~
\tilde{\boldsymbol{N}}_{j0}=
\left(
\begin{array}{c}
-\tilde{N}_{r,j0}-\frac{\mathrm{i}}{m_{j}}\left(\tilde{N}_{\theta,j0}+r\frac{\partial \tilde{N}_{\theta,j0}}{\partial r}\right)\\
-\tilde{N}_{z,j0}\\
-\tilde{N}_{T,j0}
\end{array}
\right)
,
\end{equation}

\begin{equation}
\mathcal{A}_{j0}=
\left(
\begin{array}{ccc}
\mathcal{A}_{j0}^{11} & 0 & 0\\
0 & 1 & 0\\
0 & 0 & 1\\
\end{array}
\right),~~
\mathcal{B}_{j0}=
\left(
\begin{array}{ccc}
\mathcal{B}_{j0}^{11} & 0 & 0\\
0 & \mathcal{B}_{j0}^{22} & \mathcal{B}_{j0}^{23}\\
0 & \mathcal{B}_{j0}^{32} & \mathcal{B}_{j0}^{33}\\
\end{array}
\right),
\end{equation}
where
\begin{equation}
\mathcal{A}_{j0}^{11}=1-\frac{1}{m_{j}^{2}}\left(r^{2}\frac{\partial^{2}}{\partial r^{2}}+3r\frac{\partial}{\partial r}+1\right)
\end{equation}
\begin{eqnarray}
\mathcal{B}_{j0}^{11}&=&\frac{\mathrm{i}}{m_{j}}\left[r^{2}\Omega\frac{\partial^{2}}{\partial r^{2}}+3r\Omega\frac{\partial}{\partial r}-r^{2}\frac{\mathrm{d}^{2}\Omega}{\mathrm{d}r^{2}}-3r\frac{\mathrm{d}\Omega}{\mathrm{d}r}+(1-m_{j}^{2})\Omega\right]\nonumber\\
&&-\frac{1}{Rem_{j}^{2}}\left[\frac{\partial}{\partial r}\left(r\tilde{\nabla}_{j0}^{2}+r\tilde{\nabla}_{j0}^{2}\left(r\frac{\partial}{\partial r}\right)\right)-m_{j}^{2}\tilde{\nabla}_{j0}^{2}-\frac{\partial^{2}}{\partial r^{2}}-\frac{1}{r}\frac{\partial}{\partial r}+\frac{1-3m_{j}^{2}}{r^{2}}\right],\nonumber\\
\mathcal{B}_{j0}^{22}&=&-\mathrm{i}m_{j}\Omega+\frac{1}{Re}\tilde{\nabla}_{j0}^{2},~
\mathcal{B}_{j0}^{23}=N^{2},~
\mathcal{B}_{j0}^{32}=-1,~
\mathcal{B}_{j0}^{33}=-\mathrm{i}m_{j}\Omega+\frac{1}{RePr}\tilde{\nabla}_{j0}^{2}.
\end{eqnarray}
For the case with $j=l=0$, we consider in (\ref{eq:evolution_slm}) the following vectors and matrices
\begin{equation}
\tilde{\boldsymbol{q}}_{00}=
\left(
\begin{array}{c}
\tilde{v}_{00}\\
\tilde{w}_{00}\\
\tilde{T}_{00}
\end{array}
\right)
,~
\tilde{\boldsymbol{N}}_{00}=
\left(
\begin{array}{c}
-\tilde{N}_{\theta,00}\\
-\tilde{N}_{z,00}\\
-\tilde{N}_{T,00}
\end{array}
\right)
,
\end{equation}
\begin{equation}
\mathcal{A}_{00}=
\left(
\begin{array}{ccc}
1 & 0 & 0\\
0 & 1 & 0\\
0 & 0 & 1\\
\end{array}
\right),~
\mathcal{B}_{00}=
\left(
\begin{array}{ccc}
\frac{1}{Re}\left(\tilde{\nabla}^{2}_{00}-\frac{1}{r^{2}}\right) & 0 & 0\\
0 & \frac{1}{Re}\tilde{\nabla}^{2}_{00} & N^{2}\\
0 & -1 & \frac{1}{RePr}\tilde{\nabla}^{2}_{00}\\
\end{array}
\right).
\end{equation}
In the eigenvalue problem (\ref{eq:evp}), the eigenfunction $\hat{\boldsymbol{q}}$ and operator matrices $\mathcal{A}$ and $\mathcal{B}$ are the same as $\tilde{\boldsymbol{q}}_{11}$, $\mathcal{A}_{11}$ and $\mathcal{B}_{11}$ for given $m$ and $k$ (i.e. $j=l=1$). 
\subsection{Operator matrices in the bi-global stability analysis}
\label{app:matrices_2dlsa}
For the two-dimensional base state $\bar{\boldsymbol{Q}}(r,z)=\left(\bar{U},\bar{V},\bar{W},\bar{\mathcal{T}}\right)$, we consider the following linearised perturbation equations 
\begin{equation}
\label{eq:lpe_2D_continuity}
\frac{\partial \bar{u}_{r}}{\partial r}+\frac{\bar{u}_{r}}{r}+\frac{1}{r}\frac{\partial\bar{u}_{\theta}}{\partial\theta}+\frac{\partial \bar{u}_{z}}{\partial z}=0,
\end{equation}
\begin{eqnarray}
\label{eq:lpe_2D_momentum_r}
&&\frac{\partial\bar{u}_{r}}{\partial t}+\left(\bar{U}\frac{\partial }{\partial r}+\frac{\bar{V}}{r}\frac{\partial}{\partial\theta}+\bar{W}\frac{\partial }{\partial z}+\frac{\partial\bar{U}}{\partial r}\right)\bar{u}_{r}-\frac{2\bar{V}}{r}\bar{u}_{\theta}+\frac{\partial \bar{U}}{\partial z}\bar{u}_{z}=-\frac{\partial \bar{p}}{\partial r}+\frac{1}{Re}\left[\left({\nabla}^{2}-\frac{1}{r^{2}}\right)\bar{u}_{r}-\frac{2}{r^{2}}\frac{\partial\bar{u}_{\theta}}{\partial\theta}\right],\nonumber\\
\end{eqnarray}
\begin{equation}
\label{eq:lpe_2D_momentum_theta}
\frac{\partial\bar{u}_{\theta}}{\partial t}+\left(\frac{\partial \bar{V}}{\partial r}+\frac{\bar{V}}{r}\right)\bar{u}_{r}+\left(\bar{U}\frac{\partial }{\partial r}+\frac{\bar{V}}{r}\frac{\partial}{\partial\theta}+\bar{W}\frac{\partial }{\partial z}+\frac{\bar{U}}{r}\right)\bar{u}_{\theta}+\frac{\partial \bar{V}}{\partial z}\bar{u}_{z}=-\frac{1}{r}\frac{\partial\bar{p}}{\partial\theta}+\frac{1}{Re}\left[\left({\nabla}^{2}-\frac{1}{r^{2}}\right)\bar{u}_{\theta}+\frac{2}{r^{2}}\frac{\partial\bar{u}_{r}}{\partial\theta}\right],
\end{equation}
\begin{equation}
\label{eq:lpe_2D_momentum_z}
\frac{\partial\bar{u}_{z}}{\partial t}+\frac{\partial\bar{W}}{\partial r}\bar{u}_{r}+\left(\bar{U}\frac{\partial }{\partial r}+\frac{\bar{V}}{r}\frac{\partial}{\partial\theta}+\bar{W}\frac{\partial }{\partial z}+\frac{\partial\bar{W}}{\partial z}\right)\bar{u}_{z}=-\frac{\partial \bar{p}}{\partial z}+N^{2}\bar{T}+\frac{1}{Re}{\nabla}_{m}^{2}\bar{u}_{z},
\end{equation}
\begin{equation}
\label{eq:lpe_2D_energy}
\frac{\partial\bar{T}}{\partial t}+\frac{\partial \bar{\mathcal{T}}}{\partial r}\bar{u}_{r}+\left(\frac{\partial\bar{\mathcal{T}}}{\partial z}+1\right)\bar{u}_{z}+\left(\bar{U}\frac{\partial }{\partial r}+\frac{\bar{V}}{r}\frac{\partial}{\partial\theta}+\bar{W}\frac{\partial }{\partial z}\right)\bar{T}=\frac{1}{RePr}{\nabla}^{2}\bar{T},
\end{equation}

By applying the normal mode (\ref{eq:normal_mode_global}), we obtain the equations in modal form as follows:
\begin{equation}
\frac{\partial \hat{u}_{m}}{\partial r}+\frac{\hat{u}_{m}}{r}+\frac{\mathrm{i}m\hat{v}_{m}}{r}+\frac{\partial \hat{w}_{m}}{\partial z}=0,
\end{equation}
\begin{eqnarray}
&&-\mathrm{i}\omega_{m}\hat{u}_{m}+\left(\mathcal{L}+\frac{\partial\bar{U}}{\partial r}\right)\hat{u}_{m}-\frac{2\bar{V}}{r}\hat{v}_{m}+\frac{\partial \bar{U}}{\partial z}\hat{w}_{m}=-\frac{\partial \hat{p}_{m}}{\partial r}+\frac{1}{Re}\left[\left(\hat{\nabla}_{m}^{2}-\frac{1}{r^{2}}\right)\hat{u}_{m}-\frac{2\mathrm{i}m\hat{v}_{m}}{r^{2}}\right],\nonumber\\
\end{eqnarray}
\begin{equation}
-\mathrm{i}\omega_{m}\hat{v}_{m}+\left(\frac{\partial \bar{V}}{\partial r}+\frac{\bar{V}}{r}\right)\hat{u}_{m}+\left(\mathcal{L}+\frac{\bar{U}}{r}\right)\hat{v}_{m}+\frac{\partial \bar{V}}{\partial z}\hat{w}_{m}=-\frac{\mathrm{i}m\hat{p}_{m}}{r}+\frac{1}{Re}\left[\left(\hat{\nabla}_{m}^{2}-\frac{1}{r^{2}}\right)\hat{v}_{m}+\frac{2\mathrm{i}m\hat{u}_{m}}{r^{2}}\right],
\end{equation}
\begin{equation}
-\mathrm{i}\omega_{m}\hat{w}_{m}+\frac{\partial\bar{W}}{\partial r}\hat{u}_{m}+\left(\mathcal{L}+\frac{\partial\bar{W}}{\partial z}\right)\hat{w}_{m}=-\frac{\partial \hat{p}_{m}}{\partial z}+N^{2}\hat{T}_{m}+\frac{1}{Re}\hat{\nabla}_{m}^{2}\hat{w}_{m},
\end{equation}
\begin{equation}
-\mathrm{i}\omega_{m}\hat{T}_{m}+\frac{\partial \bar{\mathcal{T}}}{\partial r}\hat{u}_{m}+\left(\frac{\partial\bar{\mathcal{T}}}{\partial z}+1\right)\hat{w}_{m}+\mathcal{L}\hat{T}_{m}=\frac{1}{RePr}\hat{\nabla}_{m}^{2}\hat{T}_{m},
\end{equation}
where $\mathcal{L}=\bar{U}\frac{\partial }{\partial r}+\frac{\mathrm{i}m\bar{V}}{r}+\bar{W}\frac{\partial }{\partial z}$ is the advection operator and $\hat{\nabla}_{m}^{2}=\partial^{2}/\partial r^{2}+(1/r)(\partial/\partial r)+m^{2}/r^{2}+\partial^{2}/\partial z^{2}$ is the modal Laplacian operator. 
After eliminating the pressure $\hat{p}_{m}$, the above equations can be simplified as the eigenvalue problem (\ref{eq:evp_biglobal}) where the operator matrices $\mathcal{A}_{m}$ and $\mathcal{B}_{m}$ are defined as
\begin{equation}
\mathcal{A}_{m}=\left[
\begin{array}{ccc}
\mathcal{A}_{11,m} & \mathcal{A}_{12,m} & 0\\
\mathcal{A}_{21,m} & \mathcal{A}_{22,m} & 0\\
0 & 0 & 1
\end{array}
\right],~
\mathcal{B}_{m}=\left[
\begin{array}{ccc}
\mathcal{B}_{11,m} & \mathcal{B}_{12,m} & 0\\
\mathcal{B}_{21,m} & \mathcal{B}_{22,m} & N^{2}\\
-\frac{\partial\bar{\mathcal{T}}}{\partial r} & -1-\frac{\partial\bar{\mathcal{T}}}{\partial z} & -\mathcal{L}+\frac{\hat{\nabla}_{m}^{2}}{RePr}
\end{array}
\right],~
\end{equation}
where
\begin{eqnarray}
\mathcal{A}_{11,m}&=&1-\frac{1}{m^{2}}\frac{\partial}{\partial r}\left(r+r^{2}\frac{\partial}{\partial r}\right),~
\mathcal{A}_{12,m}=-\frac{1}{m^{2}}\frac{\partial}{\partial r}\left(r^{2}\frac{\partial}{\partial z}\right),\nonumber\\
\mathcal{A}_{21,m}&=&-\frac{1}{m^{2}}\frac{\partial}{\partial z}\left(r+r^{2}\frac{\partial}{\partial r}\right),~
\mathcal{A}_{22,m}=1-\frac{1}{m^{2}}\frac{\partial}{\partial z}\left(r^{2}\frac{\partial}{\partial z}\right),
\end{eqnarray}
\begin{eqnarray}
\mathcal{B}_{11,m}&=&-\mathcal{L}-\frac{\partial \bar{U}}{\partial r}+\frac{1}{Re}\left(\hat{\nabla}_{m}^{2}+\frac{1}{r^{2}}+\frac{2}{r}\frac{\partial}{\partial r}\right)+\frac{2\mathrm{i}}{m}\left(\frac{\bar{V}}{r}+\bar{V}\frac{\partial}{\partial r}\right)\nonumber\\
&&+\frac{1}{m^{2}}\frac{\partial}{\partial r}\left[\left(r\mathcal{L}+\bar{U}-\frac{r\hat{\nabla}_{m}^{2}}{Re}+\frac{1}{Rer}\right)\left(1+r\frac{\partial}{\partial r}\right)\right]-\frac{\mathrm{i}}{m}\frac{\partial}{\partial r}\left(\bar{V}+r\frac{\partial\bar{V}}{\partial r}-\frac{2\mathrm{i}m}{Rer}\right),\nonumber\\
\mathcal{B}_{12,m}&=&\frac{\mathrm{i}}{m}\left[2\bar{V}\frac{\partial}{\partial z}-\frac{\partial}{\partial r}\left(r\frac{\partial\bar{V}}{\partial z}\right)\right]-\frac{\partial\bar{U}}{\partial z}+\frac{2}{Rer}\frac{\partial}{\partial z}+\frac{1}{m^{2}}\frac{\partial}{\partial r}\left[\left(r\mathcal{L}+\bar{U}-\frac{r\hat{\nabla}_{m}^{2}}{Re}+\frac{1}{Rer}\right)\left(r\frac{\partial}{\partial z}\right)\right],\nonumber\\
\mathcal{B}_{21,m}&=&-\frac{\partial\bar{W}}{\partial r}+\frac{1}{m^{2}}\frac{\partial}{\partial z}\left[\left(r\mathcal{L}+\bar{U}-\frac{r\hat{\nabla}_{m}^{2}}{Re}+\frac{1}{Rer}\right)\left(1+r\frac{\partial}{\partial r}\right)\right]-\frac{\mathrm{i}}{m}\frac{\partial}{\partial z}\left(\bar{V}+r\frac{\partial\bar{V}}{\partial r}-\frac{2\mathrm{i}m}{Rer}\right),\nonumber\\
\mathcal{B}_{22,m}&=&-\mathcal{L}-\frac{\partial\bar{W}}{\partial z}+\frac{\hat{\nabla}_{m}^{2}}{Re}+\frac{1}{m^{2}}\frac{\partial}{\partial z}\left[\left(r\mathcal{L}+\bar{U}-\frac{r\hat{\nabla}_{m}^{2}}{Re}+\frac{1}{Rer}\right)\left(r\frac{\partial}{\partial z}\right)\right]-\frac{\mathrm{i}}{m}\frac{\partial}{\partial z}\left(r\frac{\partial\bar{V}}{\partial z}\right).
\end{eqnarray}

\section{Validation}
\label{sec:App_validation}
\begin{figure}
  \centerline{
  \includegraphics[height=5cm]{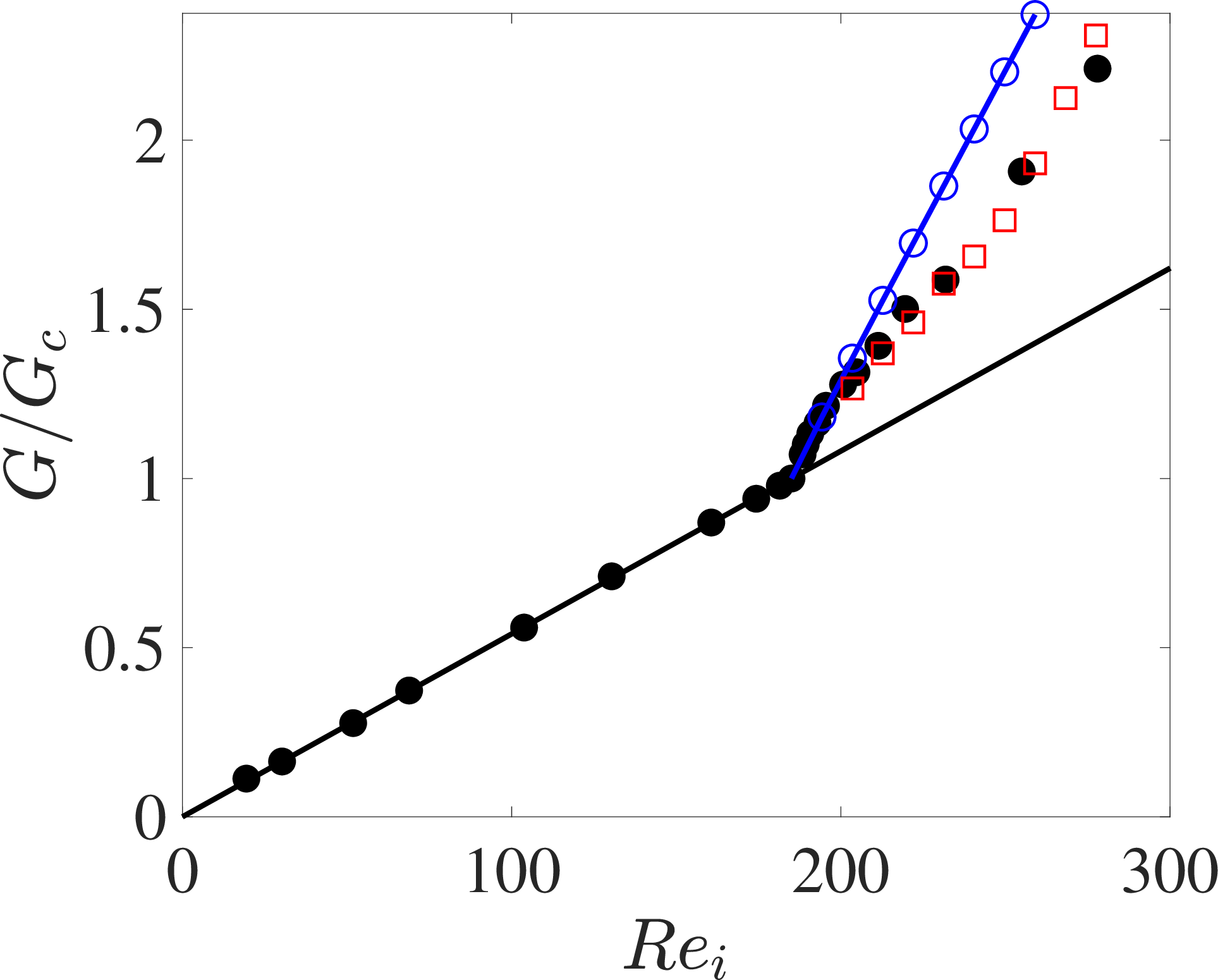}
    \includegraphics[height=4.95cm]{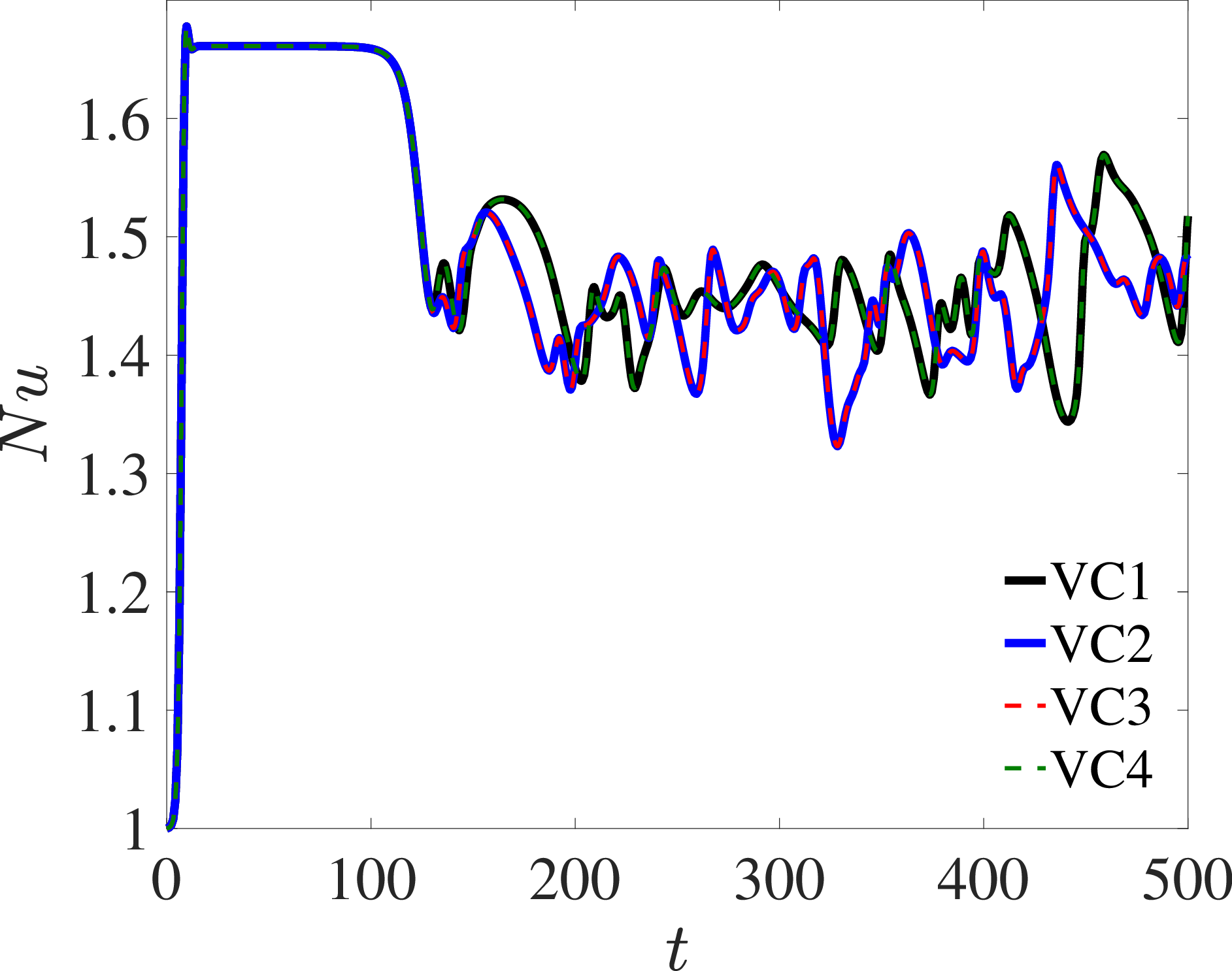}
    }
  \caption{(Left) Normalised torque $G/G_{c}$ at the inner cylinder versus the Reynolds number $Re_{i}$ for an unstratified case $N=0$ at $\mu=0$ and $\eta=0.95$. Results are from the laminar solution (black solid line), experiments (black filled circles) from \citet{Donnelly1960}, axisymmetric 2D DNS (blue line with empty circles) and non-axisymmetric 3D DNS (red empty squares). 
  (Right) The Nusselt number $Nu$ versus time $t$ for different Validation Cases (VCs). }
\label{fig:validation}
\end{figure}
\subsection{Validation against experiments}
To validate the current DNS code, 2D and 3D DNS are conducted for unstratified cases with $N=0$, $\mu=0$ and $\eta=0.95$. 
The DNS results are compared with experimental results from \citet{Donnelly1960}, which details an empirical relation for the torque measured at the inner cylinder. 
In this case, the critical Reynolds number is $Re_{i,c}=185$ for the axisymmetric mode with $k_{d}=3.128$ \citep[according to][as well as our 1D LSA computation]{DiPrima1984}.
We conduct two types of DNS: an axisymmetric 2D DNS by considering only the axisymmetric modes with $m=0$ (e.g. $M=0$ and $N_{\theta}=1$ in the DNS) and a non-axisymmetric 3D DNS by considering both axisymmetric and non-axisymmetric modes. 
For both DNS, one periodic length $\lambda_{z}=2\pi/k$ is considered as the axial domain length. 
At each $Re_{i}$, the most unstable axisymmetric mode with $k_{d}=3.128$ with its modal energy $\tilde{E}_{01}=5\times10^{-7}$ is considered as an initial condition in the axisymmetric DNS. 
In the non-axisymmetric DNS, a non-axisymmetric mode with $(m,k_{d})=(1,3.128)$ with a smaller energy $\tilde{E}_{11}=5\times10^{-9}$ is added to the initial condition.
From experiments, \citet{Donnelly1960} presents the relation between the torque and the Reynolds number $Re_{i}$. 
In Figure \ref{fig:validation} (left), we plot the non-dimensional torque $G$ normlised by the torque $G_{c}$ at the critical Reynolds number $Re_{i,c}=185$ to facilitate the comparison between their experiments and our DNS.
In the range $1<\mathcal{R}_{c}<1.1$ (i.e. $185<Re_{i}<203.5$), the axisymmetric DNS demonstrating the saturation of centrifugal instability with axisymmetric Taylor vortices agree well the experimental results on the torque. 
For $\mathcal{R}_{c}>1.1$ (i.e. $Re_{i}>203.5$), non-axisymmetric modes also develop and generate more complex interaction between modes and base state, leading to a new saturated state in oscillatory motion known as the wavy Taylor vortices. 
This explains the torque difference between the axisymmetric 2D and non-axisymmetric 3D DNS cases. 
As shown in Figure \ref{fig:validation} (left), the torque measured from experiments agrees well with the prediction from the DNS.    
\subsection{Validation of numerical resolution and domain size}
\begin{table}
  \begin{center}
\def~{\hphantom{0}}
  \begin{tabular}{cccccccccccc}
      Case & $N_{r}$ & $N_{\theta}$ & $N_{z}$ & $L_{z}$  \\[3pt]
       VC1 & 60 & 33 & 33 & $2\pi/k$ \\
       VC2 & 120 & 65 & 65 & $2\pi/k$\\
       VC3 & 120 & 65 & 129 & $4\pi/k$\\
       VC4 & 60 & 33 & 129 & $8\pi/k$\\
  \end{tabular}
  \caption{Numerical parameters for different Validation Cases (VCs) at $Re_{i}=200$, $N=1$, $Pr=1$, $k=30.6$.}
  \label{tab:VC}
  \end{center}
\end{table}
For Case 3 with $(Re_{i},N,Pr)=(200,1,1)$, which is the case where the flow becomes chaotic, different numerical resolutions and domain sizes are tested as detailed in Table \ref{tab:VC} where Validation Case (VC) 2 is the same as Case 3 and other VCs have lower resolutions or longer domain lengths than VC2. 
As shown in Figure \ref{fig:validation}(right), the Nusselt number $Nu$, which is an averaged quantity over the axial domain length, is the same for every case at saturation and before the flow becomes chaotic. 
The fluctuation behavior of $Nu$ is the same for VC1 and VC4 (low resolution cases) and for VC2 and VC3 (high resolution cases).
This implies that the choice of the axial domain size as one periodic axial length is validated for Case 3. 
Although the fluctuation is different between low and high resolution cases, their time-averaged $Nu$ do not vary significantly, thus the mean Nusselt numbers like the ones in Figure \ref{fig:torque_all} may vary insignificantly with resolutions. 
This aspect should, however, be further validated if the flow becomes fully turbulent at higher Reynolds numbers. 

\bibliography{jfm}

\bibliographystyle{jfm}


\end{document}